\documentclass[aps,pra,showpacs,amssymb,nofootinbib,superscriptaddress,twocolumn,longbibliography]{revtex4-1}
\usepackage[T1]{fontenc}
\usepackage[english]{babel}
\usepackage[utf8]{inputenc}
\usepackage[colorlinks,breaklinks]{hyperref}
\usepackage{float}
\usepackage{tabularx}
\usepackage{graphicx}
\usepackage{amsthm}
\usepackage{bbm}
\usepackage{tikz}
\usepackage[toc,page]{appendix}
\usepackage{titlesec}
\usepackage{xcolor}

\makeatletter
\newcommand\org@hypertarget{}
\let\org@hypertarget\hypertarget
\renewcommand\hypertarget[2]{%
  \Hy@raisedlink{\org@hypertarget{#1}{}}#2%
  }
\makeatother
\hypersetup{
	bookmarksnumbered,
	pdfstartview={FitH},
	citecolor={darkgreen},
	linkcolor={darkred},
	urlcolor={darkblue},
	pdfpagemode={UseOutlines}}
\definecolor{darkgreen}{RGB}{50,190,50}
\definecolor{darkblue}{RGB}{0,0,190}
\definecolor{darkred}{RGB}{238,0,0}

\usepackage[framemethod=tikz]{mdframed}
\definecolor{mycolor}{RGB}{12, 104, 180}
\definecolor{mycolor2}{RGB}{168, 188, 204}
\newmdenv[innerlinewidth=0.5pt, roundcorner=4pt,linecolor=mycolor,innerleftmargin=6pt,
innerrightmargin=6pt,innertopmargin=6pt,innerbottommargin=6pt]{mybox}

\usepackage{paracol}
\usepackage{tcolorbox}

\newcommand\tabletitlefontsize{\fontsize{9pt}{11pt}\selectfont}

\tcbset{colback=mycolor2!5 , colframe=mycolor, fonttitle=\bfseries\tabletitlefontsize,float=htb}
\newtcolorbox[blend into=figures]{boxfigure}[3][]
{ float*=ht,width=\textwidth,lower separated=false, center upper,
title={#2},label= fig:#3,#1}
\newtcolorbox[blend into=figures]{smallboxfigure}[3][]
{float=ht,lower separated=false, blend before title=colon hang,
title={#2}, label= fig:#3 ,#1}
\newtcolorbox{smallbox}[3][]
{float=ht,lower separated=false, blend before title=colon hang,
title={#2}, label= fig:#3 ,#1}
\newtcolorbox[blend into=tables]{smallboxtable}[3][]
{float=h,lower separated=false, blend before title=colon,left=0pt,
title={#2}, label=table:#3 ,#1}
\newtcolorbox[blend into=tables]{bigboxtable}[3][]
{float*=t,lower separated=false, blend before title=colon hang, width = 2\linewidth,
title={#2}, label= table:#3 ,#1}
\newcolumntype{Z}{|>{\centering\arraybackslash}X}

\usepackage{MnSymbol}
\usepackage{enumerate}
\hypersetup{
	bookmarksnumbered,
	pdfstartview={FitH},
	citecolor={darkgreen},
	linkcolor={darkred},
	urlcolor={darkblue},
	pdfpagemode={UseOutlines}}
\definecolor{darkgreen}{RGB}{50,190,50}
\definecolor{darkblue}{RGB}{0,0,190}
\definecolor{darkred}{RGB}{238,0,0}
\usepackage{soul}

\newcommand{\be}{\begin{equation}}
\newcommand{\ee}{\end{equation}}
\newcommand{\ben}{\begin{equation*}}
\newcommand{\een}{\end{equation*}}
\newcommand{\bea}{\begin{eqnarray}}
\newcommand{\eea}{\end{eqnarray}}

\newcommand{\tr}{\textnormal{Tr}}

\newcommand{\ket}[1]{\ensuremath{\left|\right.\!{#1}\!\left.\right\rangle}}

\newcommand{\bra}[1]{\ensuremath{\left\langle\right.\!{#1}\!\left.\right|}}

\newcommand{\ketbra}[2]{\ensuremath{|{#1}\rangle\langle{#2}|}}

\newcommand{\djj}{d\kern-0.4em\char"16\kern-0.1em}

\definecolor{magenta}{rgb}{1.0, 0.0, 0.56}

\newcommand{\bibunitbreak}[1]{}

\usepackage{multirow}

\begin{document}
\title{Inverse-design of high-dimensional quantum optical circuits in a complex medium}

\author{Suraj Goel{$ ^\text{a}$}}
    \email[Email address: ]{gs74@hw.ac.uk}
    \affiliation{Institute of Photonics and Quantum Sciences, Heriot-Watt University, Edinburgh, UK}

\author{Saroch Leedumrongwatthanakun{$ ^\text{a}$}}
    \affiliation{Institute of Photonics and Quantum Sciences, Heriot-Watt University, Edinburgh, UK}

\author{Natalia Herrera Valencia}
    \affiliation{Institute of Photonics and Quantum Sciences, Heriot-Watt University, Edinburgh, UK}
    
\author{Will McCutcheon}
    \affiliation{Institute of Photonics and Quantum Sciences, Heriot-Watt University, Edinburgh, UK}
    
\author{Armin Tavakoli}
    \affiliation{Physics Department, Lund University, Box 118, 22100 Lund, Sweden}
    
\author{Claudio Conti}
    \affiliation{Department of Physics, University Sapienza, Piazzale Aldo Moro 2, 00185 Rome, Italy}
    
\author{Pepijn W. H. Pinkse}
    \affiliation{MESA+ Institute for Nanotechnology, University of Twente, P.O. Box 217, 7500 AE Enschede, The Netherlands}
    
\author{Mehul Malik}
    \email[Email address: ]{m.malik@hw.ac.uk}
    \homepage[Website: ]{http://bbqlab.org}
    \affiliation{Institute of Photonics and Quantum Sciences, Heriot-Watt University, Edinburgh, UK}

\footnotetext{These authors contributed equally to this work}

\begin{abstract}
Programmable optical circuits are an important tool in developing quantum technologies such as transceivers for quantum communication and integrated photonic chips for quantum information processing. Maintaining precise control over every individual component becomes challenging at large scales, leading to a reduction in the quality of the operations performed. In parallel, minor imperfections in circuit fabrication are amplified in this regime, dramatically inhibiting their performance. Here we use inverse-design techniques to embed optical circuits in the higher-dimensional space of a large, ambient mode-mixer such as a commercial multi-mode fibre. This approach allows us to forgo control over each individual circuit element, while retaining a high degree of programmability. We use our circuits as quantum gates to manipulate high-dimensional spatial-mode entanglement in up to seven dimensions. Their programmability allows us to turn a multi-mode fibre into a generalised multi-outcome measurement device, allowing us to both transport and certify entanglement within the transmission channel. With the support of numerical simulations, we show our method is a scalable approach to obtaining high circuit fidelity with a low circuit depth by harnessing the resource of a high-dimensional mode-mixer. 
\end{abstract}
\maketitle
A programmable optical circuit is an essential element for applications in fields as diverse as sensing, communication, neuromorphic computing, artificial intelligence, and quantum information processing~\cite{Shastri2021,Bogaerts2020,Wetzstein2020}. The production of large, reprogrammable circuits is of paramount importance for coherently processing information encoded in light. However, there remain many challenges associated with the design, manufacture, and control of such circuits, which normally require a sophisticated mesh of interferometers constructed with bulk or integrated optics~\cite{Harris2018,Bogaerts2020}. Conventional construction of these circuits exploits universal programmability on two-dimensional unitary spaces to construct arbitrary high-dimensional unitary transformations~\cite{Reck1994,Miller2013a,Clements2016,Kumar2020}, herein referred to as the ``bottom-up'' technique (Fig.~1a). Over the past two decades, the technological development of integrated programmable circuits has enabled universal programmability in up to 20 path-encoded modes, containing a few hundreds of optical components on the same chip~\cite{Carolan2015,Wang2019,Tang2021,Taballione2022}.

\begin{figure*}[htp]
\centering\includegraphics[width=\textwidth]{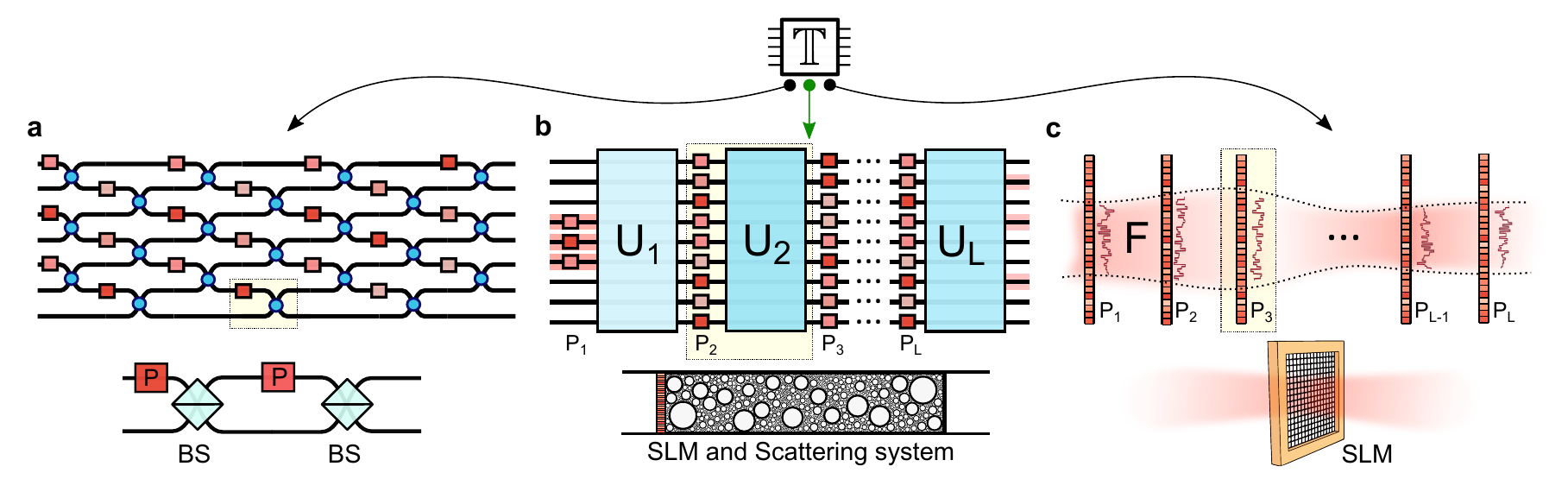}
\caption{\textbf{Figure 1: Design of programmable optical circuits:} A general linear transformation $\mathbb{T}$ can be implemented via (a) the conventional ``bottom-up'' approach, where the circuit is constructed from units consisting of beamsplitters (BS) and phase shifters (P), or (b) the herein proposed ``top-down'' approach, where a target $d$-dimensional linear circuit is embedded within a large ambient mode-mixer with dimension $n>d$, where $n-d$ auxiliary modes serve as an additional resource. The technique harnesses random unitaries $U_j$ (such as a complex scattering system) interspersed with controllable phase planes $P_j$ implemented via spatial light modulators (SLMs), which provide programmability over the target circuit. A similar approach is used in (c) multi-plane light converters, where the random unitaries are replaced with free-space propagation $F$.}
\label{fig:concept}
\end{figure*}

Imperfections in optical circuits such as scattering loss, unbalanced mode-mixing, and undesired cross-talk between modes are problematic as they reduce the accuracy and success probability of the implemented circuit~\cite{Miller2015,Burgwal2017,Pai2018,Fang2019,Hamerly2021}. These issues become increasingly challenging in large dimensions, as the number of optical elements grows quadratically with the size of the circuit~\cite{Harris2018,Bogaerts2020}. Such imperfections can be addressed to some extent by increasing the depth of the circuit through the introduction of additional phase shifters and beamsplitters~\cite{Miller2015,Burgwal2017,Pai2018,Fldzhyan2020,Tanomura2021}. However, these additional components necessitate additional control, further increasing the demands associated with circuit complexity. 

In this work, we present an alternative solution where the optical circuit is embedded in a higher-dimensional space of a large ambient mode-mixer such as a random scattering medium, placed between reprogrammable phase planes (Fig.~1b). This ``top-down'' approach harnesses the complicated scattering process within a large mode-mixer to forgo control over each individual circuit element. Instead, an inverse-design approach is used that employs algorithmic techniques to program an optical circuit with a desired functionality within the random scattering medium \cite{Molesky2018,Marcucci2020}. Similar approaches based on inverse-design have been used in multi-plane light converters (MPLC) for spatial-mode manipulation~\cite{Morizur2010,Labroille2014,Fontaine2019}, where free-space propagation is commonly used in place of a random scattering medium (Fig.~1c). Furthermore, inverse-design techniques have also enabled a variety of optical circuits tailored towards specific functionalities, ranging from designs of on-chip photonic devices~\cite{Hashimoto2005} to arrangements of bulk optical elements for fundamental quantum experiments~\cite{OAMGHZ,Krenn2016b,Melnikov2018,Krenn2020a}.

The capability to manipulate quantum states of light using large-scale programmable circuits promises a myriad of applications in quantum information science, ranging from the demonstration of computational advantage~\cite{Zhong2020} to the realisation of quantum networks~\cite{Llewellyn2020}. In this regard, high-dimensional quantum systems offer significant advantages in terms of increased information capacity and noise-resistance in quantum communication~\cite{malikboyd2014,Hu2018c,Ecker-Huber2019,Zhu2021b}, reducing multi-qubit circuit complexity~\cite{Gao2022}, while also enabling more practical tests of quantum nonlocality~\cite{Vertesi:2010bq,Srivastav2022}. While methods for the transport \cite{Valencia2020, Cao2020} and certification \cite{Bavaresco:2018gw,Friis2019, HerreraValencia2020} of high-dimensional entanglement have seen rapid progress over the past few years, scalable techniques for its precise manipulation and measurement are still lacking. As an alternative to the bottom-up approach normally implemented on integrated platforms~\cite{Wang2019}, inverse-design techniques have been used for realising quantum gates in dimensions up to $d=4$ using bulk optical interferometers \cite{Babazadeh2017,Wang2017} and in $d=5$ with multi-plane light conversion \cite{Brandt:2020er,Lib2021}. In parallel, recent advances in control over light scattering in complex media~\cite{Rotter2017,Cao2022} have enabled linear optical circuits for classical light~\cite{Huisman2015,Matthes2019} and demonstrations of programmable two-photon quantum interference \cite{Wolterink2016, Defienne2016,Leedumrongwatthanakun2020}, showing their clear potential to serve as a high-dimensional quantum photonics platform.

In this article, we harness light scattering through an off-the-shelf multi-mode fibre to program generalised quantum circuits for transverse spatial photonic modes in dimensions up to seven. We apply these circuits for the manipulation of high-dimensional entangled states of light in multiple spatial-mode bases, demonstrating high-dimensional Pauli-$\mathbb{Z}$ and $\mathbb{X}$ gates, discrete Fourier transforms, and random unitaries in the macro-pixel and OAM spatial-mode bases~\cite{HerreraValencia2020,Valencia2021}. Furthermore, in contrast with single-outcome projective measurements that are inherently inefficient~\cite{Bouchard:2018hr}, our technique realises generalised transformations to a spatially localised ``pixel'' basis, effectively turning the channel itself into a generalised multi-outcome measurement device. By harnessing this functionality, we show how the on-demand programmability of our gates enables us to both transport and certify entanglement within the same complex medium. Such multi-outcome measurements can be easily integrated with next-generation single-photon-detector arrays~\cite{Allmaras2020} and provide a key functionality in many quantum information applications, such as allowing one to overcome fair-sampling assumptions~\cite{Designolle2021}.

\begin{figure*}[htp]
\centering\includegraphics[width=0.92\textwidth]{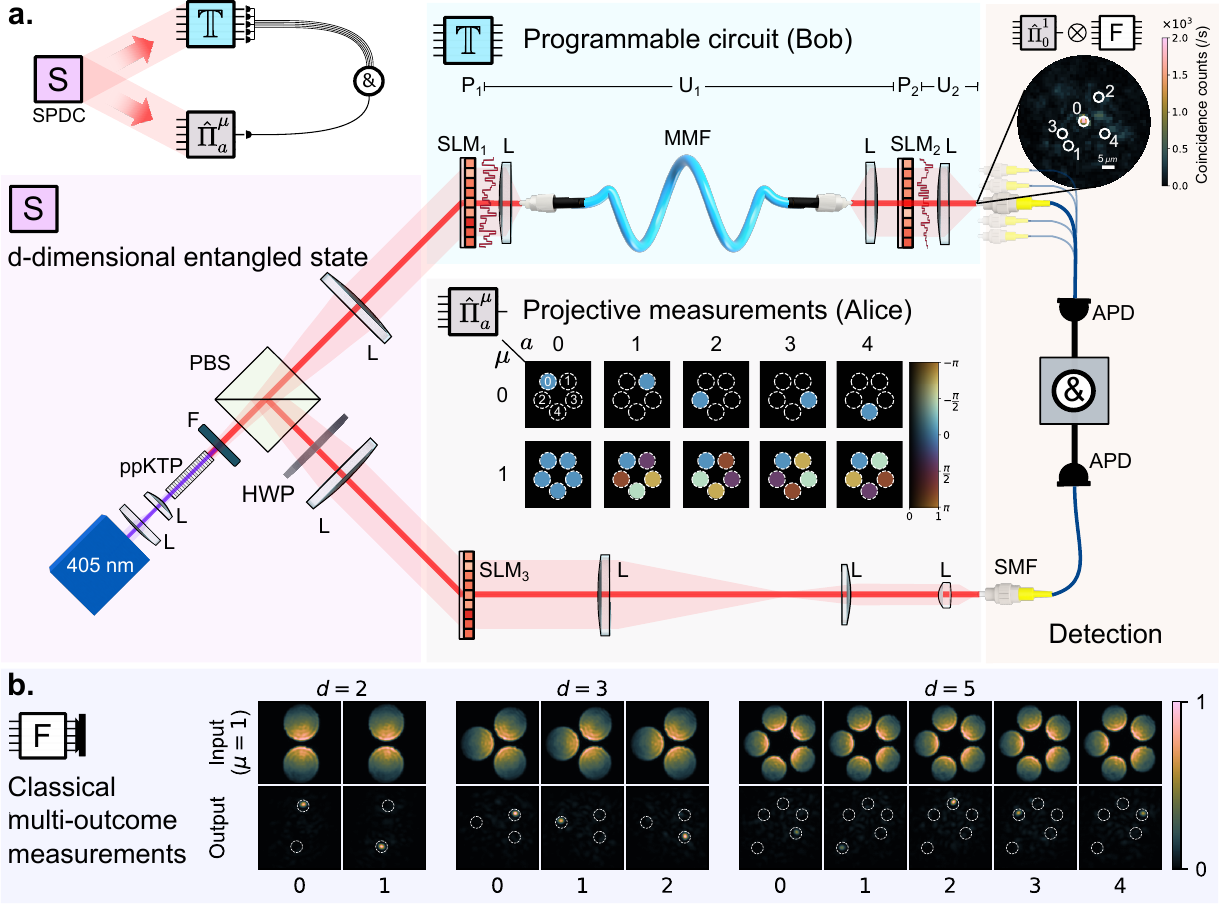}
\caption{\textbf{Figure 2: Experimental setup.} a) A high-dimensional spatially entangled two-photon state is generated via Type-II spontaneous parametric downconversion (SPDC) in a periodically poled Potassium Titanyl Phosphate (ppKTP) crystal. The two photons are spatially separated by a polarising beamsplitter (PBS) and sent to two parties, Alice and Bob. Alice performs single-outcome projective measurements $\hat{\Pi}^{\mu}_{a}$ that measure whether a photon is carrying spatial mode $a$ from modal basis $\mu$. These are performed by a combination of a spatial light modulator (SLM$_3$), single-mode fibre (SMF), and a single-photon avalanche photodiode (APD). Bob implements a top-down programmable circuit that is constructed from a multi-mode optical fibre (MMF) placed between two programmable spatial light modulators (SLM$_{1,2}$). The circuit is used to program a variety of high-dimensional quantum gates and serves as a generalised multi-outcome device. (Circular inset) A coincidence image depicting a five-outcome measurement in basis $\mu=1$ performed with the Fourier-$\mathbb{F}$ gate at Bob. The image is obtained by scanning a detector across the output of the circuit, conditioned on a measurement of $\hat{\Pi}^{\mu=1}_{a=0}$ at Alice, and shows a large intensity in mode 0 due to strong spatial-mode correlations. Coincidence detection events between Alice and Bob are registered by time-tagging electronics. b) CCD images demonstrating the operation of the Fourier-$\mathbb{F}$ gate as a multi-outcome measurement of classical macro-pixel modes prepared in basis $\mu=1$ in dimensions $d=\{2,3,5\}$. Note that while the input modes have the same amplitude for a given $d$, they are orthogonal in phase (not seen in intensity images). (L: lens, F: filter, HWP: half-wave plate).}
\label{fig:setup}
\end{figure*}

\section{\label{concept}Top-Down Programmable Circuits}

An optical circuit is described by a linear transformation $\mathbb{T}$ that maps a set of input optical modes onto a set of output modes~\cite{Miller2012a, Miller2019}. The linear circuit $\mathbb{T}$ of dimension $d$ is built from a cascade of optical mode mixers, $U$, and phase shifters, $P$. A deterministic construction can be based on a cascade of reconfigurable Mach–Zehnder interferometers, wherein $U_j$ represents the embedded balanced two-mode mixer, i.e. a 50:50 beam splitter, and the $P_j$ are phase shifters (Fig.1a)~\cite{Reck1994,Clements2016}.
As an alternative to this deterministic \textit{bottom-up} construction, the \textit{top-down} design presented here relies on the capability to harness large, complex, inter-modal mode mixers $U_j$ of dimension $n$ $(U_j \in \mathcal{U}(\mathbf{n}))$ and reconfigurable phase planes $P_j=\text{diag} (e^{i\boldsymbol\theta})$ to construct a programmable target circuit $\mathbb{T}$ of a smaller size ($d \leqslant n$) embedded within the larger mode-mixers (Fig.1b). The decomposition of such a top-down programmable circuit is represented as
\be
    \mathbf{T}\approx\prod_{j=1}^{L\leqslant \mathcal{O}(\mathbf{d})}U_j P_j,
    \label{eq1}
\ee
where $L$ is the depth of the circuit (the number of layers), and the target circuit $\mathbb{T}$ is embedded in the total transfer matrix of the system $\mathbf{T}$. Optimal choices for the large mode-mixer dimension ($n$) and the circuit depth ($L$) for a given target circuit dimension ($d$) are discussed in Section \ref{Results2}.

We experimentally construct the programmable optical circuit with a 2-metre-long graded-index multi-mode fibre (MMF) positioned between two programmable phase planes, $P_1$ and $P_2$, implemented on spatial light modulators (SLMs) as depicted in Fig.~2a. The MMF serves as a large, complex mode-mixer with dimension $n\approx200$ that provides complicated inter-modal coupling~\cite{Carpenter2014a,Ploschner2015,Xiong2018}, while the SLMs provide programmability over the circuit to be implemented. The circuit can be decomposed as $\mathbf{T}=U_2 P_2 U_1 P_1$, where $U_1$ represents the transfer matrix of the optical system consisting of the MMF and the associated coupling optics and $U_2$ is the $2f$ lens system. To construct the circuit, we begin by characterising $U_1$ in a referenceless manner via our developed technique \cite{goel2023referenceless}. Random phase patterns are displayed on the planes $P_1$ and $P_2$ and the resulting intensity speckle images are measured at the output. The data set is then used to optimise the machine learning model that describes the optical system of our experiment using a gradient descent method, which takes $\sim1$ minute to be calculated on a GPU. Once we have complete knowledge of the mode-mixer $U_1$, a given target circuit is then programmed using a solution of phase patterns obtained from the wavefront-matching (WFM) algorithm~\cite{Hashimoto2005,Sakamaki2007,Fontaine2017}. The WFM algorithm is an inverse-design technique that calculates the reconfigurable phase planes by iterating through each of them in order to maximise the overlap between a set of input fields with the desired output ones, and repeating this procedure several times \cite{goel2023referenceless}. The process of finding the optimal phase patterns takes a few seconds for a $2$-dimensional circuit to $\sim1$ minute for a $7$-dimensional circuit. For further details on transfer matrix acquisition or circuit construction, please refer to Methods. 

Using the WFM algorithm, we implement a variety of different target circuits $\mathbb{T}$ on our system including the identity-$\mathbb{I}$, high-dimensional analogues of Pauli-$\mathbb{Z}$ and $\mathbb{X}$, Fourier-$\mathbb{F}$, and random unitaries-$\mathbb{R}$ in dimensions $d=\{2,3,5,7\}$ for two different input transverse spatial mode bases (macro-pixel \cite{HerreraValencia2020} and orbital-angular-momentum \cite{Valencia2021}). Our circuits perform generalised basis transformations to a localised output ``pixel'' basis with a maximum theoretical efficiency of unity (see Section \ref{Results2}), enabling \textit{multi-outcome} measurements in any given basis. The target output modes are randomly selected from the set of all possible foci at the output of the circuit. As an example of this, Fig.~2b shows intensity images demonstrating the operation of the Fourier $\mathbb{F}$ circuit in $d=\{2,3,5\}$. This circuit simultaneously transforms the first mutually unbiased basis ($\mu=1$) of input macro-pixel modes (indistinguishable by their amplitude) into a basis of spatially localised output modes that can be detected on a camera or suitable single-photon detector array. This can be compared with conventional \textit{single-outcome} projective measurements $\hat{\Pi}^{\mu}_{a}$ shown in Fig.~2a (Alice), which project a photon in a particular mode $a$ in basis $\mu$ via the combination of a holographic spatial light modulator (SLM$_3$) and single mode fibre \cite{Bouchard:2018hr,Qassim:2014fp}. Such measurements require one to perform $d$ projections to realise a complete measurement, giving them a maximum effective efficiency of $1/d$, which is normally even lower due to device loss. We characterise our circuits by preparing a tomographically complete set of classical input modes and performing quantum process tomography (QPT, see Methods), allowing us to calculate their purity and fidelity to an ideal circuit. As a representative example, the Fourier-$\mathbb{F}$ circuit achieves fidelities of $\mathcal{F}=\{96.9\%,90.5\%,89.3\%,81.4\%\}$ in macro-pixel dimensions $d=\{2,3,5,7\}$ (see Supplementary Information for extended results).

\begin{figure*}[htp]
\centering\includegraphics[width=\textwidth]{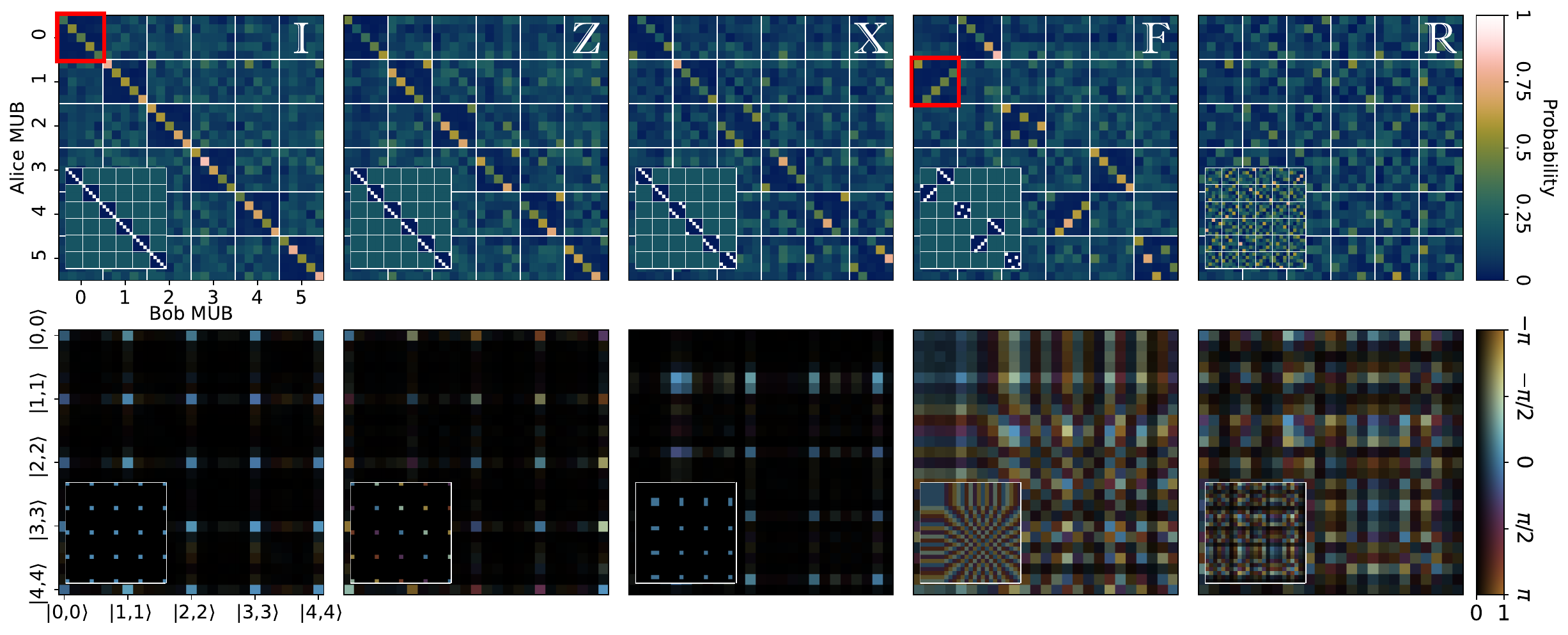}
\caption{\textbf{Figure 3: Manipulation and certification of high-dimensional entanglement}: Normalised two-photon coincidence counts (top row) and reconstructed Choi state density matrices (bottom row) corresponding to operation of the Identity $\mathbb{I}$, Pauli-$\mathbb{Z}$, Pauli-$\mathbb{X}$, Fourier $\mathbb{F}$, and random unitary $\mathbb{R}$ gates on an input two-photon, five-dimensional entangled state in the macro-pixel basis. The two-photon coincidences are measured in all six mutually unbiased bases (MUBs $\mu\in\{0...5\}$) via projective measurements at Alice and Bob. Due to state-channel duality, measurements on the input and output two-photon states can be used to perform ancilla-assisted process tomography of the gates themselves \cite{Valencia2020, Altepeter2003}. The fidelities of the reconstructed Choi states are reported in Table~\ref{table:F}. The Identity and Fourier gates enable our circuit to be used as a multi-outcome measurement device in MUBs $\mu\in\{0,1\}$ (red squares), allowing us to certify five-dimensional entanglement using the channel itself as a measurement device (see main text for details). The density matrix legend captures both the amplitude (brightness, normalised for clarity) and phase (colour).}
\label{fig:result1}
\end{figure*}

\section{\label{Results1}Applications of Quantum Gates}

We utilise our programmable circuits to manipulate and certify spatially entangled two-photon states in a range of dimensions, $d=\{2,3,5,7\}$. As shown in Fig.~2a, the two photons are generated via the process of spontaneous parametric down-conversion (SPDC) and sent to two parties, Alice and Bob. Bob's photon is locally manipulated by the programmable circuit $\mathbb{T}$, while Alice's photon is detected via single-outcome projective measurements $\hat{\Pi}^\mu_{a}$. While we have verified the operation of our circuits $\mathbb{T}$ with classical light, it is important to repeat this process with our entangled source due to slight differences between their spectral/spatial modes and experimental alignment. Here, instead of using standard quantum process tomography (QPT), we use the lesser-known method of ancilla-assisted quantum process tomography (AA-QPT) \cite{Altepeter2003,Valencia2020}. AA-QPT uses a single, well-characterised, and sufficiently strongly correlated state supported on an extended Hilbert space, along with a tomographically complete measurement, to fully characterise a process. This exploits channel-state duality, in which a channel, acting only on one photon from an ideal maximally entangled biphoton state, is fully characterised by the resultant output state, called the Choi state. In the case of non-ideal input states, the Choi state of the channel can still be recovered \cite{DAriano2003}. For more details on QST and AA-QPT, please see Methods.

We first characterise the input state in each dimension using quantum state tomography (QST).  We then use these states to recover the Choi states of the processes corresponding to the programmed circuits via AA-QPT. 
We also perform QST of the output states after operation by the circuits, directly enabling certification of properties contingent on both the input state and process, such as their entanglement and purity. Once the input states, output states, and Choi states of the processes are characterised, one can verify and quantify their performance via their fidelity to the ideal states (as defined in Methods). In this manner, our process fidelities quantify how close to ideal our circuits are, taking into account the non-maximally entangled nature of the input entangled state, whilst our output state fidelities allow independent verification that they preserve high-dimensional entanglement.
To quantify the effect of measurement imperfections on our tomographic procedures, we perform an independent characterisation of the measurement apparatus, as well as a Monte-Carlo sampling of these imperfections in QST and AA-QPT (See Supplementary Information) to arrive at realistic error bounds on these fidelities.

To showcase the versatility of our platform, we program 296 instances of all gates, sampling from different output foci in different dimensions and input bases. Fig.~3 (top row) shows examples of normalised two-photon coincidence count data in all mutually unbiased bases (MUBs) for five-dimensional $\mathbb{I}$, $\mathbb{Z}$, $\mathbb{X}$, $\mathbb{F}$, and $\mathbb{R}$ gates programmed for the macro-pixel basis. Fig.~3 (bottom row) shows the density matrices of the Choi states reconstructed from these data, presenting a clear agreement with the theoretical prediction (insets).
The fidelity of the processes, quantified through the fidelity of the experimental Choi states to the ideal Choi states is reported in Table~\ref{table:F} for all dimensions in the macro-pixel basis. The fidelity of the output biphoton states to an ideal transformed state, their purity, and certified entanglement dimensionality are presented in the Supplementary Information, along with an estimation of the systematic errors arising from the measurement apparatus. It is noteworthy that all output states are seen to preserve high-dimensional entanglement after operation by the gates. For instance, the output state after the identity gate $\mathbb{I}$ has a fidelity of $\mathcal{F}(\rho_o,\ketbra{\Phi^+}{\Phi^+})= 85.3 \pm 0.7 \%$ to the maximally entangled state, which exceeds the bound of $80\%$ necessary to certify five-dimensional entanglement (the fidelity of any state $\rho$ to a maximally entangled target state $F(\rho,\ketbra{\Phi^+}{\Phi^+})>k/d$ implies an entanglement dimensionality of at least $k+1$~\cite{Bavaresco:2018gw}). Results for gates implemented in the OAM basis are also presented in the Supplementary Information.

While we have verified that these gates are able to manipulate and preserve high-dimensional entanglement, we now demonstrate how they can be used for the certification of high-dimensional entanglement. Our programmable circuit functions as a generalised multi-outcome measurement device, allowing measurements to be made both in the pixel basis $\mu=0$ (via the identity gate $\mathbb{I}$), as well as its first mutually unbiased basis (MUB) $\mu=1$ (via the Fourier gate $\mathbb{F}$) with an appropriate single-photon detector array. The coincidence counts data corresponding to a two-basis measurement obtained by programming the circuit in this manner are highlighted in Fig.~3 (red squares). However, as can be seen in Table~\ref{table:F}, our gates are not perfect. Recent work has shown how even slight imperfections in a measurement can compromise entanglement witnesses that normally assume perfect measurements \cite{Morelli2022}. Thus, it becomes necessary to take these imperfections into account when investigating the detection of entanglement. As described in the Methods, we employ a computational approach using semidefinite programming to show that our experimentally measured data can only be reproduced by a quantum model that relies on five-dimensional entanglement. Our method takes reasonable statistical and systematic errors into account and yields a quantitative certificate of the entanglement dimension that is also robust to noise. In this manner, we demonstrate how the multi-mode optical fibre can be used to both \textit{transport} as well as \textit{certify} high-dimensional entanglement. It is important to note that in general, the programmability of our circuit allows for measurements in multiple or even ``tilted'' MUBs \cite{Bavaresco:2018gw} corresponding to a non-maximally entangled target state, which would lead to increased fidelities and robustness to noise.

The coincidence image in Fig.~2 (inset) also provides information about scattering loss outside the output modes of interest. This allows us to measure the success probability of the gate operation, which is defined as the ratio of coincidence counts in the target output modes over the total coincidence counts integrated over all outputs in one polarisation channel. We perform this measurement on three randomly chosen implementations of the $\mathbb{F}$ gate in $2,3$, and $5$ dimensions and measure a success probability of $0.36\pm0.01$, $0.27\pm0.03$, and $0.18\pm0.04$ respectively (see Methods for additional details). It is worth noting that for this proof-of-concept demonstration, we only control a single polarisation channel of the multi-mode fibre, thus reducing the success probability by about half. Controlling both polarisation channels of the multi-mode fibre can increase the success probability by almost a factor of two as well as improve the fidelity of the implemented circuits. Errors in the fidelities are calculated by taking into account photon counting statistics as well as systematic errors due to misalignments. A detailed analysis of misalignments in the measurement apparatus and their effect on the reported fidelities~\cite{Rosset2012} for both QST and AA-QPT is presented in the Supplementary Information.

\begin{smallboxtable}{Quantum process fidelities of inverse-designed experimental gates to the ideal gates in the macro-pixel basis.}{F}
\begin{tabular}{c|c|c|c|c}
    \hline
   Gate & $d = 2$ & $d = 3$ & $d = 5$ & $d = 7$ \\
   \hline
$\mathbb{I}$ & $96.7 \pm 0.9\%$ & $97.4 \pm 0.7\%$ & $88.0 \pm 0.7\%$ & $71.9 \pm 1.1\%$\\
$\mathbb{Z}$ & $97.7 \pm 0.9\%$ & $96.1 \pm 0.5\%$ & $80.6 \pm 1.0\%$ & $65.2 \pm 1.0\%$\\
$\mathbb{X}$ & $97.6 \pm 0.8\%$ & $95.0 \pm 0.7\%$ & $79.2 \pm 1.0\%$ & $60.1 \pm 1.0\%$\\
$\mathbb{F}$ & $95.7 \pm 0.9\%$ & $89.3 \pm 0.8\%$ & $76.9 \pm 1.1\%$ & $58.9 \pm 0.7\%$\\
$\mathbb{R}$ & $96.8 \pm 0.7\%$ & $91.8 \pm 0.8\%$ & $80.1 \pm 1.1\%$ & $63.5 \pm 0.7\%$\\

\end{tabular}
\vspace*{0.5pt}
\footnotesize{\\*Errors are reported to one standard deviation}
\color{black}
\end{smallboxtable}

\begin{figure*}[htp]
\centering\includegraphics[width=\textwidth]{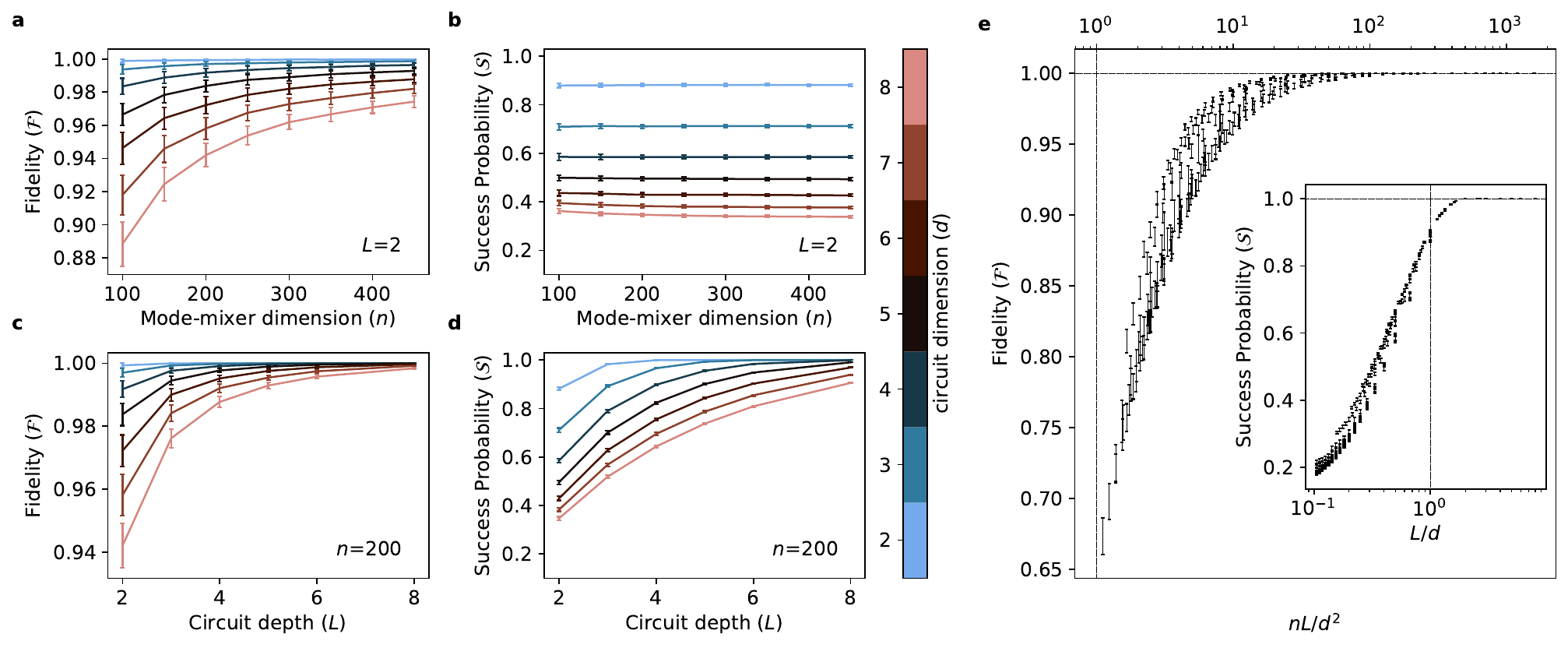}
\caption{\textbf{Figure 4: Programmability and Scalability of top-down optical circuits}: (a-b) Fidelity ($\mathcal{F}$) and success probability ($\mathcal{S}$) of a $d$-dimensional quantum optical circuit as a function of the dimension of mode-mixers ($n$) for a circuit depth of $L=2$. (c-d) $\mathcal{F}$ and $\mathcal{S}$ as a function of $L$ for $n=200$. (e) A plot of Fidelity versus the parameter $nL/d^2$ in the regime where $d/n<0.1$. The plot shows a converging trend towards unit Fidelity ($\mathcal{F}=1$), demonstrating that full programmability can be achieved by increasing either the mode-mixer dimension $n$ or the circuit depth $L$. (Inset) A plot of Success probability versus the parameter $L/d$ shows convergence to unity when the number of phase planes approaches $L\approx\mathcal{O}(d)$. A sample size of 200 is used to derive the statistics of each data-point. The data is presented as mean values $\pm$1 standard deviation.}
\label{fig:simulation_res}
\end{figure*}

\section{\label{Results2}Programmability and Scalability}

We have successfully demonstrated the ability to perform various gates in multiple spatial-mode bases in dimensions $2,3,5$, and $7$. However, as the dimension of the target gate increases, maintaining high fidelities and success probabilities becomes increasingly challenging.
It is thus imperative to examine the programmability and scalablility of our design and address ways to improve its performance and target practical experimental regimes to work in. We investigate these numerically by simulating a circuit based on Eq.~\ref{eq1} and varying three major design parameters---the dimension of the mode mixers, $n$, the dimension of the target gate, $d$, and the depth of the circuit, $L$. Multiple implementations of the $\mathbb{I}$, $\mathbb{X}$, $\mathbb{Z}$, $\mathbb{F}$, and $\mathbb{R}$ gates are simulated for specific values of the design parameters by changing the random-unitary mode mixers and the sets of input and output modes for each instance. For each gate, we calculate the fidelity with respect to the ideal target gate and the success probability. 

The key design considerations we address are: what dimension of mode mixers ($n$) should be chosen? And how many layers ($L$) is it practical to use, given the optical losses usually present at interfaces, and the experimental overhead involved? 
Figs.~4a-d depict simulation results showing how the fidelity and success probability of the top-down design ($n>d$) scale as a function of either the dimension of mode mixers ($n$) or the depth of circuit ($L$), while keeping other design parameters constant. The first observation is that increasing the size of the mode mixers ($n$) increases the circuit fidelity (Fig.~4a). This demonstrates that even when implementing practical low depth circuits, such as we have presented here, high fidelities can be reached, with high-dimensional mode mixers serving as a key resource for this behaviour. Furthermore, for $d/n<0.1$ the success probability is approximately constant with $n$ (Fig.~4b), allowing high-dimensional mode mixers to be employed without affecting the success probability.

Alongside this, we observe that increasing the depth of the circuit ($L$) increases both fidelity and success probability (Figs.~4c-d), generalising recent results into the $n>d$ regime~\cite{Saygin2020,Pereira2020}.
In practise, however, scaling up the circuit depth introduces experimental overheads in the form of propagation and interface losses and accumulation of errors. 
In general, we observe that the fidelity increases and converges to unity (Fig.~4e for $d/n < 0.1$) when the total number of reconfigurable elements $nL$ exceeds the requirement for parameterising a $d$-dimensional unitary transform $\mathcal{O}(d^2)$, thereby showing a high level of programmability of the top-down design. Furthermore, the convergence of the success probability to unity occurs when the circuit depth is approximately twice the circuit dimension ($L\approx2d$). This is because there are $d(n-d)$ amplitudes in the transformation which correspond to scattering to modes outside of the target output modes. To achieve unit success probability these must all vanish, requiring at least these many controllable parameters. The top-down approach thus presents a powerful route towards realising high-fidelity circuits by harnessing the resource of a high-dimensional mode-mixing space ($n>d$), while operating in a practical, low circuit depth regime ($L\leqslant \mathcal{O}(d)$). Full details of these simulations are presented in the Supplementary Information.

\section{\label{Conclusion}Discussion}
We have demonstrated that programmable optical circuits in the transverse-spatial domain can be reliably implemented using the top-down approach that incorporates complex scattering processes between reconfigurable phase planes. 
We verified that these gates preserve quantum coherence by certifying that high-dimensional entanglement persists after gate operations. We also demonstrate how our gates function as a generalised multi-outcome measurement device, enabling the MMF channel to both transport, manipulate, and certify high-dimensional entanglement. 
While our numerical simulations demonstrate the scalability of our technique, finding practical physical platforms for its implementation remains an open challenge. Fundamental aspects of circuit design also present important questions, such as proving that the technique can be used for the universal implementation of unitary transformations, deterministic calculations for setting phase shifters, and optimising these circuits for better performance.

Beyond the transverse spatial degree-of-freedom, our methods readily generalise to other platforms where phase shifters and mode-mixers can be realised. For instance, implementations of top-down designs in integrated optics will be forthcoming as random-mixed waveguides develop and low-loss reconfigurable phase shifters become available~\cite{Bruck2016,Wang2019c,Dinsdale2021}. Further developments must also address practical issues including modal dispersion and spatio-temporal mixing that are present in long multi-mode fibres and thick scattering media. These obstacles, however, also enable the extension of the top-down circuit design into the spectral-temporal domain~\cite{jha2008,Lukens2016,Mounaix2016,Lu2018,Mounaix2020,Lu2022}. By demonstrating the practical realisation of high-dimensional programmable optical circuits---within the transmission channel itself---our work overcomes a significant hurdle facing the adoption of high-dimensional encoding in quantum communication systems, and paves the way for practical implementations of programmable optical circuits in various near-term photonic and quantum technologies including sensing and computation.\\


\begin{acknowledgements}
  This work was made possible by financial support from the QuantERA ERA-NET Co-fund (FWF Project I3773-N36). M.M., S.G., S.L., N.H.V., and W.M. acknowledge financial support from the UK Engineering and Physical Sciences Research Council (EPSRC) (EP/P024114/1) and European Research Council (ERC) Starting grant PIQUaNT (950402). A.T. acknowledges financial support from the Wenner-Gren Foundation and the Knut and Alice Wallenberg Foundation through the Wallenberg Centre for Quantum Technology (WACQT). Colormaps for all the figures in this article are adopted from~\cite{crameri_fabio_2021_5501399}. 
\end{acknowledgements}

\section*{Author Contributions}
M.M.~conceived the research and supervised the project. M.M.~and S.L.~designed the experiment. S.G., S.L.~and N.H.V.~performed the experiment. S.G., S.L., N.H.V., W.M., A.T., and C.C. developed theoretical methods. S.G.~and S.L.~performed the simulations and numerical study. S.G., S.L., N.H.V., W.M., and A.T.~analysed the data. All authors contributed to writing the manuscript.

\section*{Competing interests}
The authors declare no competing interests.


\bibliographystyle{apsrev4-1fixed_with_article_titles_full_names}
\bibliography{LogicGatesv2}

\begin{thebibliography}{102}%
\makeatletter
\providecommand \@ifxundefined [1]{%
 \@ifx{#1\undefined}
}%
\providecommand \@ifnum [1]{%
 \ifnum #1\expandafter \@firstoftwo
 \else \expandafter \@secondoftwo
 \fi
}%
\providecommand \@ifx [1]{%
 \ifx #1\expandafter \@firstoftwo
 \else \expandafter \@secondoftwo
 \fi
}%
\providecommand \natexlab [1]{#1}%
\providecommand \enquote  [1]{#1}%
\providecommand \bibnamefont  [1]{#1}%
\providecommand \bibfnamefont [1]{#1}%
\providecommand \citenamefont [1]{#1}%
\providecommand \href@noop [0]{\@secondoftwo}%
\providecommand \href [0]{\begingroup \@sanitize@url \@href}%
\providecommand \@href[1]{\@@startlink{#1}\@@href}%
\providecommand \@@href[1]{\endgroup#1\@@endlink}%
\providecommand \@sanitize@url [0]{\catcode `\\12\catcode `\$12\catcode
  `\&12\catcode `\#12\catcode `\^12\catcode `\_12\catcode `\%12\relax}%
\providecommand \@@startlink[1]{}%
\providecommand \@@endlink[0]{}%
\providecommand \url  [0]{\begingroup\@sanitize@url \@url }%
\providecommand \@url [1]{\endgroup\@href {#1}{\urlprefix }}%
\providecommand \urlprefix  [0]{URL }%
\providecommand \Eprint [0]{\href }%
\providecommand \doibase [0]{http://dx.doi.org/}%
\providecommand \selectlanguage [0]{\@gobble}%
\providecommand \bibinfo  [0]{\@secondoftwo}%
\providecommand \bibfield  [0]{\@secondoftwo}%
\providecommand \translation [1]{[#1]}%
\providecommand \BibitemOpen [0]{}%
\providecommand \bibitemStop [0]{}%
\providecommand \bibitemNoStop [0]{.\EOS\space}%
\providecommand \EOS [0]{\spacefactor3000\relax}%
\providecommand \BibitemShut  [1]{\csname bibitem#1\endcsname}%
\let\auto@bib@innerbib\@empty
\bibitem [{\citenamefont {Shastri}\ \emph {et~al.}(2021)\citenamefont
  {Shastri}, \citenamefont {Tait}, \citenamefont {{Ferreira de Lima}},
  \citenamefont {Pernice}, \citenamefont {Bhaskaran}, \citenamefont {Wright},\
  and\ \citenamefont {Prucnal}}]{Shastri2021}%
  \BibitemOpen
  \bibfield  {author} {\bibinfo {author} {\bibfnamefont {Bhavin~J}\
  \bibnamefont {Shastri}}, \bibinfo {author} {\bibfnamefont {Alexander~N}\
  \bibnamefont {Tait}}, \bibinfo {author} {\bibfnamefont {T.}~\bibnamefont
  {{Ferreira de Lima}}}, \bibinfo {author} {\bibfnamefont {Wolfram H~P}\
  \bibnamefont {Pernice}}, \bibinfo {author} {\bibfnamefont {Harish}\
  \bibnamefont {Bhaskaran}}, \bibinfo {author} {\bibfnamefont {C~D}\
  \bibnamefont {Wright}}, \ and\ \bibinfo {author} {\bibfnamefont {Paul~R}\
  \bibnamefont {Prucnal}},\ }\emph {\enquote {\bibinfo {title} {{Photonics for
  artificial intelligence and neuromorphic computing}},}\ }\href {\doibase
  10.1038/s41566-020-00754-y} {\bibfield  {journal} {\bibinfo  {journal}
  {Nature Photonics}\ }\textbf {\bibinfo {volume} {15}},\ \bibinfo {pages}
  {102} (\bibinfo {year} {2021})}\BibitemShut {NoStop}%
\bibitem [{\citenamefont {Bogaerts}\ \emph {et~al.}(2020)\citenamefont
  {Bogaerts}, \citenamefont {P{\'{e}}rez}, \citenamefont {Capmany},
  \citenamefont {Miller}, \citenamefont {Poon}, \citenamefont {Englund},
  \citenamefont {Morichetti},\ and\ \citenamefont {Melloni}}]{Bogaerts2020}%
  \BibitemOpen
  \bibfield  {author} {\bibinfo {author} {\bibfnamefont {Wim}\ \bibnamefont
  {Bogaerts}}, \bibinfo {author} {\bibfnamefont {Daniel}\ \bibnamefont
  {P{\'{e}}rez}}, \bibinfo {author} {\bibfnamefont {Jos{\'{e}}}\ \bibnamefont
  {Capmany}}, \bibinfo {author} {\bibfnamefont {David A~B}\ \bibnamefont
  {Miller}}, \bibinfo {author} {\bibfnamefont {Joyce}\ \bibnamefont {Poon}},
  \bibinfo {author} {\bibfnamefont {Dirk}\ \bibnamefont {Englund}}, \bibinfo
  {author} {\bibfnamefont {Francesco}\ \bibnamefont {Morichetti}}, \ and\
  \bibinfo {author} {\bibfnamefont {Andrea}\ \bibnamefont {Melloni}},\ }\emph
  {\enquote {\bibinfo {title} {{Programmable photonic circuits}},}\ }\href
  {\doibase 10.1038/s41586-020-2764-0} {\bibfield  {journal} {\bibinfo
  {journal} {Nature}\ }\textbf {\bibinfo {volume} {586}},\ \bibinfo {pages}
  {207} (\bibinfo {year} {2020})}\BibitemShut {NoStop}%
\bibitem [{\citenamefont {Wetzstein}\ \emph {et~al.}(2020)\citenamefont
  {Wetzstein}, \citenamefont {Ozcan}, \citenamefont {Gigan}, \citenamefont
  {Fan}, \citenamefont {Englund}, \citenamefont {Solja{\v{c}}i{\'{c}}},
  \citenamefont {Denz}, \citenamefont {Miller},\ and\ \citenamefont
  {Psaltis}}]{Wetzstein2020}%
  \BibitemOpen
  \bibfield  {author} {\bibinfo {author} {\bibfnamefont {Gordon}\ \bibnamefont
  {Wetzstein}}, \bibinfo {author} {\bibfnamefont {Aydogan}\ \bibnamefont
  {Ozcan}}, \bibinfo {author} {\bibfnamefont {Sylvain}\ \bibnamefont {Gigan}},
  \bibinfo {author} {\bibfnamefont {Shanhui}\ \bibnamefont {Fan}}, \bibinfo
  {author} {\bibfnamefont {Dirk}\ \bibnamefont {Englund}}, \bibinfo {author}
  {\bibfnamefont {Marin}\ \bibnamefont {Solja{\v{c}}i{\'{c}}}}, \bibinfo
  {author} {\bibfnamefont {Cornelia}\ \bibnamefont {Denz}}, \bibinfo {author}
  {\bibfnamefont {David A~B}\ \bibnamefont {Miller}}, \ and\ \bibinfo {author}
  {\bibfnamefont {Demetri}\ \bibnamefont {Psaltis}},\ }\emph {\enquote
  {\bibinfo {title} {{Inference in artificial intelligence with deep optics and
  photonics}},}\ }\href {\doibase 10.1038/s41586-020-2973-6} {\bibfield
  {journal} {\bibinfo  {journal} {Nature}\ }\textbf {\bibinfo {volume} {588}},\
  \bibinfo {pages} {39} (\bibinfo {year} {2020})}\BibitemShut {NoStop}%
\bibitem [{\citenamefont {Harris}\ \emph {et~al.}(2018)\citenamefont {Harris},
  \citenamefont {Carolan}, \citenamefont {Bunandar}, \citenamefont {Prabhu},
  \citenamefont {Hochberg}, \citenamefont {Baehr-Jones}, \citenamefont {Fanto},
  \citenamefont {Smith}, \citenamefont {Tison}, \citenamefont {Alsing},\ and\
  \citenamefont {Englund}}]{Harris2018}%
  \BibitemOpen
  \bibfield  {author} {\bibinfo {author} {\bibfnamefont {Nicholas~C.}\
  \bibnamefont {Harris}}, \bibinfo {author} {\bibfnamefont {Jacques}\
  \bibnamefont {Carolan}}, \bibinfo {author} {\bibfnamefont {Darius}\
  \bibnamefont {Bunandar}}, \bibinfo {author} {\bibfnamefont {Mihika}\
  \bibnamefont {Prabhu}}, \bibinfo {author} {\bibfnamefont {Michael}\
  \bibnamefont {Hochberg}}, \bibinfo {author} {\bibfnamefont {Tom}\
  \bibnamefont {Baehr-Jones}}, \bibinfo {author} {\bibfnamefont {Michael~L.}\
  \bibnamefont {Fanto}}, \bibinfo {author} {\bibfnamefont {A.~Matthew}\
  \bibnamefont {Smith}}, \bibinfo {author} {\bibfnamefont {Christopher~C.}\
  \bibnamefont {Tison}}, \bibinfo {author} {\bibfnamefont {Paul~M.}\
  \bibnamefont {Alsing}}, \ and\ \bibinfo {author} {\bibfnamefont {Dirk}\
  \bibnamefont {Englund}},\ }\emph {\enquote {\bibinfo {title} {{Linear
  programmable nanophotonic processors}},}\ }\href {\doibase
  10.1364/OPTICA.5.001623} {\bibfield  {journal} {\bibinfo  {journal} {Optica}\
  }\textbf {\bibinfo {volume} {5}},\ \bibinfo {pages} {1623} (\bibinfo {year}
  {2018})}\BibitemShut {NoStop}%
\bibitem [{\citenamefont {Reck}\ \emph {et~al.}(1994)\citenamefont {Reck},
  \citenamefont {Zeilinger}, \citenamefont {Bernstein},\ and\ \citenamefont
  {Bertani}}]{Reck1994}%
  \BibitemOpen
  \bibfield  {author} {\bibinfo {author} {\bibfnamefont {Michael}\ \bibnamefont
  {Reck}}, \bibinfo {author} {\bibfnamefont {Anton}\ \bibnamefont {Zeilinger}},
  \bibinfo {author} {\bibfnamefont {Herbert~J}\ \bibnamefont {Bernstein}}, \
  and\ \bibinfo {author} {\bibfnamefont {Philip}\ \bibnamefont {Bertani}},\
  }\emph {\enquote {\bibinfo {title} {{Experimental realization of any discrete
  unitary operator}},}\ }\href {\doibase 10.1103/PhysRevLett.73.58} {\bibfield
  {journal} {\bibinfo  {journal} {Physical Review Letters}\ }\textbf {\bibinfo
  {volume} {73}},\ \bibinfo {pages} {58} (\bibinfo {year} {1994})}\BibitemShut
  {NoStop}%
\bibitem [{\citenamefont {Miller}(2013)}]{Miller2013a}%
  \BibitemOpen
  \bibfield  {author} {\bibinfo {author} {\bibfnamefont {David A~B}\
  \bibnamefont {Miller}},\ }\emph {\enquote {\bibinfo {title} {{How complicated
  must an optical component be?}}}\ }\href {\doibase 10.1364/JOSAA.30.000238}
  {\bibfield  {journal} {\bibinfo  {journal} {Journal of the Optical Society of
  America A}\ }\textbf {\bibinfo {volume} {30}},\ \bibinfo {pages} {238}
  (\bibinfo {year} {2013})}\BibitemShut {NoStop}%
\bibitem [{\citenamefont {Clements}\ \emph {et~al.}(2016)\citenamefont
  {Clements}, \citenamefont {Humphreys}, \citenamefont {Metcalf}, \citenamefont
  {Kolthammer}, \citenamefont {Walsmley},\ and\ \citenamefont
  {Walmsley}}]{Clements2016}%
  \BibitemOpen
  \bibfield  {author} {\bibinfo {author} {\bibfnamefont {William~R.}\
  \bibnamefont {Clements}}, \bibinfo {author} {\bibfnamefont {Peter~C.}\
  \bibnamefont {Humphreys}}, \bibinfo {author} {\bibfnamefont {Benjamin~J.}\
  \bibnamefont {Metcalf}}, \bibinfo {author} {\bibfnamefont {W.~Steven}\
  \bibnamefont {Kolthammer}}, \bibinfo {author} {\bibfnamefont {Ian~A.}\
  \bibnamefont {Walsmley}}, \ and\ \bibinfo {author} {\bibfnamefont {Ian~A.}\
  \bibnamefont {Walmsley}},\ }\emph {\enquote {\bibinfo {title} {{Optimal
  design for universal multiport interferometers}},}\ }\href {\doibase
  10.1364/OPTICA.3.001460} {\bibfield  {journal} {\bibinfo  {journal} {Optica}\
  }\textbf {\bibinfo {volume} {3}},\ \bibinfo {pages} {1460} (\bibinfo {year}
  {2016})},\ \Eprint {http://arxiv.org/abs/1603.08788}
  {arXiv:1603.08788}\BibitemShut {NoStop}%
\bibitem [{\citenamefont {Kumar}\ and\ \citenamefont
  {Dhand}(2021)}]{Kumar2020}%
  \BibitemOpen
  \bibfield  {author} {\bibinfo {author} {\bibfnamefont {Shreya~P}\
  \bibnamefont {Kumar}}\ and\ \bibinfo {author} {\bibfnamefont {Ish}\
  \bibnamefont {Dhand}},\ }\emph {\enquote {\bibinfo {title} {{Unitary matrix
  decompositions for optimal and modular linear optics architectures}},}\
  }\href {\doibase 10.1088/1751-8121/abd4ae} {\bibfield  {journal} {\bibinfo
  {journal} {Journal of Physics A: Mathematical and Theoretical}\ }\textbf
  {\bibinfo {volume} {54}},\ \bibinfo {pages} {045301} (\bibinfo {year}
  {2021})}\BibitemShut {NoStop}%
\bibitem [{\citenamefont {Carolan}\ \emph {et~al.}(2015)\citenamefont
  {Carolan}, \citenamefont {Harrold}, \citenamefont {Sparrow}, \citenamefont
  {Martin-Lopez}, \citenamefont {Russell}, \citenamefont {Silverstone},
  \citenamefont {Shadbolt}, \citenamefont {Matsuda}, \citenamefont {Oguma},
  \citenamefont {Itoh}, \citenamefont {Marshall}, \citenamefont {Thompson},
  \citenamefont {Matthews}, \citenamefont {Hashimoto}, \citenamefont
  {O'Brien},\ and\ \citenamefont {Laing}}]{Carolan2015}%
  \BibitemOpen
  \bibfield  {author} {\bibinfo {author} {\bibfnamefont {Jacques}\ \bibnamefont
  {Carolan}}, \bibinfo {author} {\bibfnamefont {Christopher}\ \bibnamefont
  {Harrold}}, \bibinfo {author} {\bibfnamefont {Chris}\ \bibnamefont
  {Sparrow}}, \bibinfo {author} {\bibfnamefont {E.}~\bibnamefont
  {Martin-Lopez}}, \bibinfo {author} {\bibfnamefont {Nicholas~J.}\ \bibnamefont
  {Russell}}, \bibinfo {author} {\bibfnamefont {Joshua~W.}\ \bibnamefont
  {Silverstone}}, \bibinfo {author} {\bibfnamefont {Peter~J.}\ \bibnamefont
  {Shadbolt}}, \bibinfo {author} {\bibfnamefont {Nobuyuki}\ \bibnamefont
  {Matsuda}}, \bibinfo {author} {\bibfnamefont {Manabu}\ \bibnamefont {Oguma}},
  \bibinfo {author} {\bibfnamefont {Mikitaka}\ \bibnamefont {Itoh}}, \bibinfo
  {author} {\bibfnamefont {Graham~D.}\ \bibnamefont {Marshall}}, \bibinfo
  {author} {\bibfnamefont {Mark~G.}\ \bibnamefont {Thompson}}, \bibinfo
  {author} {\bibfnamefont {Jonathan C.F.~F.}\ \bibnamefont {Matthews}},
  \bibinfo {author} {\bibfnamefont {Toshikazu}\ \bibnamefont {Hashimoto}},
  \bibinfo {author} {\bibfnamefont {Jeremy~L.}\ \bibnamefont {O'Brien}}, \ and\
  \bibinfo {author} {\bibfnamefont {Anthony}\ \bibnamefont {Laing}},\ }\emph
  {\enquote {\bibinfo {title} {{Universal linear optics}},}\ }\href {\doibase
  10.1126/science.aab3642} {\bibfield  {journal} {\bibinfo  {journal}
  {Science}\ }\textbf {\bibinfo {volume} {349}},\ \bibinfo {pages} {711}
  (\bibinfo {year} {2015})},\ \Eprint {http://arxiv.org/abs/1505.01182}
  {arXiv:1505.01182}\BibitemShut {NoStop}%
\bibitem [{\citenamefont {Wang}\ \emph {et~al.}(2020)\citenamefont {Wang},
  \citenamefont {Sciarrino}, \citenamefont {Laing},\ and\ \citenamefont
  {Thompson}}]{Wang2019}%
  \BibitemOpen
  \bibfield  {author} {\bibinfo {author} {\bibfnamefont {Jianwei}\ \bibnamefont
  {Wang}}, \bibinfo {author} {\bibfnamefont {Fabio}\ \bibnamefont {Sciarrino}},
  \bibinfo {author} {\bibfnamefont {Anthony}\ \bibnamefont {Laing}}, \ and\
  \bibinfo {author} {\bibfnamefont {Mark~G}\ \bibnamefont {Thompson}},\ }\emph
  {\enquote {\bibinfo {title} {{Integrated photonic quantum technologies}},}\
  }\href {\doibase 10.1038/s41566-019-0532-1} {\bibfield  {journal} {\bibinfo
  {journal} {Nature Photonics}\ }\textbf {\bibinfo {volume} {14}},\ \bibinfo
  {pages} {273} (\bibinfo {year} {2020})}\BibitemShut {NoStop}%
\bibitem [{\citenamefont {Tang}\ \emph {et~al.}(2021)\citenamefont {Tang},
  \citenamefont {Tanomura}, \citenamefont {Tanemura},\ and\ \citenamefont
  {Nakano}}]{Tang2021}%
  \BibitemOpen
  \bibfield  {author} {\bibinfo {author} {\bibfnamefont {Rui}\ \bibnamefont
  {Tang}}, \bibinfo {author} {\bibfnamefont {Ryota}\ \bibnamefont {Tanomura}},
  \bibinfo {author} {\bibfnamefont {Takuo}\ \bibnamefont {Tanemura}}, \ and\
  \bibinfo {author} {\bibfnamefont {Yoshiaki}\ \bibnamefont {Nakano}},\ }\emph
  {\enquote {\bibinfo {title} {{Ten-Port Unitary Optical Processor on a Silicon
  Photonic Chip}},}\ }\href {\doibase 10.1021/acsphotonics.1c00419} {\bibfield
  {journal} {\bibinfo  {journal} {ACS Photonics}\ }\textbf {\bibinfo {volume}
  {8}},\ \bibinfo {pages} {2074} (\bibinfo {year} {2021})}\BibitemShut
  {NoStop}%
\bibitem [{\citenamefont {Taballione}\ \emph {et~al.}(2022)\citenamefont
  {Taballione}, \citenamefont {Anguita}, \citenamefont {de~Goede},
  \citenamefont {Venderbosch}, \citenamefont {Kassenberg}, \citenamefont
  {Snijders}, \citenamefont {Smith}, \citenamefont {Epping}, \citenamefont
  {van~der Meer}, \citenamefont {Pinkse} \emph {et~al.}}]{Taballione2022}%
  \BibitemOpen
  \bibfield  {author} {\bibinfo {author} {\bibfnamefont {Caterina}\
  \bibnamefont {Taballione}}, \bibinfo {author} {\bibfnamefont
  {Malaquias~Correa}\ \bibnamefont {Anguita}}, \bibinfo {author} {\bibfnamefont
  {Michiel}\ \bibnamefont {de~Goede}}, \bibinfo {author} {\bibfnamefont {Pim}\
  \bibnamefont {Venderbosch}}, \bibinfo {author} {\bibfnamefont {Ben}\
  \bibnamefont {Kassenberg}}, \bibinfo {author} {\bibfnamefont {Henk}\
  \bibnamefont {Snijders}}, \bibinfo {author} {\bibfnamefont {Devin}\
  \bibnamefont {Smith}}, \bibinfo {author} {\bibfnamefont {J{\"o}rn~P}\
  \bibnamefont {Epping}}, \bibinfo {author} {\bibfnamefont {Reinier}\
  \bibnamefont {van~der Meer}}, \bibinfo {author} {\bibfnamefont {Pepijn~WH}\
  \bibnamefont {Pinkse}},  \emph {et~al.},\ }\emph {\enquote {\bibinfo {title}
  {20-mode universal quantum photonic processor},}\ }\href@noop {} {\
  (\bibinfo {year} {2022})},\ \Eprint {http://arxiv.org/abs/2203.01801}
  {arXiv:2203.01801}\BibitemShut {NoStop}%
\bibitem [{\citenamefont {Miller}(2015)}]{Miller2015}%
  \BibitemOpen
  \bibfield  {author} {\bibinfo {author} {\bibfnamefont {David A.~B.}\
  \bibnamefont {Miller}},\ }\emph {\enquote {\bibinfo {title} {{Perfect optics
  with imperfect components}},}\ }\href {\doibase 10.1364/OPTICA.2.000747}
  {\bibfield  {journal} {\bibinfo  {journal} {Optica}\ }\textbf {\bibinfo
  {volume} {2}},\ \bibinfo {pages} {747} (\bibinfo {year} {2015})}\BibitemShut
  {NoStop}%
\bibitem [{\citenamefont {Burgwal}\ \emph {et~al.}(2017)\citenamefont
  {Burgwal}, \citenamefont {Clements}, \citenamefont {Smith}, \citenamefont
  {Gates}, \citenamefont {Kolthammer}, \citenamefont {Renema},\ and\
  \citenamefont {Walmsley}}]{Burgwal2017}%
  \BibitemOpen
  \bibfield  {author} {\bibinfo {author} {\bibfnamefont {Roel}\ \bibnamefont
  {Burgwal}}, \bibinfo {author} {\bibfnamefont {William~R.}\ \bibnamefont
  {Clements}}, \bibinfo {author} {\bibfnamefont {Devin~H.}\ \bibnamefont
  {Smith}}, \bibinfo {author} {\bibfnamefont {James~C.}\ \bibnamefont {Gates}},
  \bibinfo {author} {\bibfnamefont {W.~Steven}\ \bibnamefont {Kolthammer}},
  \bibinfo {author} {\bibfnamefont {Jelmer~J.}\ \bibnamefont {Renema}}, \ and\
  \bibinfo {author} {\bibfnamefont {Ian~A.}\ \bibnamefont {Walmsley}},\ }\emph
  {\enquote {\bibinfo {title} {{Using an imperfect photonic network to
  implement random unitaries}},}\ }\href {\doibase 10.1364/OE.25.028236}
  {\bibfield  {journal} {\bibinfo  {journal} {Optics Express}\ }\textbf
  {\bibinfo {volume} {25}},\ \bibinfo {pages} {28236} (\bibinfo {year}
  {2017})},\ \Eprint {http://arxiv.org/abs/1612.01199}
  {arXiv:1612.01199}\BibitemShut {NoStop}%
\bibitem [{\citenamefont {Pai}\ \emph {et~al.}(2018)\citenamefont {Pai},
  \citenamefont {Bartlett}, \citenamefont {Solgaard},\ and\ \citenamefont
  {Miller}}]{Pai2018}%
  \BibitemOpen
  \bibfield  {author} {\bibinfo {author} {\bibfnamefont {Sunil}\ \bibnamefont
  {Pai}}, \bibinfo {author} {\bibfnamefont {Ben}\ \bibnamefont {Bartlett}},
  \bibinfo {author} {\bibfnamefont {Olav}\ \bibnamefont {Solgaard}}, \ and\
  \bibinfo {author} {\bibfnamefont {David A.~B.}\ \bibnamefont {Miller}},\
  }\emph {\enquote {\bibinfo {title} {{Matrix Optimization on Universal Unitary
  Photonic Devices}},}\ }\href {\doibase 10.1103/physrevapplied.11.064044}
  {\bibfield  {journal} {\bibinfo  {journal} {Physical Review Applied}\
  }\textbf {\bibinfo {volume} {11}},\ \bibinfo {pages} {064044} (\bibinfo
  {year} {2018})},\ \Eprint {http://arxiv.org/abs/1808.00458}
  {1808.00458}\BibitemShut {NoStop}%
\bibitem [{\citenamefont {Fang}\ \emph {et~al.}(2019)\citenamefont {Fang},
  \citenamefont {Manipatruni}, \citenamefont {Wierzynski}, \citenamefont
  {Khosrowshahi},\ and\ \citenamefont {DeWeese}}]{Fang2019}%
  \BibitemOpen
  \bibfield  {author} {\bibinfo {author} {\bibfnamefont {Michael Y.-S.}\
  \bibnamefont {Fang}}, \bibinfo {author} {\bibfnamefont {Sasikanth}\
  \bibnamefont {Manipatruni}}, \bibinfo {author} {\bibfnamefont {Casimir}\
  \bibnamefont {Wierzynski}}, \bibinfo {author} {\bibfnamefont {Amir}\
  \bibnamefont {Khosrowshahi}}, \ and\ \bibinfo {author} {\bibfnamefont
  {Michael~R.}\ \bibnamefont {DeWeese}},\ }\emph {\enquote {\bibinfo {title}
  {{Design of optical neural networks with component imprecisions}},}\ }\href
  {\doibase 10.1364/OE.27.014009} {\bibfield  {journal} {\bibinfo  {journal}
  {Optics Express}\ }\textbf {\bibinfo {volume} {27}},\ \bibinfo {pages}
  {14009} (\bibinfo {year} {2019})}\BibitemShut {NoStop}%
\bibitem [{\citenamefont {Hamerly}\ \emph {et~al.}(2021)\citenamefont
  {Hamerly}, \citenamefont {Bandyopadhyay},\ and\ \citenamefont
  {Englund}}]{Hamerly2021}%
  \BibitemOpen
  \bibfield  {author} {\bibinfo {author} {\bibfnamefont {Ryan}\ \bibnamefont
  {Hamerly}}, \bibinfo {author} {\bibfnamefont {Saumil}\ \bibnamefont
  {Bandyopadhyay}}, \ and\ \bibinfo {author} {\bibfnamefont {Dirk}\
  \bibnamefont {Englund}},\ }\emph {\enquote {\bibinfo {title} {{Stability of
  Self-Configuring Large Multiport Interferometers}},}\ }\href
  {http://arxiv.org/abs/2106.04363} {\  (\bibinfo {year} {2021})},\ \Eprint
  {http://arxiv.org/abs/2106.04363} {arXiv:2106.04363}\BibitemShut {NoStop}%
\bibitem [{\citenamefont {Fldzhyan}\ \emph {et~al.}(2020)\citenamefont
  {Fldzhyan}, \citenamefont {Saygin},\ and\ \citenamefont
  {Kulik}}]{Fldzhyan2020}%
  \BibitemOpen
  \bibfield  {author} {\bibinfo {author} {\bibfnamefont {S.~A.}\ \bibnamefont
  {Fldzhyan}}, \bibinfo {author} {\bibfnamefont {M.~Yu.}\ \bibnamefont
  {Saygin}}, \ and\ \bibinfo {author} {\bibfnamefont {S.~P.}\ \bibnamefont
  {Kulik}},\ }\emph {\enquote {\bibinfo {title} {{Optimal design of
  error-tolerant reprogrammable multiport interferometers}},}\ }\href {\doibase
  10.1364/ol.385433} {\bibfield  {journal} {\bibinfo  {journal} {Optics
  Letters}\ }\textbf {\bibinfo {volume} {45}},\ \bibinfo {pages} {2632}
  (\bibinfo {year} {2020})}\BibitemShut {NoStop}%
\bibitem [{\citenamefont {Tanomura}\ \emph {et~al.}(2022)\citenamefont
  {Tanomura}, \citenamefont {Tang}, \citenamefont {Umezaki}, \citenamefont
  {Soma}, \citenamefont {Tanemura},\ and\ \citenamefont
  {Nakano}}]{Tanomura2021}%
  \BibitemOpen
  \bibfield  {author} {\bibinfo {author} {\bibfnamefont {Ryota}\ \bibnamefont
  {Tanomura}}, \bibinfo {author} {\bibfnamefont {Rui}\ \bibnamefont {Tang}},
  \bibinfo {author} {\bibfnamefont {Toshikazu}\ \bibnamefont {Umezaki}},
  \bibinfo {author} {\bibfnamefont {Go}~\bibnamefont {Soma}}, \bibinfo {author}
  {\bibfnamefont {Takuo}\ \bibnamefont {Tanemura}}, \ and\ \bibinfo {author}
  {\bibfnamefont {Yoshiaki}\ \bibnamefont {Nakano}},\ }\emph {\enquote
  {\bibinfo {title} {{Scalable and Robust Photonic Integrated Unitary Converter
  Based on Multiplane Light Conversion}},}\ }\href {\doibase
  10.1103/physrevapplied.17.024071} {\bibfield  {journal} {\bibinfo  {journal}
  {Physical Review Applied}\ }\textbf {\bibinfo {volume} {17}},\ \bibinfo
  {pages} {024071} (\bibinfo {year} {2022})},\ \Eprint
  {http://arxiv.org/abs/2103.14782} {2103.14782}\BibitemShut {NoStop}%
\bibitem [{\citenamefont {Molesky}\ \emph {et~al.}(2018)\citenamefont
  {Molesky}, \citenamefont {Lin}, \citenamefont {Piggott}, \citenamefont {Jin},
  \citenamefont {Vuckovi{\'{c}}},\ and\ \citenamefont
  {Rodriguez}}]{Molesky2018}%
  \BibitemOpen
  \bibfield  {author} {\bibinfo {author} {\bibfnamefont {Sean}\ \bibnamefont
  {Molesky}}, \bibinfo {author} {\bibfnamefont {Zin}\ \bibnamefont {Lin}},
  \bibinfo {author} {\bibfnamefont {Alexander~Y}\ \bibnamefont {Piggott}},
  \bibinfo {author} {\bibfnamefont {Weiliang}\ \bibnamefont {Jin}}, \bibinfo
  {author} {\bibfnamefont {Jelena}\ \bibnamefont {Vuckovi{\'{c}}}}, \ and\
  \bibinfo {author} {\bibfnamefont {Alejandro~W}\ \bibnamefont {Rodriguez}},\
  }\emph {\enquote {\bibinfo {title} {{Inverse design in nanophotonics}},}\
  }\href {\doibase 10.1038/s41566-018-0246-9} {\bibfield  {journal} {\bibinfo
  {journal} {Nature Photonics}\ }\textbf {\bibinfo {volume} {12}},\ \bibinfo
  {pages} {659} (\bibinfo {year} {2018})}\BibitemShut {NoStop}%
\bibitem [{\citenamefont {Marcucci}\ \emph {et~al.}(2020)\citenamefont
  {Marcucci}, \citenamefont {Pierangeli}, \citenamefont {Pinkse}, \citenamefont
  {Malik},\ and\ \citenamefont {Conti}}]{Marcucci2020}%
  \BibitemOpen
  \bibfield  {author} {\bibinfo {author} {\bibfnamefont {Giulia}\ \bibnamefont
  {Marcucci}}, \bibinfo {author} {\bibfnamefont {Davide}\ \bibnamefont
  {Pierangeli}}, \bibinfo {author} {\bibfnamefont {Pepijn W.~H.}\ \bibnamefont
  {Pinkse}}, \bibinfo {author} {\bibfnamefont {Mehul}\ \bibnamefont {Malik}}, \
  and\ \bibinfo {author} {\bibfnamefont {Claudio}\ \bibnamefont {Conti}},\
  }\emph {\enquote {\bibinfo {title} {{Programming multi-level quantum gates in
  disordered computing reservoirs via machine learning}},}\ }\href {\doibase
  10.1364/OE.389432} {\bibfield  {journal} {\bibinfo  {journal} {Optics
  Express}\ }\textbf {\bibinfo {volume} {28}},\ \bibinfo {pages} {14018}
  (\bibinfo {year} {2020})},\ \Eprint {http://arxiv.org/abs/1905.05264}
  {arXiv:1905.05264}\BibitemShut {NoStop}%
\bibitem [{\citenamefont {Morizur}\ \emph {et~al.}(2010)\citenamefont
  {Morizur}, \citenamefont {Nicholls}, \citenamefont {Jian}, \citenamefont
  {Armstrong}, \citenamefont {Treps}, \citenamefont {Hage}, \citenamefont
  {Hsu}, \citenamefont {Bowen}, \citenamefont {Janousek},\ and\ \citenamefont
  {Bachor}}]{Morizur2010}%
  \BibitemOpen
  \bibfield  {author} {\bibinfo {author} {\bibfnamefont {Jean-Fran{\c{c}}ois}\
  \bibnamefont {Morizur}}, \bibinfo {author} {\bibfnamefont {Lachlan}\
  \bibnamefont {Nicholls}}, \bibinfo {author} {\bibfnamefont {Pu}~\bibnamefont
  {Jian}}, \bibinfo {author} {\bibfnamefont {Seiji}\ \bibnamefont {Armstrong}},
  \bibinfo {author} {\bibfnamefont {Nicolas}\ \bibnamefont {Treps}}, \bibinfo
  {author} {\bibfnamefont {Boris}\ \bibnamefont {Hage}}, \bibinfo {author}
  {\bibfnamefont {Magnus}\ \bibnamefont {Hsu}}, \bibinfo {author}
  {\bibfnamefont {Warwick}\ \bibnamefont {Bowen}}, \bibinfo {author}
  {\bibfnamefont {Jiri}\ \bibnamefont {Janousek}}, \ and\ \bibinfo {author}
  {\bibfnamefont {Hans-A}\ \bibnamefont {Bachor}},\ }\emph {\enquote {\bibinfo
  {title} {{Programmable unitary spatial mode manipulation}},}\ }\href
  {\doibase 10.1364/JOSAA.27.002524} {\bibfield  {journal} {\bibinfo  {journal}
  {Journal of the Optical Society of America A}\ }\textbf {\bibinfo {volume}
  {27}},\ \bibinfo {pages} {2524} (\bibinfo {year} {2010})},\ \Eprint
  {http://arxiv.org/abs/1005.3366} {arXiv:1005.3366}\BibitemShut {NoStop}%
\bibitem [{\citenamefont {Labroille}\ \emph {et~al.}(2014)\citenamefont
  {Labroille}, \citenamefont {Denolle}, \citenamefont {Jian}, \citenamefont
  {Genevaux}, \citenamefont {Treps}, \citenamefont {Morizur}, \citenamefont
  {Genevaux},\ and\ \citenamefont {Treps}}]{Labroille2014}%
  \BibitemOpen
  \bibfield  {author} {\bibinfo {author} {\bibfnamefont {Guillaume}\
  \bibnamefont {Labroille}}, \bibinfo {author} {\bibfnamefont {Bertrand}\
  \bibnamefont {Denolle}}, \bibinfo {author} {\bibfnamefont {Pu}~\bibnamefont
  {Jian}}, \bibinfo {author} {\bibfnamefont {Philippe}\ \bibnamefont
  {Genevaux}}, \bibinfo {author} {\bibfnamefont {Nicolas}\ \bibnamefont
  {Treps}}, \bibinfo {author} {\bibfnamefont
  {Jean-Fran{\c{c}}ois~Fran{\c{c}}ois}\ \bibnamefont {Morizur}}, \bibinfo
  {author} {\bibfnamefont {Philippe}\ \bibnamefont {Genevaux}}, \ and\ \bibinfo
  {author} {\bibfnamefont {Nicolas}\ \bibnamefont {Treps}},\ }\emph {\enquote
  {\bibinfo {title} {{Efficient and mode selective spatial mode multiplexer
  based on multi-plane light conversion}},}\ }\href {\doibase
  10.1364/OE.22.015599} {\bibfield  {journal} {\bibinfo  {journal} {Optics
  Express}\ }\textbf {\bibinfo {volume} {22}},\ \bibinfo {pages} {15599}
  (\bibinfo {year} {2014})},\ \Eprint {http://arxiv.org/abs/1404.6455v1}
  {arXiv:1404.6455v1}\BibitemShut {NoStop}%
\bibitem [{\citenamefont {Fontaine}\ \emph {et~al.}(2019)\citenamefont
  {Fontaine}, \citenamefont {Ryf}, \citenamefont {Chen}, \citenamefont
  {Neilson}, \citenamefont {Kim},\ and\ \citenamefont
  {Carpenter}}]{Fontaine2019}%
  \BibitemOpen
  \bibfield  {author} {\bibinfo {author} {\bibfnamefont {Nicolas~K}\
  \bibnamefont {Fontaine}}, \bibinfo {author} {\bibfnamefont {Roland}\
  \bibnamefont {Ryf}}, \bibinfo {author} {\bibfnamefont {Haoshuo}\ \bibnamefont
  {Chen}}, \bibinfo {author} {\bibfnamefont {David~T}\ \bibnamefont {Neilson}},
  \bibinfo {author} {\bibfnamefont {Kwangwoong}\ \bibnamefont {Kim}}, \ and\
  \bibinfo {author} {\bibfnamefont {Joel}\ \bibnamefont {Carpenter}},\ }\emph
  {\enquote {\bibinfo {title} {{Laguerre-Gaussian mode sorter}},}\ }\href
  {\doibase 10.1038/s41467-019-09840-4} {\bibfield  {journal} {\bibinfo
  {journal} {Nature Communications}\ }\textbf {\bibinfo {volume} {10}},\
  \bibinfo {pages} {1865} (\bibinfo {year} {2019})}\BibitemShut {NoStop}%
\bibitem [{\citenamefont {Hashimoto}\ \emph {et~al.}(2005)\citenamefont
  {Hashimoto}, \citenamefont {Saida}, \citenamefont {Ogawa}, \citenamefont
  {Kohtoku}, \citenamefont {Shibata},\ and\ \citenamefont
  {Takahashi}}]{Hashimoto2005}%
  \BibitemOpen
  \bibfield  {author} {\bibinfo {author} {\bibfnamefont {T}~\bibnamefont
  {Hashimoto}}, \bibinfo {author} {\bibfnamefont {T}~\bibnamefont {Saida}},
  \bibinfo {author} {\bibfnamefont {I}~\bibnamefont {Ogawa}}, \bibinfo {author}
  {\bibfnamefont {M}~\bibnamefont {Kohtoku}}, \bibinfo {author} {\bibfnamefont
  {T}~\bibnamefont {Shibata}}, \ and\ \bibinfo {author} {\bibfnamefont
  {H}~\bibnamefont {Takahashi}},\ }\emph {\enquote {\bibinfo {title} {{Optical
  circuit design based on a wavefront-matching method}},}\ }\href {\doibase
  10.1364/ol.30.002620} {\bibfield  {journal} {\bibinfo  {journal} {Optics
  Letters}\ }\textbf {\bibinfo {volume} {30}},\ \bibinfo {pages} {2620}
  (\bibinfo {year} {2005})}\BibitemShut {NoStop}%
\bibitem [{\citenamefont {Erhard}\ \emph {et~al.}(2018)\citenamefont {Erhard},
  \citenamefont {Malik}, \citenamefont {Krenn},\ and\ \citenamefont
  {Zeilinger}}]{OAMGHZ}%
  \BibitemOpen
  \bibfield  {author} {\bibinfo {author} {\bibfnamefont {Manuel}\ \bibnamefont
  {Erhard}}, \bibinfo {author} {\bibfnamefont {Mehul}\ \bibnamefont {Malik}},
  \bibinfo {author} {\bibfnamefont {Mario}\ \bibnamefont {Krenn}}, \ and\
  \bibinfo {author} {\bibfnamefont {Anton}\ \bibnamefont {Zeilinger}},\ }\emph
  {\enquote {\bibinfo {title} {{Experimental Greenberger–Horne–Zeilinger
  entanglement beyond qubits}},}\ }\href {\doibase 10.1038/s41566-018-0257-6}
  {\bibfield  {journal} {\bibinfo  {journal} {Nature Photonics}\ }\textbf
  {\bibinfo {volume} {12}},\ \bibinfo {pages} {759} (\bibinfo {year} {2018})},\
  \Eprint {http://arxiv.org/abs/1708.03881} {1708.03881}\BibitemShut {NoStop}%
\bibitem [{\citenamefont {Krenn}\ \emph {et~al.}(2016)\citenamefont {Krenn},
  \citenamefont {Malik}, \citenamefont {Fickler}, \citenamefont {Lapkiewicz},\
  and\ \citenamefont {Zeilinger}}]{Krenn2016b}%
  \BibitemOpen
  \bibfield  {author} {\bibinfo {author} {\bibfnamefont {Mario}\ \bibnamefont
  {Krenn}}, \bibinfo {author} {\bibfnamefont {Mehul}\ \bibnamefont {Malik}},
  \bibinfo {author} {\bibfnamefont {Robert}\ \bibnamefont {Fickler}}, \bibinfo
  {author} {\bibfnamefont {Radek}\ \bibnamefont {Lapkiewicz}}, \ and\ \bibinfo
  {author} {\bibfnamefont {Anton}\ \bibnamefont {Zeilinger}},\ }\emph {\enquote
  {\bibinfo {title} {{Automated Search for new Quantum Experiments}},}\ }\href
  {\doibase 10.1103/physrevlett.116.090405} {\bibfield  {journal} {\bibinfo
  {journal} {Physical Review Letters}\ }\textbf {\bibinfo {volume} {116}},\
  \bibinfo {pages} {090405} (\bibinfo {year} {2016})},\ \Eprint
  {http://arxiv.org/abs/1509.02749} {1509.02749}\BibitemShut {NoStop}%
\bibitem [{\citenamefont {Melnikov}\ \emph {et~al.}(2018)\citenamefont
  {Melnikov}, \citenamefont {Nautrup}, \citenamefont {Krenn}, \citenamefont
  {Dunjko}, \citenamefont {Tiersch}, \citenamefont {Zeilinger},\ and\
  \citenamefont {Briegel}}]{Melnikov2018}%
  \BibitemOpen
  \bibfield  {author} {\bibinfo {author} {\bibfnamefont {Alexey~A}\
  \bibnamefont {Melnikov}}, \bibinfo {author} {\bibfnamefont {Hendrik~Poulsen}\
  \bibnamefont {Nautrup}}, \bibinfo {author} {\bibfnamefont {Mario}\
  \bibnamefont {Krenn}}, \bibinfo {author} {\bibfnamefont {Vedran}\
  \bibnamefont {Dunjko}}, \bibinfo {author} {\bibfnamefont {Markus}\
  \bibnamefont {Tiersch}}, \bibinfo {author} {\bibfnamefont {Anton}\
  \bibnamefont {Zeilinger}}, \ and\ \bibinfo {author} {\bibfnamefont {Hans~J}\
  \bibnamefont {Briegel}},\ }\emph {\enquote {\bibinfo {title} {Active learning
  machine learns to create new quantum experiments},}\ }\href {\doibase
  10.1073/pnas.1714936115} {\bibfield  {journal} {\bibinfo  {journal}
  {Proceedings of the National Academy of Sciences}\ }\textbf {\bibinfo
  {volume} {115}},\ \bibinfo {pages} {1221} (\bibinfo {year}
  {2018})}\BibitemShut {NoStop}%
\bibitem [{\citenamefont {Krenn}\ \emph {et~al.}(2020)\citenamefont {Krenn},
  \citenamefont {Erhard},\ and\ \citenamefont {Zeilinger}}]{Krenn2020a}%
  \BibitemOpen
  \bibfield  {author} {\bibinfo {author} {\bibfnamefont {Mario}\ \bibnamefont
  {Krenn}}, \bibinfo {author} {\bibfnamefont {Manuel}\ \bibnamefont {Erhard}},
  \ and\ \bibinfo {author} {\bibfnamefont {Anton}\ \bibnamefont {Zeilinger}},\
  }\emph {\enquote {\bibinfo {title} {{Computer-inspired quantum
  experiments}},}\ }\href {\doibase 10.1038/s42254-020-0230-4} {\bibfield
  {journal} {\bibinfo  {journal} {Nature Reviews Physics}\ }\textbf {\bibinfo
  {volume} {2}},\ \bibinfo {pages} {649} (\bibinfo {year} {2020})},\ \Eprint
  {http://arxiv.org/abs/2002.09970} {arXiv:2002.09970}\BibitemShut {NoStop}%
\bibitem [{\citenamefont {Zhong}\ \emph {et~al.}(2020)\citenamefont {Zhong},
  \citenamefont {Wang}, \citenamefont {Deng}, \citenamefont {Chen},
  \citenamefont {Peng}, \citenamefont {Luo}, \citenamefont {Qin}, \citenamefont
  {Wu}, \citenamefont {Ding}, \citenamefont {Hu} \emph {et~al.}}]{Zhong2020}%
  \BibitemOpen
  \bibfield  {author} {\bibinfo {author} {\bibfnamefont {Han-Sen}\ \bibnamefont
  {Zhong}}, \bibinfo {author} {\bibfnamefont {Hui}\ \bibnamefont {Wang}},
  \bibinfo {author} {\bibfnamefont {Yu-Hao}\ \bibnamefont {Deng}}, \bibinfo
  {author} {\bibfnamefont {Ming-Cheng}\ \bibnamefont {Chen}}, \bibinfo {author}
  {\bibfnamefont {Li-Chao}\ \bibnamefont {Peng}}, \bibinfo {author}
  {\bibfnamefont {Yi-Han}\ \bibnamefont {Luo}}, \bibinfo {author}
  {\bibfnamefont {Jian}\ \bibnamefont {Qin}}, \bibinfo {author} {\bibfnamefont
  {Dian}\ \bibnamefont {Wu}}, \bibinfo {author} {\bibfnamefont {Xing}\
  \bibnamefont {Ding}}, \bibinfo {author} {\bibfnamefont {Yi}~\bibnamefont
  {Hu}},  \emph {et~al.},\ }\emph {\enquote {\bibinfo {title} {Quantum
  computational advantage using photons},}\ }\href {\doibase
  10.1126/science.abe8770} {\bibfield  {journal} {\bibinfo  {journal}
  {Science}\ }\textbf {\bibinfo {volume} {370}},\ \bibinfo {pages} {1460}
  (\bibinfo {year} {2020})}\BibitemShut {NoStop}%
\bibitem [{\citenamefont {Llewellyn}\ \emph {et~al.}(2020)\citenamefont
  {Llewellyn}, \citenamefont {Ding}, \citenamefont {Faruque}, \citenamefont
  {Paesani}, \citenamefont {Bacco}, \citenamefont {Santagati}, \citenamefont
  {Qian}, \citenamefont {Li}, \citenamefont {Xiao}, \citenamefont {Huber},
  \citenamefont {Malik}, \citenamefont {Sinclair}, \citenamefont {Zhou},
  \citenamefont {Rottwitt}, \citenamefont {O’Brien}, \citenamefont {Rarity},
  \citenamefont {Gong}, \citenamefont {Oxenlowe}, \citenamefont {Wang},\ and\
  \citenamefont {Thompson}}]{Llewellyn2020}%
  \BibitemOpen
  \bibfield  {author} {\bibinfo {author} {\bibfnamefont {Daniel}\ \bibnamefont
  {Llewellyn}}, \bibinfo {author} {\bibfnamefont {Yunhong}\ \bibnamefont
  {Ding}}, \bibinfo {author} {\bibfnamefont {Imad~I.}\ \bibnamefont {Faruque}},
  \bibinfo {author} {\bibfnamefont {Stefano}\ \bibnamefont {Paesani}}, \bibinfo
  {author} {\bibfnamefont {Davide}\ \bibnamefont {Bacco}}, \bibinfo {author}
  {\bibfnamefont {Raffaele}\ \bibnamefont {Santagati}}, \bibinfo {author}
  {\bibfnamefont {Yan-Jun}\ \bibnamefont {Qian}}, \bibinfo {author}
  {\bibfnamefont {Yan}\ \bibnamefont {Li}}, \bibinfo {author} {\bibfnamefont
  {Yun-Feng}\ \bibnamefont {Xiao}}, \bibinfo {author} {\bibfnamefont {Marcus}\
  \bibnamefont {Huber}}, \bibinfo {author} {\bibfnamefont {Mehul}\ \bibnamefont
  {Malik}}, \bibinfo {author} {\bibfnamefont {Gary~F.}\ \bibnamefont
  {Sinclair}}, \bibinfo {author} {\bibfnamefont {Xiaoqi}\ \bibnamefont {Zhou}},
  \bibinfo {author} {\bibfnamefont {Karsten}\ \bibnamefont {Rottwitt}},
  \bibinfo {author} {\bibfnamefont {Jeremy~L.}\ \bibnamefont {O’Brien}},
  \bibinfo {author} {\bibfnamefont {John~G.}\ \bibnamefont {Rarity}}, \bibinfo
  {author} {\bibfnamefont {Qihuang}\ \bibnamefont {Gong}}, \bibinfo {author}
  {\bibfnamefont {Leif~K.}\ \bibnamefont {Oxenlowe}}, \bibinfo {author}
  {\bibfnamefont {Jianwei}\ \bibnamefont {Wang}}, \ and\ \bibinfo {author}
  {\bibfnamefont {Mark~G.}\ \bibnamefont {Thompson}},\ }\emph {\enquote
  {\bibinfo {title} {{Chip-to-chip quantum teleportation and multi-photon
  entanglement in silicon}},}\ }\href {\doibase 10.1038/s41567-019-0727-x}
  {\bibfield  {journal} {\bibinfo  {journal} {Nature Physics}\ }\textbf
  {\bibinfo {volume} {16}},\ \bibinfo {pages} {148} (\bibinfo {year} {2020})},\
  \Eprint {http://arxiv.org/abs/1911.07839} {1911.07839}\BibitemShut {NoStop}%
\bibitem [{\citenamefont {Malik}\ and\ \citenamefont
  {Boyd}(2014)}]{malikboyd2014}%
  \BibitemOpen
  \bibfield  {author} {\bibinfo {author} {\bibfnamefont {Mehul}\ \bibnamefont
  {Malik}}\ and\ \bibinfo {author} {\bibfnamefont {Robert~W}\ \bibnamefont
  {Boyd}},\ }\emph {\enquote {\bibinfo {title} {Quantum imaging
  technologies},}\ }\href {\doibase 10.1393/ncr/i2014-10100-0} {\bibfield
  {journal} {\bibinfo  {journal} {Rivista Del Nuovo Cimento}\ }\textbf
  {\bibinfo {volume} {37}},\ \bibinfo {pages} {273} (\bibinfo {year}
  {2014})}\BibitemShut {NoStop}%
\bibitem [{\citenamefont {Hu}\ \emph {et~al.}(2018)\citenamefont {Hu},
  \citenamefont {Guo}, \citenamefont {Liu}, \citenamefont {Huang},
  \citenamefont {Li},\ and\ \citenamefont {Guo}}]{Hu2018c}%
  \BibitemOpen
  \bibfield  {author} {\bibinfo {author} {\bibfnamefont {Xiao-Min}\
  \bibnamefont {Hu}}, \bibinfo {author} {\bibfnamefont {Yu}~\bibnamefont
  {Guo}}, \bibinfo {author} {\bibfnamefont {Bi-Heng}\ \bibnamefont {Liu}},
  \bibinfo {author} {\bibfnamefont {Yun-Feng}\ \bibnamefont {Huang}}, \bibinfo
  {author} {\bibfnamefont {Chuan-Feng}\ \bibnamefont {Li}}, \ and\ \bibinfo
  {author} {\bibfnamefont {Guang-Can}\ \bibnamefont {Guo}},\ }\emph {\enquote
  {\bibinfo {title} {{Beating the channel capacity limit for superdense coding
  with entangled ququarts}},}\ }\href {\doibase 10.1126/sciadv.aat9304}
  {\bibfield  {journal} {\bibinfo  {journal} {Science Advances}\ }\textbf
  {\bibinfo {volume} {4}},\ \bibinfo {pages} {eaat9304} (\bibinfo {year}
  {2018})}\BibitemShut {NoStop}%
\bibitem [{\citenamefont {Ecker}\ \emph {et~al.}(2019)\citenamefont {Ecker},
  \citenamefont {Bouchard}, \citenamefont {Bulla}, \citenamefont {Brandt},
  \citenamefont {Kohout}, \citenamefont {Steinlechner}, \citenamefont
  {Fickler}, \citenamefont {Malik}, \citenamefont {Guryanova}, \citenamefont
  {Ursin},\ and\ \citenamefont {Huber}}]{Ecker-Huber2019}%
  \BibitemOpen
  \bibfield  {author} {\bibinfo {author} {\bibfnamefont {Sebastian}\
  \bibnamefont {Ecker}}, \bibinfo {author} {\bibfnamefont
  {Fr{\'{e}}d{\'{e}}ric}\ \bibnamefont {Bouchard}}, \bibinfo {author}
  {\bibfnamefont {Lukas}\ \bibnamefont {Bulla}}, \bibinfo {author}
  {\bibfnamefont {Florian}\ \bibnamefont {Brandt}}, \bibinfo {author}
  {\bibfnamefont {Oskar}\ \bibnamefont {Kohout}}, \bibinfo {author}
  {\bibfnamefont {Fabian}\ \bibnamefont {Steinlechner}}, \bibinfo {author}
  {\bibfnamefont {Robert}\ \bibnamefont {Fickler}}, \bibinfo {author}
  {\bibfnamefont {Mehul}\ \bibnamefont {Malik}}, \bibinfo {author}
  {\bibfnamefont {Yelena}\ \bibnamefont {Guryanova}}, \bibinfo {author}
  {\bibfnamefont {Rupert}\ \bibnamefont {Ursin}}, \ and\ \bibinfo {author}
  {\bibfnamefont {Marcus}\ \bibnamefont {Huber}},\ }\emph {\enquote {\bibinfo
  {title} {{Overcoming Noise in Entanglement Distribution}},}\ }\href {\doibase
  10.1103/PhysRevX.9.041042} {\bibfield  {journal} {\bibinfo  {journal}
  {Physical Review X}\ }\textbf {\bibinfo {volume} {9}},\ \bibinfo {pages}
  {041042} (\bibinfo {year} {2019})},\ \Eprint
  {http://arxiv.org/abs/1904.01552} {arXiv:1904.01552}\BibitemShut {NoStop}%
\bibitem [{\citenamefont {Zhu}\ \emph {et~al.}(2021)\citenamefont {Zhu},
  \citenamefont {Tyler}, \citenamefont {Valencia}, \citenamefont {Malik},\ and\
  \citenamefont {Leach}}]{Zhu2021b}%
  \BibitemOpen
  \bibfield  {author} {\bibinfo {author} {\bibfnamefont {Feng}\ \bibnamefont
  {Zhu}}, \bibinfo {author} {\bibfnamefont {Max}\ \bibnamefont {Tyler}},
  \bibinfo {author} {\bibfnamefont {Natalia~Herrera}\ \bibnamefont {Valencia}},
  \bibinfo {author} {\bibfnamefont {Mehul}\ \bibnamefont {Malik}}, \ and\
  \bibinfo {author} {\bibfnamefont {Jonathan}\ \bibnamefont {Leach}},\ }\emph
  {\enquote {\bibinfo {title} {{Is high-dimensional photonic entanglement
  robust to noise?}}}\ }\href {\doibase 10.1116/5.0033889} {\bibfield
  {journal} {\bibinfo  {journal} {AVS Quantum Science}\ }\textbf {\bibinfo
  {volume} {3}},\ \bibinfo {pages} {011401} (\bibinfo {year} {2021})},\ \Eprint
  {http://arxiv.org/abs/1908.08943} {arXiv:1908.08943}\BibitemShut {NoStop}%
\bibitem [{\citenamefont {Gao}\ \emph {et~al.}(2022)\citenamefont {Gao},
  \citenamefont {Appel}, \citenamefont {Friis}, \citenamefont {Ringbauer},\
  and\ \citenamefont {Huber}}]{Gao2022}%
  \BibitemOpen
  \bibfield  {author} {\bibinfo {author} {\bibfnamefont {Xiaoqin}\ \bibnamefont
  {Gao}}, \bibinfo {author} {\bibfnamefont {Paul}\ \bibnamefont {Appel}},
  \bibinfo {author} {\bibfnamefont {Nicolai}\ \bibnamefont {Friis}}, \bibinfo
  {author} {\bibfnamefont {Martin}\ \bibnamefont {Ringbauer}}, \ and\ \bibinfo
  {author} {\bibfnamefont {Marcus}\ \bibnamefont {Huber}},\ }\emph {\enquote
  {\bibinfo {title} {On the role of entanglement in qudit-based circuit
  compression},}\ }\href {https://arxiv.org/abs/2209.14584} {\  (\bibinfo
  {year} {2022})},\ \Eprint {http://arxiv.org/abs/2209.14584}
  {arXiv:2209.14584}\BibitemShut {NoStop}%
\bibitem [{\citenamefont {Vértesi}\ \emph {et~al.}(10 2)\citenamefont
  {Vértesi}, \citenamefont {Pironio},\ and\ \citenamefont
  {Brunner}}]{Vertesi:2010bq}%
  \BibitemOpen
  \bibfield  {author} {\bibinfo {author} {\bibfnamefont {Tamás}\ \bibnamefont
  {Vértesi}}, \bibinfo {author} {\bibfnamefont {Stefano}\ \bibnamefont
  {Pironio}}, \ and\ \bibinfo {author} {\bibfnamefont {Nicolas}\ \bibnamefont
  {Brunner}},\ }\emph {\enquote {\bibinfo {title} {{Closing the Detection
  Loophole in Bell Experiments Using Qudits}},}\ }\href {\doibase
  10.1103/physrevlett.104.060401} {\bibfield  {journal} {\bibinfo  {journal}
  {Physical Review Letters}\ }\textbf {\bibinfo {volume} {104}},\ \bibinfo
  {pages} {060401} (\bibinfo {year} {2010-2})},\ \Eprint
  {http://arxiv.org/abs/0909.3171} {0909.3171}\BibitemShut {NoStop}%
\bibitem [{\citenamefont {Srivastav}\ \emph
  {et~al.}(2022{\natexlab{a}})\citenamefont {Srivastav}, \citenamefont
  {Valencia}, \citenamefont {McCutcheon}, \citenamefont
  {Leedumrongwatthanakun}, \citenamefont {Designolle}, \citenamefont {Uola},
  \citenamefont {Brunner},\ and\ \citenamefont {Malik}}]{Srivastav2022}%
  \BibitemOpen
  \bibfield  {author} {\bibinfo {author} {\bibfnamefont {Vatshal}\ \bibnamefont
  {Srivastav}}, \bibinfo {author} {\bibfnamefont {Natalia~Herrera}\
  \bibnamefont {Valencia}}, \bibinfo {author} {\bibfnamefont {Will}\
  \bibnamefont {McCutcheon}}, \bibinfo {author} {\bibfnamefont {Saroch}\
  \bibnamefont {Leedumrongwatthanakun}}, \bibinfo {author} {\bibfnamefont
  {S{\'e}bastien}\ \bibnamefont {Designolle}}, \bibinfo {author} {\bibfnamefont
  {Roope}\ \bibnamefont {Uola}}, \bibinfo {author} {\bibfnamefont {Nicolas}\
  \bibnamefont {Brunner}}, \ and\ \bibinfo {author} {\bibfnamefont {Mehul}\
  \bibnamefont {Malik}},\ }\emph {\enquote {\bibinfo {title} {Quick quantum
  steering: Overcoming loss and noise with qudits},}\ }\href
  {https://journals.aps.org/prx/abstract/10.1103/PhysRevX.12.041023} {\bibfield
   {journal} {\bibinfo  {journal} {Physical Review X}\ }\textbf {\bibinfo
  {volume} {12}},\ \bibinfo {pages} {041023} (\bibinfo {year}
  {2022}{\natexlab{a}})},\ \Eprint {http://arxiv.org/abs/2202.09294}
  {arXiv:2202.09294}\BibitemShut {NoStop}%
\bibitem [{\citenamefont {Valencia}\ \emph {et~al.}(2020)\citenamefont
  {Valencia}, \citenamefont {Goel}, \citenamefont {McCutcheon}, \citenamefont
  {Defienne},\ and\ \citenamefont {Malik}}]{Valencia2020}%
  \BibitemOpen
  \bibfield  {author} {\bibinfo {author} {\bibfnamefont {Natalia~Herrera}\
  \bibnamefont {Valencia}}, \bibinfo {author} {\bibfnamefont {Suraj}\
  \bibnamefont {Goel}}, \bibinfo {author} {\bibfnamefont {Will}\ \bibnamefont
  {McCutcheon}}, \bibinfo {author} {\bibfnamefont {Hugo}\ \bibnamefont
  {Defienne}}, \ and\ \bibinfo {author} {\bibfnamefont {Mehul}\ \bibnamefont
  {Malik}},\ }\emph {\enquote {\bibinfo {title} {{Unscrambling entanglement
  through a complex medium}},}\ }\href {\doibase 10.1038/s41567-020-0970-1}
  {\bibfield  {journal} {\bibinfo  {journal} {Nature Physics}\ }\textbf
  {\bibinfo {volume} {16}},\ \bibinfo {pages} {1112} (\bibinfo {year}
  {2020})},\ \Eprint {http://arxiv.org/abs/1910.04490}
  {arXiv:1910.04490}\BibitemShut {NoStop}%
\bibitem [{\citenamefont {Cao}\ \emph {et~al.}(2020)\citenamefont {Cao},
  \citenamefont {Gao}, \citenamefont {Zhang}, \citenamefont {Wang},
  \citenamefont {He}, \citenamefont {Liu}, \citenamefont {Zhou}, \citenamefont
  {Chen}, \citenamefont {Li}, \citenamefont {Yu}, \citenamefont {Romero},
  \citenamefont {Huang}, \citenamefont {Li},\ and\ \citenamefont
  {Guo}}]{Cao2020}%
  \BibitemOpen
  \bibfield  {author} {\bibinfo {author} {\bibfnamefont {Huan}\ \bibnamefont
  {Cao}}, \bibinfo {author} {\bibfnamefont {She-Cheng}\ \bibnamefont {Gao}},
  \bibinfo {author} {\bibfnamefont {Chao}\ \bibnamefont {Zhang}}, \bibinfo
  {author} {\bibfnamefont {Jian}\ \bibnamefont {Wang}}, \bibinfo {author}
  {\bibfnamefont {De-Yong}\ \bibnamefont {He}}, \bibinfo {author}
  {\bibfnamefont {Bi-Heng}\ \bibnamefont {Liu}}, \bibinfo {author}
  {\bibfnamefont {Zheng-Wei}\ \bibnamefont {Zhou}}, \bibinfo {author}
  {\bibfnamefont {Yu-Jie}\ \bibnamefont {Chen}}, \bibinfo {author}
  {\bibfnamefont {Zhao-Hui}\ \bibnamefont {Li}}, \bibinfo {author}
  {\bibfnamefont {Si-Yuan}\ \bibnamefont {Yu}}, \bibinfo {author}
  {\bibfnamefont {Jacquiline}\ \bibnamefont {Romero}}, \bibinfo {author}
  {\bibfnamefont {Yun-Feng}\ \bibnamefont {Huang}}, \bibinfo {author}
  {\bibfnamefont {Chuan-Feng}\ \bibnamefont {Li}}, \ and\ \bibinfo {author}
  {\bibfnamefont {Guang-Can}\ \bibnamefont {Guo}},\ }\emph {\enquote {\bibinfo
  {title} {{Distribution of high-dimensional orbital angular momentum
  entanglement over a 1 km few-mode fiber}},}\ }\href {\doibase
  10.1364/optica.381403} {\bibfield  {journal} {\bibinfo  {journal} {Optica}\
  }\textbf {\bibinfo {volume} {7}},\ \bibinfo {pages} {232} (\bibinfo {year}
  {2020})}\BibitemShut {NoStop}%
\bibitem [{\citenamefont {Bavaresco}\ \emph {et~al.}(2018)\citenamefont
  {Bavaresco}, \citenamefont {Valencia}, \citenamefont {Klöckl}, \citenamefont
  {Pivoluska}, \citenamefont {Erker}, \citenamefont {Friis}, \citenamefont
  {Malik},\ and\ \citenamefont {Huber}}]{Bavaresco:2018gw}%
  \BibitemOpen
  \bibfield  {author} {\bibinfo {author} {\bibfnamefont {Jessica}\ \bibnamefont
  {Bavaresco}}, \bibinfo {author} {\bibfnamefont {Natalia~Herrera}\
  \bibnamefont {Valencia}}, \bibinfo {author} {\bibfnamefont {Claude}\
  \bibnamefont {Klöckl}}, \bibinfo {author} {\bibfnamefont {Matej}\
  \bibnamefont {Pivoluska}}, \bibinfo {author} {\bibfnamefont {Paul}\
  \bibnamefont {Erker}}, \bibinfo {author} {\bibfnamefont {Nicolai}\
  \bibnamefont {Friis}}, \bibinfo {author} {\bibfnamefont {Mehul}\ \bibnamefont
  {Malik}}, \ and\ \bibinfo {author} {\bibfnamefont {Marcus}\ \bibnamefont
  {Huber}},\ }\emph {\enquote {\bibinfo {title} {{Measurements in two bases are
  sufficient for certifying high-dimensional entanglement}},}\ }\href {\doibase
  10.1038/s41567-018-0203-z} {\bibfield  {journal} {\bibinfo  {journal} {Nature
  Physics}\ }\textbf {\bibinfo {volume} {14}},\ \bibinfo {pages} {1032}
  (\bibinfo {year} {2018})},\ \Eprint {http://arxiv.org/abs/1709.07344}
  {1709.07344}\BibitemShut {NoStop}%
\bibitem [{\citenamefont {Friis}\ \emph {et~al.}(2019)\citenamefont {Friis},
  \citenamefont {Vitagliano}, \citenamefont {Malik},\ and\ \citenamefont
  {Huber}}]{Friis2019}%
  \BibitemOpen
  \bibfield  {author} {\bibinfo {author} {\bibfnamefont {Nicolai}\ \bibnamefont
  {Friis}}, \bibinfo {author} {\bibfnamefont {Giuseppe}\ \bibnamefont
  {Vitagliano}}, \bibinfo {author} {\bibfnamefont {Mehul}\ \bibnamefont
  {Malik}}, \ and\ \bibinfo {author} {\bibfnamefont {Marcus}\ \bibnamefont
  {Huber}},\ }\emph {{\selectlanguage {English}\enquote {\bibinfo {title}
  {{Entanglement certification from theory to experiment}},}\ }}\href {\doibase
  10.1038/s42254-018-0003-5} {\bibfield  {journal} {\bibinfo  {journal} {Nature
  Reviews Physics}\ }\textbf {\bibinfo {volume} {1}},\ \bibinfo {pages} {72 }
  (\bibinfo {year} {2019})}\BibitemShut {NoStop}%
\bibitem [{\citenamefont {{Herrera Valencia}}\ \emph
  {et~al.}(2020)\citenamefont {{Herrera Valencia}}, \citenamefont {Srivastav},
  \citenamefont {Pivoluska}, \citenamefont {Huber}, \citenamefont {Friis},
  \citenamefont {McCutcheon},\ and\ \citenamefont
  {Malik}}]{HerreraValencia2020}%
  \BibitemOpen
  \bibfield  {author} {\bibinfo {author} {\bibfnamefont {Natalia}\ \bibnamefont
  {{Herrera Valencia}}}, \bibinfo {author} {\bibfnamefont {Vatshal}\
  \bibnamefont {Srivastav}}, \bibinfo {author} {\bibfnamefont {Matej}\
  \bibnamefont {Pivoluska}}, \bibinfo {author} {\bibfnamefont {Marcus}\
  \bibnamefont {Huber}}, \bibinfo {author} {\bibfnamefont {Nicolai}\
  \bibnamefont {Friis}}, \bibinfo {author} {\bibfnamefont {Will}\ \bibnamefont
  {McCutcheon}}, \ and\ \bibinfo {author} {\bibfnamefont {Mehul}\ \bibnamefont
  {Malik}},\ }\emph {\enquote {\bibinfo {title} {{High-Dimensional Pixel
  Entanglement: Efficient Generation and Certification}},}\ }\href {\doibase
  10.22331/q-2020-12-24-376} {\bibfield  {journal} {\bibinfo  {journal}
  {Quantum}\ }\textbf {\bibinfo {volume} {4}},\ \bibinfo {pages} {376}
  (\bibinfo {year} {2020})},\ \Eprint {http://arxiv.org/abs/2004.04994}
  {arXiv:2004.04994}\BibitemShut {NoStop}%
\bibitem [{\citenamefont {Babazadeh}\ \emph {et~al.}(2017)\citenamefont
  {Babazadeh}, \citenamefont {Erhard}, \citenamefont {Wang}, \citenamefont
  {Malik}, \citenamefont {Nouroozi}, \citenamefont {Krenn},\ and\ \citenamefont
  {Zeilinger}}]{Babazadeh2017}%
  \BibitemOpen
  \bibfield  {author} {\bibinfo {author} {\bibfnamefont {Amin}\ \bibnamefont
  {Babazadeh}}, \bibinfo {author} {\bibfnamefont {Manuel}\ \bibnamefont
  {Erhard}}, \bibinfo {author} {\bibfnamefont {Feiran}\ \bibnamefont {Wang}},
  \bibinfo {author} {\bibfnamefont {Mehul}\ \bibnamefont {Malik}}, \bibinfo
  {author} {\bibfnamefont {Rahman}\ \bibnamefont {Nouroozi}}, \bibinfo {author}
  {\bibfnamefont {Mario}\ \bibnamefont {Krenn}}, \ and\ \bibinfo {author}
  {\bibfnamefont {Anton}\ \bibnamefont {Zeilinger}},\ }\emph {\enquote
  {\bibinfo {title} {{High-Dimensional Single-Photon Quantum Gates: Concepts
  and Experiments}},}\ }\href {\doibase 10.1103/PhysRevLett.119.180510}
  {\bibfield  {journal} {\bibinfo  {journal} {Physical Review Letters}\
  }\textbf {\bibinfo {volume} {119}},\ \bibinfo {pages} {180510} (\bibinfo
  {year} {2017})},\ \Eprint {http://arxiv.org/abs/1702.07299}
  {arXiv:1702.07299}\BibitemShut {NoStop}%
\bibitem [{\citenamefont {Wang}\ \emph {et~al.}(2017)\citenamefont {Wang},
  \citenamefont {Erhard}, \citenamefont {Babazadeh}, \citenamefont {Malik},
  \citenamefont {Krenn},\ and\ \citenamefont {Zeilinger}}]{Wang2017}%
  \BibitemOpen
  \bibfield  {author} {\bibinfo {author} {\bibfnamefont {Feiran}\ \bibnamefont
  {Wang}}, \bibinfo {author} {\bibfnamefont {Manuel}\ \bibnamefont {Erhard}},
  \bibinfo {author} {\bibfnamefont {Amin}\ \bibnamefont {Babazadeh}}, \bibinfo
  {author} {\bibfnamefont {Mehul}\ \bibnamefont {Malik}}, \bibinfo {author}
  {\bibfnamefont {Mario}\ \bibnamefont {Krenn}}, \ and\ \bibinfo {author}
  {\bibfnamefont {Anton}\ \bibnamefont {Zeilinger}},\ }\emph {\enquote
  {\bibinfo {title} {{Generation of the complete four-dimensional Bell
  basis}},}\ }\href {\doibase 10.1364/optica.4.001462} {\bibfield  {journal}
  {\bibinfo  {journal} {Optica}\ }\textbf {\bibinfo {volume} {4}},\ \bibinfo
  {pages} {1462} (\bibinfo {year} {2017})},\ \Eprint
  {http://arxiv.org/abs/1707.05760} {arXiv:1707.05760}\BibitemShut {NoStop}%
\bibitem [{\citenamefont {Brandt}\ \emph {et~al.}(2020)\citenamefont {Brandt},
  \citenamefont {Hiekkamäki}, \citenamefont {Bouchard}, \citenamefont
  {Huber},\ and\ \citenamefont {Fickler}}]{Brandt:2020er}%
  \BibitemOpen
  \bibfield  {author} {\bibinfo {author} {\bibfnamefont {Florian}\ \bibnamefont
  {Brandt}}, \bibinfo {author} {\bibfnamefont {Markus}\ \bibnamefont
  {Hiekkamäki}}, \bibinfo {author} {\bibfnamefont {Frédéric}\ \bibnamefont
  {Bouchard}}, \bibinfo {author} {\bibfnamefont {Marcus}\ \bibnamefont
  {Huber}}, \ and\ \bibinfo {author} {\bibfnamefont {Robert}\ \bibnamefont
  {Fickler}},\ }\emph {\enquote {\bibinfo {title} {{High-dimensional quantum
  gates using full-field spatial modes of photons}},}\ }\href {\doibase
  10.1364/optica.375875} {\bibfield  {journal} {\bibinfo  {journal} {Optica}\
  }\textbf {\bibinfo {volume} {7}},\ \bibinfo {pages} {98} (\bibinfo {year}
  {2020})},\ \Eprint {http://arxiv.org/abs/1907.13002} {1907.13002}\BibitemShut
  {NoStop}%
\bibitem [{\citenamefont {Lib}\ \emph {et~al.}(2022)\citenamefont {Lib},
  \citenamefont {Sulimany},\ and\ \citenamefont {Bromberg}}]{Lib2021}%
  \BibitemOpen
  \bibfield  {author} {\bibinfo {author} {\bibfnamefont {Ohad}\ \bibnamefont
  {Lib}}, \bibinfo {author} {\bibfnamefont {Kfir}\ \bibnamefont {Sulimany}}, \
  and\ \bibinfo {author} {\bibfnamefont {Yaron}\ \bibnamefont {Bromberg}},\
  }\emph {\enquote {\bibinfo {title} {{Processing Entangled Photons in High
  Dimensions with a Programmable Light Converter}},}\ }\href {\doibase
  10.1103/physrevapplied.18.014063} {\bibfield  {journal} {\bibinfo  {journal}
  {Physical Review Applied}\ }\textbf {\bibinfo {volume} {18}},\ \bibinfo
  {pages} {014063} (\bibinfo {year} {2022})},\ \Eprint
  {http://arxiv.org/abs/2108.02258} {2108.02258}\BibitemShut {NoStop}%
\bibitem [{\citenamefont {Rotter}\ and\ \citenamefont
  {Gigan}(2017)}]{Rotter2017}%
  \BibitemOpen
  \bibfield  {author} {\bibinfo {author} {\bibfnamefont {Stefan}\ \bibnamefont
  {Rotter}}\ and\ \bibinfo {author} {\bibfnamefont {Sylvain}\ \bibnamefont
  {Gigan}},\ }\emph {\enquote {\bibinfo {title} {{Light fields in complex
  media: Mesoscopic scattering meets wave control}},}\ }\href {\doibase
  10.1103/RevModPhys.89.015005} {\bibfield  {journal} {\bibinfo  {journal}
  {Reviews of Modern Physics}\ }\textbf {\bibinfo {volume} {89}},\ \bibinfo
  {pages} {015005} (\bibinfo {year} {2017})},\ \Eprint
  {http://arxiv.org/abs/1702.05395} {arXiv:1702.05395}\BibitemShut {NoStop}%
\bibitem [{\citenamefont {Cao}\ and\ \citenamefont {Eliezer}(2022)}]{Cao2022}%
  \BibitemOpen
  \bibfield  {author} {\bibinfo {author} {\bibfnamefont {Hui}\ \bibnamefont
  {Cao}}\ and\ \bibinfo {author} {\bibfnamefont {Yaniv}\ \bibnamefont
  {Eliezer}},\ }\emph {\enquote {\bibinfo {title} {Harnessing disorder for
  photonic device applications},}\ }\href {\doibase 10.1063/5.0076318}
  {\bibfield  {journal} {\bibinfo  {journal} {Applied Physics Reviews}\
  }\textbf {\bibinfo {volume} {9}},\ \bibinfo {pages} {011309} (\bibinfo {year}
  {2022})}\BibitemShut {NoStop}%
\bibitem [{\citenamefont {Huisman}\ \emph {et~al.}(2015)\citenamefont
  {Huisman}, \citenamefont {Huisman}, \citenamefont {Wolterink}, \citenamefont
  {Mosk},\ and\ \citenamefont {Pinkse}}]{Huisman2015}%
  \BibitemOpen
  \bibfield  {author} {\bibinfo {author} {\bibfnamefont {Simon~R.}\
  \bibnamefont {Huisman}}, \bibinfo {author} {\bibfnamefont {Thomas~J.}\
  \bibnamefont {Huisman}}, \bibinfo {author} {\bibfnamefont {Tom A.~W.}\
  \bibnamefont {Wolterink}}, \bibinfo {author} {\bibfnamefont {Allard~P.}\
  \bibnamefont {Mosk}}, \ and\ \bibinfo {author} {\bibfnamefont {Pepijn W.~H.}\
  \bibnamefont {Pinkse}},\ }\emph {\enquote {\bibinfo {title} {{Programmable
  multiport optical circuits in opaque scattering materials}},}\ }\href
  {\doibase 10.1364/OE.23.003102} {\bibfield  {journal} {\bibinfo  {journal}
  {Optics Express}\ }\textbf {\bibinfo {volume} {23}},\ \bibinfo {pages} {3102}
  (\bibinfo {year} {2015})},\ \Eprint {http://arxiv.org/abs/1408.1856}
  {arXiv:1408.1856}\BibitemShut {NoStop}%
\bibitem [{\citenamefont {Matth{\`{e}}s}\ \emph {et~al.}(2019)\citenamefont
  {Matth{\`{e}}s}, \citenamefont {del Hougne}, \citenamefont {de~Rosny},
  \citenamefont {Lerosey},\ and\ \citenamefont {Popoff}}]{Matthes2019}%
  \BibitemOpen
  \bibfield  {author} {\bibinfo {author} {\bibfnamefont {Maxime~W.}\
  \bibnamefont {Matth{\`{e}}s}}, \bibinfo {author} {\bibfnamefont {Philipp}\
  \bibnamefont {del Hougne}}, \bibinfo {author} {\bibfnamefont {Julien}\
  \bibnamefont {de~Rosny}}, \bibinfo {author} {\bibfnamefont {Geoffroy}\
  \bibnamefont {Lerosey}}, \ and\ \bibinfo {author} {\bibfnamefont
  {S{\'{e}}bastien~M.}\ \bibnamefont {Popoff}},\ }\emph {\enquote {\bibinfo
  {title} {{Optical complex media as universal reconfigurable linear
  operators}},}\ }\href {\doibase 10.1364/OPTICA.6.000465} {\bibfield
  {journal} {\bibinfo  {journal} {Optica}\ }\textbf {\bibinfo {volume} {6}},\
  \bibinfo {pages} {465} (\bibinfo {year} {2019})}\BibitemShut {NoStop}%
\bibitem [{\citenamefont {Wolterink}\ \emph {et~al.}(2015)\citenamefont
  {Wolterink}, \citenamefont {Uppu}, \citenamefont {Ctistis}, \citenamefont
  {Vos}, \citenamefont {Boller},\ and\ \citenamefont {Pinkse}}]{Wolterink2016}%
  \BibitemOpen
  \bibfield  {author} {\bibinfo {author} {\bibfnamefont {Tom A.~W.}\
  \bibnamefont {Wolterink}}, \bibinfo {author} {\bibfnamefont {Ravitej}\
  \bibnamefont {Uppu}}, \bibinfo {author} {\bibfnamefont {Georgios}\
  \bibnamefont {Ctistis}}, \bibinfo {author} {\bibfnamefont {Willem~L.}\
  \bibnamefont {Vos}}, \bibinfo {author} {\bibfnamefont {Klaus~J.}\
  \bibnamefont {Boller}}, \ and\ \bibinfo {author} {\bibfnamefont {Pepijn
  W.~H.}\ \bibnamefont {Pinkse}},\ }\emph {\enquote {\bibinfo {title}
  {{Programmable two-photon quantum interference in $10^3$ channels in opaque
  scattering media}},}\ }\href {\doibase 10.1103/PhysRevA.93.053817} {\bibfield
   {journal} {\bibinfo  {journal} {Physical Review A}\ }\textbf {\bibinfo
  {volume} {93}},\ \bibinfo {pages} {053817} (\bibinfo {year} {2015})},\
  \Eprint {http://arxiv.org/abs/1511.00897} {arXiv:1511.00897}\BibitemShut
  {NoStop}%
\bibitem [{\citenamefont {Defienne}\ \emph {et~al.}(2016)\citenamefont
  {Defienne}, \citenamefont {Barbieri}, \citenamefont {Walmsley}, \citenamefont
  {Smith},\ and\ \citenamefont {Gigan}}]{Defienne2016}%
  \BibitemOpen
  \bibfield  {author} {\bibinfo {author} {\bibfnamefont {Hugo}\ \bibnamefont
  {Defienne}}, \bibinfo {author} {\bibfnamefont {Marco}\ \bibnamefont
  {Barbieri}}, \bibinfo {author} {\bibfnamefont {Ian~A.}\ \bibnamefont
  {Walmsley}}, \bibinfo {author} {\bibfnamefont {Brian~J.}\ \bibnamefont
  {Smith}}, \ and\ \bibinfo {author} {\bibfnamefont {Sylvain}\ \bibnamefont
  {Gigan}},\ }\emph {\enquote {\bibinfo {title} {{Two-photon quantum walk in a
  multimode fiber}},}\ }\href {\doibase 10.1126/sciadv.1501054} {\bibfield
  {journal} {\bibinfo  {journal} {Science Advances}\ }\textbf {\bibinfo
  {volume} {2}},\ \bibinfo {pages} {e1501054} (\bibinfo {year} {2016})},\
  \Eprint {http://arxiv.org/abs/1504.03178} {arXiv:1504.03178}\BibitemShut
  {NoStop}%
\bibitem [{\citenamefont {Leedumrongwatthanakun}\ \emph
  {et~al.}(2020)\citenamefont {Leedumrongwatthanakun}, \citenamefont
  {Innocenti}, \citenamefont {Defienne}, \citenamefont {Juffmann},
  \citenamefont {Ferraro}, \citenamefont {Paternostro},\ and\ \citenamefont
  {Gigan}}]{Leedumrongwatthanakun2020}%
  \BibitemOpen
  \bibfield  {author} {\bibinfo {author} {\bibfnamefont {Saroch}\ \bibnamefont
  {Leedumrongwatthanakun}}, \bibinfo {author} {\bibfnamefont {Luca}\
  \bibnamefont {Innocenti}}, \bibinfo {author} {\bibfnamefont {Hugo}\
  \bibnamefont {Defienne}}, \bibinfo {author} {\bibfnamefont {Thomas}\
  \bibnamefont {Juffmann}}, \bibinfo {author} {\bibfnamefont {Alessandro}\
  \bibnamefont {Ferraro}}, \bibinfo {author} {\bibfnamefont {Mauro}\
  \bibnamefont {Paternostro}}, \ and\ \bibinfo {author} {\bibfnamefont
  {Sylvain}\ \bibnamefont {Gigan}},\ }\emph {\enquote {\bibinfo {title}
  {{Programmable linear quantum networks with a multimode fibre}},}\ }\href
  {\doibase 10.1038/s41566-019-0553-9} {\bibfield  {journal} {\bibinfo
  {journal} {Nature Photonics}\ }\textbf {\bibinfo {volume} {14}},\ \bibinfo
  {pages} {139} (\bibinfo {year} {2020})},\ \Eprint
  {http://arxiv.org/abs/1902.10678} {arXiv:1902.10678}\BibitemShut {NoStop}%
\bibitem [{\citenamefont {Valencia}\ \emph {et~al.}(2021)\citenamefont
  {Valencia}, \citenamefont {Srivastav}, \citenamefont {Leedumrongwatthanakun},
  \citenamefont {McCutcheon},\ and\ \citenamefont {Malik}}]{Valencia2021}%
  \BibitemOpen
  \bibfield  {author} {\bibinfo {author} {\bibfnamefont {Natalia~Herrera}\
  \bibnamefont {Valencia}}, \bibinfo {author} {\bibfnamefont {Vatshal}\
  \bibnamefont {Srivastav}}, \bibinfo {author} {\bibfnamefont {Saroch}\
  \bibnamefont {Leedumrongwatthanakun}}, \bibinfo {author} {\bibfnamefont
  {Will}\ \bibnamefont {McCutcheon}}, \ and\ \bibinfo {author} {\bibfnamefont
  {Mehul}\ \bibnamefont {Malik}},\ }\emph {\enquote {\bibinfo {title}
  {{Entangled ripples and twists of light: radial and azimuthal
  Laguerre–Gaussian mode entanglement}},}\ }\href {\doibase
  10.1088/2040-8986/ac213c} {\bibfield  {journal} {\bibinfo  {journal} {Journal
  of Optics}\ }\textbf {\bibinfo {volume} {23}},\ \bibinfo {pages} {104001}
  (\bibinfo {year} {2021})},\ \Eprint {http://arxiv.org/abs/2104.04506}
  {arXiv:2104.04506}\BibitemShut {NoStop}%
\bibitem [{\citenamefont {Bouchard}\ \emph {et~al.}(2018)\citenamefont
  {Bouchard}, \citenamefont {Valencia}, \citenamefont {Brandt}, \citenamefont
  {Fickler}, \citenamefont {Huber},\ and\ \citenamefont
  {Malik}}]{Bouchard:2018hr}%
  \BibitemOpen
  \bibfield  {author} {\bibinfo {author} {\bibfnamefont {Fr{\'{e}}d{\'{e}}ric}\
  \bibnamefont {Bouchard}}, \bibinfo {author} {\bibfnamefont {Natalia~Herrera}\
  \bibnamefont {Valencia}}, \bibinfo {author} {\bibfnamefont {Florian}\
  \bibnamefont {Brandt}}, \bibinfo {author} {\bibfnamefont {Robert}\
  \bibnamefont {Fickler}}, \bibinfo {author} {\bibfnamefont {Marcus}\
  \bibnamefont {Huber}}, \ and\ \bibinfo {author} {\bibfnamefont {Mehul}\
  \bibnamefont {Malik}},\ }\emph {\enquote {\bibinfo {title} {{Measuring
  azimuthal and radial modes of photons}},}\ }\href {\doibase
  10.1364/OE.26.031925} {\bibfield  {journal} {\bibinfo  {journal} {Opt.
  Express}\ }\textbf {\bibinfo {volume} {26}},\ \bibinfo {pages} {31925}
  (\bibinfo {year} {2018})},\ \Eprint {http://arxiv.org/abs/arXiv:1808.03533}
  {arXiv:1808.03533}\BibitemShut {NoStop}%
\bibitem [{\citenamefont {Allmaras}\ \emph {et~al.}(2020)\citenamefont
  {Allmaras}, \citenamefont {Wollman}, \citenamefont {Beyer}, \citenamefont
  {Briggs}, \citenamefont {Korzh}, \citenamefont {Bumble},\ and\ \citenamefont
  {Shaw}}]{Allmaras2020}%
  \BibitemOpen
  \bibfield  {author} {\bibinfo {author} {\bibfnamefont {Jason~P.}\
  \bibnamefont {Allmaras}}, \bibinfo {author} {\bibfnamefont {Emma~E.}\
  \bibnamefont {Wollman}}, \bibinfo {author} {\bibfnamefont {Andrew~D.}\
  \bibnamefont {Beyer}}, \bibinfo {author} {\bibfnamefont {Ryan~M.}\
  \bibnamefont {Briggs}}, \bibinfo {author} {\bibfnamefont {Boris~A.}\
  \bibnamefont {Korzh}}, \bibinfo {author} {\bibfnamefont {Bruce}\ \bibnamefont
  {Bumble}}, \ and\ \bibinfo {author} {\bibfnamefont {Matthew~D.}\ \bibnamefont
  {Shaw}},\ }\emph {\enquote {\bibinfo {title} {{Demonstration of a Thermally
  Coupled Row-Column SNSPD Imaging Array}},}\ }\href {\doibase
  10.1021/acs.nanolett.0c00246} {\bibfield  {journal} {\bibinfo  {journal}
  {Nano Letters}\ }\textbf {\bibinfo {volume} {20}},\ \bibinfo {pages} {2163}
  (\bibinfo {year} {2020})},\ \Eprint {http://arxiv.org/abs/2002.10613}
  {2002.10613}\BibitemShut {NoStop}%
\bibitem [{\citenamefont {Designolle}\ \emph {et~al.}(2021)\citenamefont
  {Designolle}, \citenamefont {Srivastav}, \citenamefont {Uola}, \citenamefont
  {Valencia}, \citenamefont {McCutcheon}, \citenamefont {Malik},\ and\
  \citenamefont {Brunner}}]{Designolle2021}%
  \BibitemOpen
  \bibfield  {author} {\bibinfo {author} {\bibfnamefont {Sébastien}\
  \bibnamefont {Designolle}}, \bibinfo {author} {\bibfnamefont {Vatshal}\
  \bibnamefont {Srivastav}}, \bibinfo {author} {\bibfnamefont {Roope}\
  \bibnamefont {Uola}}, \bibinfo {author} {\bibfnamefont {Natalia~Herrera}\
  \bibnamefont {Valencia}}, \bibinfo {author} {\bibfnamefont {Will}\
  \bibnamefont {McCutcheon}}, \bibinfo {author} {\bibfnamefont {Mehul}\
  \bibnamefont {Malik}}, \ and\ \bibinfo {author} {\bibfnamefont {Nicolas}\
  \bibnamefont {Brunner}},\ }\emph {\enquote {\bibinfo {title} {Genuine
  high-dimensional quantum steering},}\ }\href {\doibase
  10.1103/physrevlett.126.200404} {\bibfield  {journal} {\bibinfo  {journal}
  {Physical Review Letters}\ }\textbf {\bibinfo {volume} {126}},\ \bibinfo
  {pages} {200404} (\bibinfo {year} {2021})}\BibitemShut {NoStop}%
\bibitem [{\citenamefont {Miller}(2012)}]{Miller2012a}%
  \BibitemOpen
  \bibfield  {author} {\bibinfo {author} {\bibfnamefont {David A.~B.}\
  \bibnamefont {Miller}},\ }\emph {\enquote {\bibinfo {title} {{All linear
  optical devices are mode converters}},}\ }\href {\doibase
  10.1364/OE.20.023985} {\bibfield  {journal} {\bibinfo  {journal} {Optics
  Express}\ }\textbf {\bibinfo {volume} {20}},\ \bibinfo {pages} {23985}
  (\bibinfo {year} {2012})},\ \Eprint {http://arxiv.org/abs/1209.4931}
  {arXiv:1209.4931}\BibitemShut {NoStop}%
\bibitem [{\citenamefont {Miller}(2019)}]{Miller2019}%
  \BibitemOpen
  \bibfield  {author} {\bibinfo {author} {\bibfnamefont {David A.~B.}\
  \bibnamefont {Miller}},\ }\emph {\enquote {\bibinfo {title} {{Waves, modes,
  communications, and optics: a tutorial}},}\ }\href {\doibase
  10.1364/aop.11.000679} {\bibfield  {journal} {\bibinfo  {journal} {Advances
  in Optics and Photonics}\ }\textbf {\bibinfo {volume} {11}},\ \bibinfo
  {pages} {679} (\bibinfo {year} {2019})}\BibitemShut {NoStop}%
\bibitem [{\citenamefont {Carpenter}\ \emph {et~al.}(2014)\citenamefont
  {Carpenter}, \citenamefont {Eggleton},\ and\ \citenamefont
  {Schr{\"{o}}der}}]{Carpenter2014a}%
  \BibitemOpen
  \bibfield  {author} {\bibinfo {author} {\bibfnamefont {Joel}\ \bibnamefont
  {Carpenter}}, \bibinfo {author} {\bibfnamefont {Benjamin~J.}\ \bibnamefont
  {Eggleton}}, \ and\ \bibinfo {author} {\bibfnamefont {Jochen}\ \bibnamefont
  {Schr{\"{o}}der}},\ }\emph {\enquote {\bibinfo {title} {{110X110 Optical Mode
  Transfer Matrix Inversion}},}\ }\href {\doibase 10.1364/OE.22.000096}
  {\bibfield  {journal} {\bibinfo  {journal} {Optics Express}\ }\textbf
  {\bibinfo {volume} {22}},\ \bibinfo {pages} {96} (\bibinfo {year}
  {2014})}\BibitemShut {NoStop}%
\bibitem [{\citenamefont {Pl{\"{o}}schner}\ \emph {et~al.}(2015)\citenamefont
  {Pl{\"{o}}schner}, \citenamefont {Tyc},\ and\ \citenamefont
  {{\v{C}}i{\v{z}}m{\'{a}}r}}]{Ploschner2015}%
  \BibitemOpen
  \bibfield  {author} {\bibinfo {author} {\bibfnamefont {Martin}\ \bibnamefont
  {Pl{\"{o}}schner}}, \bibinfo {author} {\bibfnamefont {Tom{\'{a}}{\v{s}}}\
  \bibnamefont {Tyc}}, \ and\ \bibinfo {author} {\bibfnamefont
  {Tom{\'{a}}{\v{s}}}\ \bibnamefont {{\v{C}}i{\v{z}}m{\'{a}}r}},\ }\emph
  {\enquote {\bibinfo {title} {{Seeing through chaos in multimode fibres}},}\
  }\href {\doibase 10.1038/nphoton.2015.112} {\bibfield  {journal} {\bibinfo
  {journal} {Nature Photonics}\ }\textbf {\bibinfo {volume} {9}},\ \bibinfo
  {pages} {529} (\bibinfo {year} {2015})}\BibitemShut {NoStop}%
\bibitem [{\citenamefont {Xiong}\ \emph {et~al.}(2018)\citenamefont {Xiong},
  \citenamefont {Hsu}, \citenamefont {Bromberg}, \citenamefont {Antonio-Lopez},
  \citenamefont {{Amezcua Correa}},\ and\ \citenamefont {Cao}}]{Xiong2018}%
  \BibitemOpen
  \bibfield  {author} {\bibinfo {author} {\bibfnamefont {Wen}\ \bibnamefont
  {Xiong}}, \bibinfo {author} {\bibfnamefont {Chia~Wei}\ \bibnamefont {Hsu}},
  \bibinfo {author} {\bibfnamefont {Yaron}\ \bibnamefont {Bromberg}}, \bibinfo
  {author} {\bibfnamefont {Jose~Enrique}\ \bibnamefont {Antonio-Lopez}},
  \bibinfo {author} {\bibfnamefont {Rodrigo}\ \bibnamefont {{Amezcua Correa}}},
  \ and\ \bibinfo {author} {\bibfnamefont {Hui}\ \bibnamefont {Cao}},\ }\emph
  {\enquote {\bibinfo {title} {{Complete polarization control in multimode
  fibers with polarization and mode coupling}},}\ }\href {\doibase
  10.1038/s41377-018-0047-4} {\bibfield  {journal} {\bibinfo  {journal} {Light:
  Science \& Applications}\ }\textbf {\bibinfo {volume} {7}},\ \bibinfo {pages}
  {54} (\bibinfo {year} {2018})},\ \Eprint {http://arxiv.org/abs/1709.01029}
  {arXiv:1709.01029}\BibitemShut {NoStop}%
\bibitem [{\citenamefont {Goel}\ \emph {et~al.}(2023)\citenamefont {Goel},
  \citenamefont {Conti}, \citenamefont {Leedumrongwatthanakun},\ and\
  \citenamefont {Malik}}]{goel2023referenceless}%
  \BibitemOpen
  \bibfield  {author} {\bibinfo {author} {\bibfnamefont {Suraj}\ \bibnamefont
  {Goel}}, \bibinfo {author} {\bibfnamefont {Claudio}\ \bibnamefont {Conti}},
  \bibinfo {author} {\bibfnamefont {Saroch}\ \bibnamefont
  {Leedumrongwatthanakun}}, \ and\ \bibinfo {author} {\bibfnamefont {Mehul}\
  \bibnamefont {Malik}},\ }\emph {\enquote {\bibinfo {title} {Referenceless
  characterisation of complex media using physics-informed neural networks},}\
  }\href {\doibase 10.1364/oe.500529} {\bibfield  {journal} {\bibinfo
  {journal} {Optics Express}\ }\textbf {\bibinfo {volume} {31}},\ \bibinfo
  {pages} {32824} (\bibinfo {year} {2023})}\BibitemShut {NoStop}%
\bibitem [{\citenamefont {Sakamaki}\ \emph {et~al.}(2007)\citenamefont
  {Sakamaki}, \citenamefont {Saida}, \citenamefont {Hashimoto},\ and\
  \citenamefont {Takahashi}}]{Sakamaki2007}%
  \BibitemOpen
  \bibfield  {author} {\bibinfo {author} {\bibfnamefont {Y.}~\bibnamefont
  {Sakamaki}}, \bibinfo {author} {\bibfnamefont {T.}~\bibnamefont {Saida}},
  \bibinfo {author} {\bibfnamefont {T.}~\bibnamefont {Hashimoto}}, \ and\
  \bibinfo {author} {\bibfnamefont {H.}~\bibnamefont {Takahashi}},\ }\emph
  {\enquote {\bibinfo {title} {{New Optical Waveguide Design Based on Wavefront
  Matching Method}},}\ }\href {\doibase 10.1109/JLT.2007.906798} {\bibfield
  {journal} {\bibinfo  {journal} {Journal of Lightwave Technology}\ }\textbf
  {\bibinfo {volume} {25}},\ \bibinfo {pages} {3511} (\bibinfo {year}
  {2007})}\BibitemShut {NoStop}%
\bibitem [{\citenamefont {Fontaine}\ \emph {et~al.}(2017)\citenamefont
  {Fontaine}, \citenamefont {Ryf}, \citenamefont {Chen}, \citenamefont
  {Neilson},\ and\ \citenamefont {Carpenter}}]{Fontaine2017}%
  \BibitemOpen
  \bibfield  {author} {\bibinfo {author} {\bibfnamefont {Nicolas~K.}\
  \bibnamefont {Fontaine}}, \bibinfo {author} {\bibfnamefont {Roland}\
  \bibnamefont {Ryf}}, \bibinfo {author} {\bibfnamefont {Haoshuo}\ \bibnamefont
  {Chen}}, \bibinfo {author} {\bibfnamefont {David}\ \bibnamefont {Neilson}}, \
  and\ \bibinfo {author} {\bibfnamefont {Joel}\ \bibnamefont {Carpenter}},\
  }\emph {\enquote {\bibinfo {title} {{Design of High Order Mode-Multiplexers
  using Multiplane Light Conversion}},}\ }in\ \href {\doibase
  10.1109/ECOC.2017.8346129} {\emph {\bibinfo {booktitle} {2017 European
  Conference on Optical Communication (ECOC)}}},\ Vol.\ \bibinfo {volume}
  {2017-Septe}\ (\bibinfo  {publisher} {IEEE},\ \bibinfo {year} {2017})\ pp.\
  \bibinfo {pages} {1--3}\BibitemShut {NoStop}%
\bibitem [{\citenamefont {Qassim}\ \emph {et~al.}(2014)\citenamefont {Qassim},
  \citenamefont {Miatto}, \citenamefont {Torres}, \citenamefont {Padgett},
  \citenamefont {Karimi},\ and\ \citenamefont {Boyd}}]{Qassim:2014fp}%
  \BibitemOpen
  \bibfield  {author} {\bibinfo {author} {\bibfnamefont {Hammam}\ \bibnamefont
  {Qassim}}, \bibinfo {author} {\bibfnamefont {Filippo~M}\ \bibnamefont
  {Miatto}}, \bibinfo {author} {\bibfnamefont {Juan~P}\ \bibnamefont {Torres}},
  \bibinfo {author} {\bibfnamefont {Miles~J}\ \bibnamefont {Padgett}}, \bibinfo
  {author} {\bibfnamefont {Ebrahim}\ \bibnamefont {Karimi}}, \ and\ \bibinfo
  {author} {\bibfnamefont {Robert~W}\ \bibnamefont {Boyd}},\ }\emph {\enquote
  {\bibinfo {title} {{Limitations to the determination of a Laguerre-Gauss
  spectrum via projective, phase-flattening measurement}},}\ }\href {\doibase
  10.1364/JOSAB.31.000A20} {\bibfield  {journal} {\bibinfo  {journal} {J. Opt.
  Soc. Am. B}\ }\textbf {\bibinfo {volume} {31}},\ \bibinfo {pages} {A20}
  (\bibinfo {year} {2014})},\ \Eprint {http://arxiv.org/abs/arXiv:1401.3512}
  {arXiv:1401.3512}\BibitemShut {NoStop}%
\bibitem [{\citenamefont {Altepeter}\ \emph {et~al.}(2003)\citenamefont
  {Altepeter}, \citenamefont {Branning}, \citenamefont {Jeffrey}, \citenamefont
  {Wei}, \citenamefont {Kwiat}, \citenamefont {Thew}, \citenamefont
  {O’Brien}, \citenamefont {Nielsen},\ and\ \citenamefont
  {White}}]{Altepeter2003}%
  \BibitemOpen
  \bibfield  {author} {\bibinfo {author} {\bibfnamefont {J~B}\ \bibnamefont
  {Altepeter}}, \bibinfo {author} {\bibfnamefont {D}~\bibnamefont {Branning}},
  \bibinfo {author} {\bibfnamefont {E}~\bibnamefont {Jeffrey}}, \bibinfo
  {author} {\bibfnamefont {T~C}\ \bibnamefont {Wei}}, \bibinfo {author}
  {\bibfnamefont {P~G}\ \bibnamefont {Kwiat}}, \bibinfo {author} {\bibfnamefont
  {R~T}\ \bibnamefont {Thew}}, \bibinfo {author} {\bibfnamefont {J~L}\
  \bibnamefont {O’Brien}}, \bibinfo {author} {\bibfnamefont {M~A}\
  \bibnamefont {Nielsen}}, \ and\ \bibinfo {author} {\bibfnamefont {A.~G.}\
  \bibnamefont {White}},\ }\emph {\enquote {\bibinfo {title} {Ancilla-assisted
  quantum process tomography},}\ }\href {\doibase
  10.1103/physrevlett.90.193601} {\bibfield  {journal} {\bibinfo  {journal}
  {Physical Review Letters}\ }\textbf {\bibinfo {volume} {90}},\ \bibinfo
  {pages} {193601} (\bibinfo {year} {2003})}\BibitemShut {NoStop}%
\bibitem [{\citenamefont {D’Ariano}\ and\ \citenamefont
  {Presti}(2003)}]{DAriano2003}%
  \BibitemOpen
  \bibfield  {author} {\bibinfo {author} {\bibfnamefont {Giacomo~Mauro}\
  \bibnamefont {D’Ariano}}\ and\ \bibinfo {author} {\bibfnamefont
  {Paoloplacido~Lo}\ \bibnamefont {Presti}},\ }\emph {\enquote {\bibinfo
  {title} {{Imprinting Complete Information about a Quantum Channel on its
  Output State}},}\ }\href {\doibase 10.1103/physrevlett.91.047902} {\bibfield
  {journal} {\bibinfo  {journal} {Physical Review Letters}\ }\textbf {\bibinfo
  {volume} {91}},\ \bibinfo {pages} {047902} (\bibinfo {year} {2003})},\
  \Eprint {http://arxiv.org/abs/quant-ph/0211133}
  {quant-ph/0211133}\BibitemShut {NoStop}%
\bibitem [{\citenamefont {Morelli}\ \emph {et~al.}(2022)\citenamefont
  {Morelli}, \citenamefont {Yamasaki}, \citenamefont {Huber},\ and\
  \citenamefont {Tavakoli}}]{Morelli2022}%
  \BibitemOpen
  \bibfield  {author} {\bibinfo {author} {\bibfnamefont {Simon}\ \bibnamefont
  {Morelli}}, \bibinfo {author} {\bibfnamefont {Hayata}\ \bibnamefont
  {Yamasaki}}, \bibinfo {author} {\bibfnamefont {Marcus}\ \bibnamefont
  {Huber}}, \ and\ \bibinfo {author} {\bibfnamefont {Armin}\ \bibnamefont
  {Tavakoli}},\ }\emph {\enquote {\bibinfo {title} {Entanglement detection with
  imprecise measurements},}\ }\href {\doibase 10.1103/PhysRevLett.128.250501}
  {\bibfield  {journal} {\bibinfo  {journal} {Phys. Rev. Lett.}\ }\textbf
  {\bibinfo {volume} {128}},\ \bibinfo {pages} {250501} (\bibinfo {year}
  {2022})}\BibitemShut {NoStop}%
\bibitem [{\citenamefont {Rosset}\ \emph {et~al.}(2012)\citenamefont {Rosset},
  \citenamefont {Ferretti-Sch\"obitz}, \citenamefont {Bancal}, \citenamefont
  {Gisin},\ and\ \citenamefont {Liang}}]{Rosset2012}%
  \BibitemOpen
  \bibfield  {author} {\bibinfo {author} {\bibfnamefont {Denis}\ \bibnamefont
  {Rosset}}, \bibinfo {author} {\bibfnamefont {Raphael}\ \bibnamefont
  {Ferretti-Sch\"obitz}}, \bibinfo {author} {\bibfnamefont {Jean-Daniel}\
  \bibnamefont {Bancal}}, \bibinfo {author} {\bibfnamefont {Nicolas}\
  \bibnamefont {Gisin}}, \ and\ \bibinfo {author} {\bibfnamefont
  {Yeong-Cherng}\ \bibnamefont {Liang}},\ }\emph {\enquote {\bibinfo {title}
  {Imperfect measurement settings: Implications for quantum state tomography
  and entanglement witnesses},}\ }\href {\doibase 10.1103/PhysRevA.86.062325}
  {\bibfield  {journal} {\bibinfo  {journal} {Phys. Rev. A}\ }\textbf {\bibinfo
  {volume} {86}},\ \bibinfo {pages} {062325} (\bibinfo {year}
  {2012})}\BibitemShut {NoStop}%
\bibitem [{\citenamefont {Saygin}\ \emph {et~al.}(2020)\citenamefont {Saygin},
  \citenamefont {Kondratyev}, \citenamefont {Dyakonov}, \citenamefont
  {Mironov}, \citenamefont {Straupe},\ and\ \citenamefont
  {Kulik}}]{Saygin2020}%
  \BibitemOpen
  \bibfield  {author} {\bibinfo {author} {\bibfnamefont {M.~Yu.}\ \bibnamefont
  {Saygin}}, \bibinfo {author} {\bibfnamefont {I.~V.}\ \bibnamefont
  {Kondratyev}}, \bibinfo {author} {\bibfnamefont {I.~V.}\ \bibnamefont
  {Dyakonov}}, \bibinfo {author} {\bibfnamefont {S.~A.}\ \bibnamefont
  {Mironov}}, \bibinfo {author} {\bibfnamefont {S.~S.}\ \bibnamefont
  {Straupe}}, \ and\ \bibinfo {author} {\bibfnamefont {S.~P.}\ \bibnamefont
  {Kulik}},\ }\emph {\enquote {\bibinfo {title} {{Robust Architecture for
  Programmable Universal Unitaries}},}\ }\href {\doibase
  10.1103/physrevlett.124.010501} {\bibfield  {journal} {\bibinfo  {journal}
  {Physical Review Letters}\ }\textbf {\bibinfo {volume} {124}},\ \bibinfo
  {pages} {010501} (\bibinfo {year} {2020})},\ \Eprint
  {http://arxiv.org/abs/1906.06748} {1906.06748}\BibitemShut {NoStop}%
\bibitem [{\citenamefont {Pereira}\ \emph {et~al.}(2020)\citenamefont
  {Pereira}, \citenamefont {Rojas}, \citenamefont {Ca{\~{n}}as}, \citenamefont
  {Lima}, \citenamefont {Delgado},\ and\ \citenamefont
  {Cabello}}]{Pereira2020}%
  \BibitemOpen
  \bibfield  {author} {\bibinfo {author} {\bibfnamefont {Luciano}\ \bibnamefont
  {Pereira}}, \bibinfo {author} {\bibfnamefont {Alejandro}\ \bibnamefont
  {Rojas}}, \bibinfo {author} {\bibfnamefont {Gustavo}\ \bibnamefont
  {Ca{\~{n}}as}}, \bibinfo {author} {\bibfnamefont {Gustavo}\ \bibnamefont
  {Lima}}, \bibinfo {author} {\bibfnamefont {Aldo}\ \bibnamefont {Delgado}}, \
  and\ \bibinfo {author} {\bibfnamefont {Ad{\'{a}}n}\ \bibnamefont {Cabello}},\
  }\emph {\enquote {\bibinfo {title} {{Universal multi-port interferometers
  with minimal optical depth}},}\ }\href {http://arxiv.org/abs/2002.01371} {\
  (\bibinfo {year} {2020})},\ \Eprint {http://arxiv.org/abs/2002.01371}
  {arXiv:2002.01371}\BibitemShut {NoStop}%
\bibitem [{\citenamefont {Bruck}\ \emph {et~al.}(2016)\citenamefont {Bruck},
  \citenamefont {Vynck}, \citenamefont {Lalanne}, \citenamefont {Mills},
  \citenamefont {Thomson}, \citenamefont {Mashanovich}, \citenamefont {Reed},\
  and\ \citenamefont {Muskens}}]{Bruck2016}%
  \BibitemOpen
  \bibfield  {author} {\bibinfo {author} {\bibfnamefont {Roman}\ \bibnamefont
  {Bruck}}, \bibinfo {author} {\bibfnamefont {Kevin}\ \bibnamefont {Vynck}},
  \bibinfo {author} {\bibfnamefont {Philippe}\ \bibnamefont {Lalanne}},
  \bibinfo {author} {\bibfnamefont {Ben}\ \bibnamefont {Mills}}, \bibinfo
  {author} {\bibfnamefont {David~J.}\ \bibnamefont {Thomson}}, \bibinfo
  {author} {\bibfnamefont {Goran~Z.}\ \bibnamefont {Mashanovich}}, \bibinfo
  {author} {\bibfnamefont {Graham~T.}\ \bibnamefont {Reed}}, \ and\ \bibinfo
  {author} {\bibfnamefont {Otto~L.}\ \bibnamefont {Muskens}},\ }\emph {\enquote
  {\bibinfo {title} {{All-optical spatial light modulator for reconfigurable
  silicon photonic circuits}},}\ }\href {\doibase 10.1364/OPTICA.3.000396}
  {\bibfield  {journal} {\bibinfo  {journal} {Optica}\ }\textbf {\bibinfo
  {volume} {3}},\ \bibinfo {pages} {396} (\bibinfo {year} {2016})},\ \Eprint
  {http://arxiv.org/abs/1601.06679} {arXiv:1601.06679}\BibitemShut {NoStop}%
\bibitem [{\citenamefont {Wang}\ \emph {et~al.}(2019)\citenamefont {Wang},
  \citenamefont {Li}, \citenamefont {Soman}, \citenamefont {Mao}, \citenamefont
  {Kananen},\ and\ \citenamefont {Gu}}]{Wang2019c}%
  \BibitemOpen
  \bibfield  {author} {\bibinfo {author} {\bibfnamefont {Zi}~\bibnamefont
  {Wang}}, \bibinfo {author} {\bibfnamefont {Tiantian}\ \bibnamefont {Li}},
  \bibinfo {author} {\bibfnamefont {Anishkumar}\ \bibnamefont {Soman}},
  \bibinfo {author} {\bibfnamefont {Dun}\ \bibnamefont {Mao}}, \bibinfo
  {author} {\bibfnamefont {Thomas}\ \bibnamefont {Kananen}}, \ and\ \bibinfo
  {author} {\bibfnamefont {Tingyi}\ \bibnamefont {Gu}},\ }\emph {\enquote
  {\bibinfo {title} {{On-chip wavefront shaping with dielectric
  metasurface}},}\ }\href {\doibase 10.1038/s41467-019-11578-y} {\bibfield
  {journal} {\bibinfo  {journal} {Nature Communications}\ }\textbf {\bibinfo
  {volume} {10}},\ \bibinfo {pages} {3547} (\bibinfo {year}
  {2019})}\BibitemShut {NoStop}%
\bibitem [{\citenamefont {Dinsdale}\ \emph {et~al.}(2021)\citenamefont
  {Dinsdale}, \citenamefont {Wiecha}, \citenamefont {Delaney}, \citenamefont
  {Reynolds}, \citenamefont {Ebert}, \citenamefont {Zeimpekis}, \citenamefont
  {Thomson}, \citenamefont {Reed}, \citenamefont {Lalanne}, \citenamefont
  {Vynck},\ and\ \citenamefont {Muskens}}]{Dinsdale2021}%
  \BibitemOpen
  \bibfield  {author} {\bibinfo {author} {\bibfnamefont {Nicholas~J.}\
  \bibnamefont {Dinsdale}}, \bibinfo {author} {\bibfnamefont {Peter~R.}\
  \bibnamefont {Wiecha}}, \bibinfo {author} {\bibfnamefont {Matthew}\
  \bibnamefont {Delaney}}, \bibinfo {author} {\bibfnamefont {Jamie}\
  \bibnamefont {Reynolds}}, \bibinfo {author} {\bibfnamefont {Martin}\
  \bibnamefont {Ebert}}, \bibinfo {author} {\bibfnamefont {Ioannis}\
  \bibnamefont {Zeimpekis}}, \bibinfo {author} {\bibfnamefont {David~J.}\
  \bibnamefont {Thomson}}, \bibinfo {author} {\bibfnamefont {Graham~T.}\
  \bibnamefont {Reed}}, \bibinfo {author} {\bibfnamefont {Philippe}\
  \bibnamefont {Lalanne}}, \bibinfo {author} {\bibfnamefont {Kevin}\
  \bibnamefont {Vynck}}, \ and\ \bibinfo {author} {\bibfnamefont {Otto~L.}\
  \bibnamefont {Muskens}},\ }\emph {\enquote {\bibinfo {title} {{Deep Learning
  Enabled Design of Complex Transmission Matrices for Universal Optical
  Components}},}\ }\href {\doibase 10.1021/acsphotonics.0c01481} {\bibfield
  {journal} {\bibinfo  {journal} {ACS Photonics}\ }\textbf {\bibinfo {volume}
  {8}},\ \bibinfo {pages} {283} (\bibinfo {year} {2021})},\ \Eprint
  {http://arxiv.org/abs/2009.11810} {arXiv:2009.11810}\BibitemShut {NoStop}%
\bibitem [{\citenamefont {Jha}\ \emph {et~al.}(2008)\citenamefont {Jha},
  \citenamefont {Malik},\ and\ \citenamefont {Boyd}}]{jha2008}%
  \BibitemOpen
  \bibfield  {author} {\bibinfo {author} {\bibfnamefont {Anand~Kumar}\
  \bibnamefont {Jha}}, \bibinfo {author} {\bibfnamefont {Mehul}\ \bibnamefont
  {Malik}}, \ and\ \bibinfo {author} {\bibfnamefont {Robert~W}\ \bibnamefont
  {Boyd}},\ }\emph {{\selectlanguage {English}\enquote {\bibinfo {title}
  {Exploring energy-time entanglement using geometric phase},}\ }}\href
  {\doibase 10.1103/physrevlett.101.180405} {\bibfield  {journal} {\bibinfo
  {journal} {Physical Review Letters}\ }\textbf {\bibinfo {volume} {101}},\
  \bibinfo {pages} {180405} (\bibinfo {year} {2008})}\BibitemShut {NoStop}%
\bibitem [{\citenamefont {Lukens}\ and\ \citenamefont
  {Lougovski}(2016)}]{Lukens2016}%
  \BibitemOpen
  \bibfield  {author} {\bibinfo {author} {\bibfnamefont {Joseph~M}\
  \bibnamefont {Lukens}}\ and\ \bibinfo {author} {\bibfnamefont {Pavel}\
  \bibnamefont {Lougovski}},\ }\emph {\enquote {\bibinfo {title}
  {{Frequency-encoded photonic qubits for scalable quantum information
  processing}},}\ }\href {\doibase 10.1364/optica.4.000008} {\bibfield
  {journal} {\bibinfo  {journal} {Optica}\ }\textbf {\bibinfo {volume} {4}},\
  \bibinfo {pages} {8} (\bibinfo {year} {2016})},\ \Eprint
  {http://arxiv.org/abs/1612.03131} {1612.03131}\BibitemShut {NoStop}%
\bibitem [{\citenamefont {Mounaix}\ \emph {et~al.}(2016)\citenamefont
  {Mounaix}, \citenamefont {Andreoli}, \citenamefont {Defienne}, \citenamefont
  {Volpe}, \citenamefont {Katz}, \citenamefont {Grésillon},\ and\
  \citenamefont {Gigan}}]{Mounaix2016}%
  \BibitemOpen
  \bibfield  {author} {\bibinfo {author} {\bibfnamefont {Mickael}\ \bibnamefont
  {Mounaix}}, \bibinfo {author} {\bibfnamefont {Daria}\ \bibnamefont
  {Andreoli}}, \bibinfo {author} {\bibfnamefont {Hugo}\ \bibnamefont
  {Defienne}}, \bibinfo {author} {\bibfnamefont {Giorgio}\ \bibnamefont
  {Volpe}}, \bibinfo {author} {\bibfnamefont {Ori}\ \bibnamefont {Katz}},
  \bibinfo {author} {\bibfnamefont {Samuel}\ \bibnamefont {Grésillon}}, \ and\
  \bibinfo {author} {\bibfnamefont {Sylvain}\ \bibnamefont {Gigan}},\ }\emph
  {\enquote {\bibinfo {title} {{Spatiotemporal Coherent Control of Light
  through a Multiple Scattering Medium with the Multispectral Transmission
  Matrix}},}\ }\href {\doibase 10.1103/physrevlett.116.253901} {\bibfield
  {journal} {\bibinfo  {journal} {Physical Review Letters}\ }\textbf {\bibinfo
  {volume} {116}},\ \bibinfo {pages} {253901} (\bibinfo {year} {2016})},\
  \Eprint {http://arxiv.org/abs/1512.07753} {1512.07753}\BibitemShut {NoStop}%
\bibitem [{\citenamefont {Lu}\ \emph {et~al.}(2018)\citenamefont {Lu},
  \citenamefont {Lukens}, \citenamefont {Peters}, \citenamefont {Odele},
  \citenamefont {Leaird}, \citenamefont {Weiner},\ and\ \citenamefont
  {Lougovski}}]{Lu2018}%
  \BibitemOpen
  \bibfield  {author} {\bibinfo {author} {\bibfnamefont {Hsuan-Hao}\
  \bibnamefont {Lu}}, \bibinfo {author} {\bibfnamefont {Joseph~M.}\
  \bibnamefont {Lukens}}, \bibinfo {author} {\bibfnamefont {Nicholas~A.}\
  \bibnamefont {Peters}}, \bibinfo {author} {\bibfnamefont {Ogaga~D.}\
  \bibnamefont {Odele}}, \bibinfo {author} {\bibfnamefont {Daniel~E.}\
  \bibnamefont {Leaird}}, \bibinfo {author} {\bibfnamefont {Andrew~M.}\
  \bibnamefont {Weiner}}, \ and\ \bibinfo {author} {\bibfnamefont {Pavel}\
  \bibnamefont {Lougovski}},\ }\emph {\enquote {\bibinfo {title}
  {{Electro-Optic Frequency Beam Splitters and Tritters for High-Fidelity
  Photonic Quantum Information Processing}},}\ }\href {\doibase
  10.1103/physrevlett.120.030502} {\bibfield  {journal} {\bibinfo  {journal}
  {Physical Review Letters}\ }\textbf {\bibinfo {volume} {120}},\ \bibinfo
  {pages} {030502} (\bibinfo {year} {2018})},\ \Eprint
  {http://arxiv.org/abs/1712.03992} {1712.03992}\BibitemShut {NoStop}%
\bibitem [{\citenamefont {Mounaix}\ \emph {et~al.}(2020)\citenamefont
  {Mounaix}, \citenamefont {Fontaine}, \citenamefont {Neilson}, \citenamefont
  {Ryf}, \citenamefont {Chen}, \citenamefont {Alvarado-Zacarias},\ and\
  \citenamefont {Carpenter}}]{Mounaix2020}%
  \BibitemOpen
  \bibfield  {author} {\bibinfo {author} {\bibfnamefont {Mickael}\ \bibnamefont
  {Mounaix}}, \bibinfo {author} {\bibfnamefont {Nicolas~K.}\ \bibnamefont
  {Fontaine}}, \bibinfo {author} {\bibfnamefont {David~T.}\ \bibnamefont
  {Neilson}}, \bibinfo {author} {\bibfnamefont {Roland}\ \bibnamefont {Ryf}},
  \bibinfo {author} {\bibfnamefont {Haoshuo}\ \bibnamefont {Chen}}, \bibinfo
  {author} {\bibfnamefont {Juan~Carlos}\ \bibnamefont {Alvarado-Zacarias}}, \
  and\ \bibinfo {author} {\bibfnamefont {Joel}\ \bibnamefont {Carpenter}},\
  }\emph {\enquote {\bibinfo {title} {{Time reversed optical waves by arbitrary
  vector spatiotemporal field generation}},}\ }\href {\doibase
  10.1038/s41467-020-19601-3} {\bibfield  {journal} {\bibinfo  {journal}
  {Nature Communications}\ }\textbf {\bibinfo {volume} {11}},\ \bibinfo {pages}
  {5813} (\bibinfo {year} {2020})}\BibitemShut {NoStop}%
\bibitem [{\citenamefont {Lu}\ \emph {et~al.}(2022)\citenamefont {Lu},
  \citenamefont {Lingaraju}, \citenamefont {Leaird}, \citenamefont {Weiner},\
  and\ \citenamefont {Lukens}}]{Lu2022}%
  \BibitemOpen
  \bibfield  {author} {\bibinfo {author} {\bibfnamefont {Hsuan-Hao}\
  \bibnamefont {Lu}}, \bibinfo {author} {\bibfnamefont {Navin~B}\ \bibnamefont
  {Lingaraju}}, \bibinfo {author} {\bibfnamefont {Daniel~E}\ \bibnamefont
  {Leaird}}, \bibinfo {author} {\bibfnamefont {Andrew~M}\ \bibnamefont
  {Weiner}}, \ and\ \bibinfo {author} {\bibfnamefont {Joseph~M}\ \bibnamefont
  {Lukens}},\ }\emph {\enquote {\bibinfo {title} {{High-dimensional discrete
  Fourier transform gates with a quantum frequency processor}},}\ }\href
  {\doibase 10.1364/oe.454677} {\bibfield  {journal} {\bibinfo  {journal}
  {Optics Express}\ }\textbf {\bibinfo {volume} {30}},\ \bibinfo {pages}
  {10126} (\bibinfo {year} {2022})},\ \Eprint {http://arxiv.org/abs/2201.10973}
  {2201.10973}\BibitemShut {NoStop}%
\bibitem [{\citenamefont {Crameri}(2021)}]{crameri_fabio_2021_5501399}%
  \BibitemOpen
  \bibfield  {author} {\bibinfo {author} {\bibfnamefont {Fabio}\ \bibnamefont
  {Crameri}},\ }\href {\doibase 10.5281/zenodo.5501399} {\emph {\enquote
  {\bibinfo {title} {Scientific colour maps},}\ }} (\bibinfo {year}
  {2021})\BibitemShut {NoStop}%
\bibitem [{\citenamefont {Caramazza}\ \emph {et~al.}(2019)\citenamefont
  {Caramazza}, \citenamefont {Moran}, \citenamefont {Murray-Smith},\ and\
  \citenamefont {Faccio}}]{Caramazza2019}%
  \BibitemOpen
  \bibfield  {author} {\bibinfo {author} {\bibfnamefont {Piergiorgio}\
  \bibnamefont {Caramazza}}, \bibinfo {author} {\bibfnamefont {Oisín}\
  \bibnamefont {Moran}}, \bibinfo {author} {\bibfnamefont {Roderick}\
  \bibnamefont {Murray-Smith}}, \ and\ \bibinfo {author} {\bibfnamefont
  {Daniele}\ \bibnamefont {Faccio}},\ }\emph {\enquote {\bibinfo {title}
  {{Transmission of natural scene images through a multimode fibre}},}\ }\href
  {\doibase 10.1038/s41467-019-10057-8} {\bibfield  {journal} {\bibinfo
  {journal} {Nature Communications}\ }\textbf {\bibinfo {volume} {10}},\
  \bibinfo {pages} {2029} (\bibinfo {year} {2019})},\ \Eprint
  {http://arxiv.org/abs/1904.11985} {1904.11985}\BibitemShut {NoStop}%
\bibitem [{\citenamefont {Wootters}\ and\ \citenamefont
  {Fields}(1989)}]{Wootters1989}%
  \BibitemOpen
  \bibfield  {author} {\bibinfo {author} {\bibfnamefont {William~K}\
  \bibnamefont {Wootters}}\ and\ \bibinfo {author} {\bibfnamefont {Brian~D}\
  \bibnamefont {Fields}},\ }\emph {\enquote {\bibinfo {title} {{Optimal
  state-determination by mutually unbiased measurements}},}\ }\href {\doibase
  10.1016/0003-4916(89)90322-9} {\bibfield  {journal} {\bibinfo  {journal}
  {Annals of Physics}\ }\textbf {\bibinfo {volume} {191}},\ \bibinfo {pages}
  {363} (\bibinfo {year} {1989})}\BibitemShut {NoStop}%
\bibitem [{\citenamefont {Giovannini}\ \emph {et~al.}(2013)\citenamefont
  {Giovannini}, \citenamefont {Romero}, \citenamefont {Leach}, \citenamefont
  {Dudley}, \citenamefont {Forbes},\ and\ \citenamefont
  {Padgett}}]{Giovannini2013}%
  \BibitemOpen
  \bibfield  {author} {\bibinfo {author} {\bibfnamefont {D.}~\bibnamefont
  {Giovannini}}, \bibinfo {author} {\bibfnamefont {J.}~\bibnamefont {Romero}},
  \bibinfo {author} {\bibfnamefont {J.}~\bibnamefont {Leach}}, \bibinfo
  {author} {\bibfnamefont {A.}~\bibnamefont {Dudley}}, \bibinfo {author}
  {\bibfnamefont {A.}~\bibnamefont {Forbes}}, \ and\ \bibinfo {author}
  {\bibfnamefont {M.~J.}\ \bibnamefont {Padgett}},\ }\emph {\enquote {\bibinfo
  {title} {{Characterization of High-Dimensional Entangled Systems via Mutually
  Unbiased Measurements}},}\ }\href {\doibase 10.1103/physrevlett.110.143601}
  {\bibfield  {journal} {\bibinfo  {journal} {Physical Review Letters}\
  }\textbf {\bibinfo {volume} {110}},\ \bibinfo {pages} {143601} (\bibinfo
  {year} {2013})},\ \Eprint {http://arxiv.org/abs/1212.5825}
  {1212.5825}\BibitemShut {NoStop}%
\bibitem [{\citenamefont {Bongioanni}\ \emph {et~al.}(2010)\citenamefont
  {Bongioanni}, \citenamefont {Sansoni}, \citenamefont {Sciarrino},
  \citenamefont {Vallone},\ and\ \citenamefont {Mataloni}}]{Bongioanni2010}%
  \BibitemOpen
  \bibfield  {author} {\bibinfo {author} {\bibfnamefont {Irene}\ \bibnamefont
  {Bongioanni}}, \bibinfo {author} {\bibfnamefont {Linda}\ \bibnamefont
  {Sansoni}}, \bibinfo {author} {\bibfnamefont {Fabio}\ \bibnamefont
  {Sciarrino}}, \bibinfo {author} {\bibfnamefont {Giuseppe}\ \bibnamefont
  {Vallone}}, \ and\ \bibinfo {author} {\bibfnamefont {Paolo}\ \bibnamefont
  {Mataloni}},\ }\emph {\enquote {\bibinfo {title} {{Experimental quantum
  process tomography of non-trace-preserving maps}},}\ }\href {\doibase
  10.1103/physreva.82.042307} {\bibfield  {journal} {\bibinfo  {journal}
  {Physical Review A}\ }\textbf {\bibinfo {volume} {82}},\ \bibinfo {pages}
  {042307} (\bibinfo {year} {2010})},\ \Eprint {http://arxiv.org/abs/1008.5334}
  {1008.5334}\BibitemShut {NoStop}%
\bibitem [{\citenamefont {Jebarathinam}\ \emph {et~al.}(2020)\citenamefont
  {Jebarathinam}, \citenamefont {Home},\ and\ \citenamefont
  {Sinha}}]{Jebarathinam2020}%
  \BibitemOpen
  \bibfield  {author} {\bibinfo {author} {\bibfnamefont {C.}~\bibnamefont
  {Jebarathinam}}, \bibinfo {author} {\bibfnamefont {Dipankar}\ \bibnamefont
  {Home}}, \ and\ \bibinfo {author} {\bibfnamefont {Urbasi}\ \bibnamefont
  {Sinha}},\ }\emph {\enquote {\bibinfo {title} {Pearson correlation
  coefficient as a measure for certifying and quantifying high-dimensional
  entanglement},}\ }\href {\doibase 10.1103/PhysRevA.101.022112} {\bibfield
  {journal} {\bibinfo  {journal} {Phys. Rev. A}\ }\textbf {\bibinfo {volume}
  {101}},\ \bibinfo {pages} {022112} (\bibinfo {year} {2020})}\BibitemShut
  {NoStop}%
\bibitem [{\citenamefont {Sadana}\ \emph {et~al.}(2022)\citenamefont {Sadana},
  \citenamefont {Kanjilal}, \citenamefont {Home},\ and\ \citenamefont
  {Sinha}}]{sadana2022}%
  \BibitemOpen
  \bibfield  {author} {\bibinfo {author} {\bibfnamefont {Simanraj}\
  \bibnamefont {Sadana}}, \bibinfo {author} {\bibfnamefont {Som}\ \bibnamefont
  {Kanjilal}}, \bibinfo {author} {\bibfnamefont {Dipankar}\ \bibnamefont
  {Home}}, \ and\ \bibinfo {author} {\bibfnamefont {Urbasi}\ \bibnamefont
  {Sinha}},\ }\emph {\enquote {\bibinfo {title} {Relating an entanglement
  measure with statistical correlators for two-qudit mixed states using only a
  pair of complementary observables},}\ }\href@noop {} {\  (\bibinfo {year}
  {2022})},\ \Eprint {http://arxiv.org/abs/2201.06188} {arXiv:2201.06188
  [quant-ph]}\BibitemShut {NoStop}%
\bibitem [{\citenamefont {Tomiyama}(1985)}]{Tomiyama1985}%
  \BibitemOpen
  \bibfield  {author} {\bibinfo {author} {\bibfnamefont {Jun}\ \bibnamefont
  {Tomiyama}},\ }\emph {\enquote {\bibinfo {title} {On the geometry of positive
  maps in matrix algebras. ii},}\ }\href {\doibase
  https://doi.org/10.1016/0024-3795(85)90074-6} {\bibfield  {journal} {\bibinfo
   {journal} {Linear Algebra and its Applications}\ }\textbf {\bibinfo {volume}
  {69}},\ \bibinfo {pages} {169} (\bibinfo {year} {1985})}\BibitemShut
  {NoStop}%
\bibitem [{\citenamefont {Terhal}\ and\ \citenamefont
  {Horodecki}(2000)}]{Terhal2000}%
  \BibitemOpen
  \bibfield  {author} {\bibinfo {author} {\bibfnamefont {Barbara~M.}\
  \bibnamefont {Terhal}}\ and\ \bibinfo {author} {\bibfnamefont {Pawe\l{}}\
  \bibnamefont {Horodecki}},\ }\emph {\enquote {\bibinfo {title} {Schmidt
  number for density matrices},}\ }\href {\doibase 10.1103/PhysRevA.61.040301}
  {\bibfield  {journal} {\bibinfo  {journal} {Phys. Rev. A}\ }\textbf {\bibinfo
  {volume} {61}},\ \bibinfo {pages} {040301} (\bibinfo {year}
  {2000})}\BibitemShut {NoStop}%
\bibitem [{\citenamefont {Goel}\ \emph {et~al.}(2022)\citenamefont {Goel},
  \citenamefont {Leedumrongwatthanakun}, \citenamefont {Herrera~Valencia},
  \citenamefont {McCutcheon}, \citenamefont {Conti}, \citenamefont {Pinkse},\
  and\ \citenamefont {Malik}}]{Goel_Simulation_Codes_for_2022}%
  \BibitemOpen
  \bibfield  {author} {\bibinfo {author} {\bibfnamefont {Suraj}\ \bibnamefont
  {Goel}}, \bibinfo {author} {\bibfnamefont {Saroch}\ \bibnamefont
  {Leedumrongwatthanakun}}, \bibinfo {author} {\bibfnamefont {Natalia}\
  \bibnamefont {Herrera~Valencia}}, \bibinfo {author} {\bibfnamefont {Will}\
  \bibnamefont {McCutcheon}}, \bibinfo {author} {\bibfnamefont {Claudio}\
  \bibnamefont {Conti}}, \bibinfo {author} {\bibfnamefont {Pepijn W.~H.}\
  \bibnamefont {Pinkse}}, \ and\ \bibinfo {author} {\bibfnamefont {Mehul}\
  \bibnamefont {Malik}},\ }\href
  {https://github.com/BBQuantum/simulations_top_down_design} {\emph {\enquote
  {\bibinfo {title} {{Simulation codes for : Inverse-design of high-dimensional
  quantum optical circuits in a complex medium}},}\ }} (\bibinfo {year}
  {2022})\BibitemShut {NoStop}%
\bibitem [{\citenamefont {Srivastav}\ \emph
  {et~al.}(2022{\natexlab{b}})\citenamefont {Srivastav}, \citenamefont
  {Valencia}, \citenamefont {Leedumrongwatthanakun}, \citenamefont
  {McCutcheon},\ and\ \citenamefont {Malik}}]{Srivastav2021}%
  \BibitemOpen
  \bibfield  {author} {\bibinfo {author} {\bibfnamefont {Vatshal}\ \bibnamefont
  {Srivastav}}, \bibinfo {author} {\bibfnamefont {Natalia~Herrera}\
  \bibnamefont {Valencia}}, \bibinfo {author} {\bibfnamefont {Saroch}\
  \bibnamefont {Leedumrongwatthanakun}}, \bibinfo {author} {\bibfnamefont
  {Will}\ \bibnamefont {McCutcheon}}, \ and\ \bibinfo {author} {\bibfnamefont
  {Mehul}\ \bibnamefont {Malik}},\ }\emph {\enquote {\bibinfo {title}
  {Characterizing and tailoring spatial correlations in multimode parametric
  down-conversion},}\ }\href
  {https://journals.aps.org/prapplied/abstract/10.1103/PhysRevApplied.18.054006}
  {\bibfield  {journal} {\bibinfo  {journal} {Physical Review Applied}\
  }\textbf {\bibinfo {volume} {18}},\ \bibinfo {pages} {054006} (\bibinfo
  {year} {2022}{\natexlab{b}})},\ \Eprint {http://arxiv.org/abs/2110.03462}
  {arXiv:2110.03462}\BibitemShut {NoStop}%
\bibitem [{\citenamefont {Klyshko}(1988)}]{klyshko1988simple}%
  \BibitemOpen
  \bibfield  {author} {\bibinfo {author} {\bibfnamefont {D~N}\ \bibnamefont
  {Klyshko}},\ }\emph {\enquote {\bibinfo {title} {{A simple method of
  preparing pure states of an optical field, of implementing the
  Einstein–Podolsky–Rosen experiment, and of demonstrating the
  complementarity principle}},}\ }\href {\doibase
  10.1070/pu1988v031n01abeh002537} {\bibfield  {journal} {\bibinfo  {journal}
  {Soviet Physics Uspekhi}\ }\textbf {\bibinfo {volume} {31}},\ \bibinfo
  {pages} {74} (\bibinfo {year} {1988})}\BibitemShut {NoStop}%
\bibitem [{\citenamefont {Unter}\ \emph {et~al.}(1981)\citenamefont {Unter},
  \citenamefont {Borevich},\ and\ \citenamefont {Krupetskii}}]{Borevich1981}%
  \BibitemOpen
  \bibfield  {author} {\bibinfo {author} {\bibnamefont {Unter}}, \bibinfo
  {author} {\bibfnamefont {Z.~I.}\ \bibnamefont {Borevich}}, \ and\ \bibinfo
  {author} {\bibfnamefont {S.~L.}\ \bibnamefont {Krupetskii}},\ }\emph
  {\enquote {\bibinfo {title} {{Subgroups of the unitary group that contain the
  group of diagonal matrices}},}\ }\href {\doibase 10.1007/BF01465451}
  {\bibfield  {journal} {\bibinfo  {journal} {Journal of Soviet Mathematics}\
  }\textbf {\bibinfo {volume} {17}},\ \bibinfo {pages} {1951} (\bibinfo {year}
  {1981})}\BibitemShut {NoStop}%
\bibitem [{\citenamefont {Boucher}\ \emph {et~al.}(2020)\citenamefont
  {Boucher}, \citenamefont {Goetschy}, \citenamefont {Sorelli}, \citenamefont
  {Walschaers},\ and\ \citenamefont {Treps}}]{Boucher2020}%
  \BibitemOpen
  \bibfield  {author} {\bibinfo {author} {\bibfnamefont {Pauline}\ \bibnamefont
  {Boucher}}, \bibinfo {author} {\bibfnamefont {Arthur}\ \bibnamefont
  {Goetschy}}, \bibinfo {author} {\bibfnamefont {Giacomo}\ \bibnamefont
  {Sorelli}}, \bibinfo {author} {\bibfnamefont {Mattia}\ \bibnamefont
  {Walschaers}}, \ and\ \bibinfo {author} {\bibfnamefont {Nicolas}\
  \bibnamefont {Treps}},\ }\emph {\enquote {\bibinfo {title} {{Full
  characterization of the transmission properties of a multi-plane light
  converter}},}\ }\href {\doibase 10.1103/physrevresearch.3.023226} {\bibfield
  {journal} {\bibinfo  {journal} {Physical Review Research}\ }\textbf {\bibinfo
  {volume} {3}},\ \bibinfo {pages} {023226} (\bibinfo {year} {2020})},\ \Eprint
  {http://arxiv.org/abs/2005.11982} {2005.11982}\BibitemShut {NoStop}%
\bibitem [{\citenamefont {Schmid}\ \emph {et~al.}(2000)\citenamefont {Schmid},
  \citenamefont {Steinwandt}, \citenamefont {M{\"{u}}ller-Quade}, \citenamefont
  {R{\"{o}}tteler},\ and\ \citenamefont {Beth}}]{Schmid2000}%
  \BibitemOpen
  \bibfield  {author} {\bibinfo {author} {\bibfnamefont {Michael}\ \bibnamefont
  {Schmid}}, \bibinfo {author} {\bibfnamefont {Rainer}\ \bibnamefont
  {Steinwandt}}, \bibinfo {author} {\bibfnamefont {J{\"{o}}rn}\ \bibnamefont
  {M{\"{u}}ller-Quade}}, \bibinfo {author} {\bibfnamefont {Martin}\
  \bibnamefont {R{\"{o}}tteler}}, \ and\ \bibinfo {author} {\bibfnamefont
  {Thomas}\ \bibnamefont {Beth}},\ }\emph {\enquote {\bibinfo {title}
  {{Decomposing a matrix into circulant and diagonal factors}},}\ }\href
  {\doibase 10.1016/S0024-3795(99)00250-5} {\bibfield  {journal} {\bibinfo
  {journal} {Linear Algebra and Its Applications}\ }\textbf {\bibinfo {volume}
  {306}},\ \bibinfo {pages} {131} (\bibinfo {year} {2000})}\BibitemShut
  {NoStop}%
\bibitem [{\citenamefont {Huhtanen}\ and\ \citenamefont
  {Per{\"{a}}m{\"{a}}ki}(2015)}]{Huhtanen2015}%
  \BibitemOpen
  \bibfield  {author} {\bibinfo {author} {\bibfnamefont {Marko}\ \bibnamefont
  {Huhtanen}}\ and\ \bibinfo {author} {\bibfnamefont {Allan}\ \bibnamefont
  {Per{\"{a}}m{\"{a}}ki}},\ }\emph {\enquote {\bibinfo {title} {{Factoring
  Matrices into the Product of Circulant and Diagonal Matrices}},}\ }\href
  {\doibase 10.1007/s00041-015-9395-0} {\bibfield  {journal} {\bibinfo
  {journal} {Journal of Fourier Analysis and Applications}\ }\textbf {\bibinfo
  {volume} {21}},\ \bibinfo {pages} {1018} (\bibinfo {year}
  {2015})}\BibitemShut {NoStop}%
\bibitem [{\citenamefont {Idel}\ and\ \citenamefont {Wolf}(2015)}]{Idel2015}%
  \BibitemOpen
  \bibfield  {author} {\bibinfo {author} {\bibfnamefont {Martin}\ \bibnamefont
  {Idel}}\ and\ \bibinfo {author} {\bibfnamefont {Michael~M.}\ \bibnamefont
  {Wolf}},\ }\emph {\enquote {\bibinfo {title} {{Sinkhorn normal form for
  unitary matrices}},}\ }\href {\doibase 10.1016/j.laa.2014.12.031} {\bibfield
  {journal} {\bibinfo  {journal} {Linear Algebra and Its Applications}\
  }\textbf {\bibinfo {volume} {471}},\ \bibinfo {pages} {76} (\bibinfo {year}
  {2015})},\ \Eprint {http://arxiv.org/abs/1408.5728}
  {arXiv:1408.5728}\BibitemShut {NoStop}%
\bibitem [{\citenamefont {{L{\'{o}}pez Pastor}}\ \emph
  {et~al.}(2021)\citenamefont {{L{\'{o}}pez Pastor}}, \citenamefont {Lundeen},\
  and\ \citenamefont {Marquardt}}]{LopezPastor2021}%
  \BibitemOpen
  \bibfield  {author} {\bibinfo {author} {\bibfnamefont {V{\'{i}}ctor}\
  \bibnamefont {{L{\'{o}}pez Pastor}}}, \bibinfo {author} {\bibfnamefont
  {Jeff}\ \bibnamefont {Lundeen}}, \ and\ \bibinfo {author} {\bibfnamefont
  {Florian}\ \bibnamefont {Marquardt}},\ }\emph {\enquote {\bibinfo {title}
  {{Arbitrary optical wave evolution with Fourier transforms and phase
  masks}},}\ }\href {\doibase 10.1364/OE.432787} {\bibfield  {journal}
  {\bibinfo  {journal} {Optics Express}\ }\textbf {\bibinfo {volume} {29}},\
  \bibinfo {pages} {38441} (\bibinfo {year} {2021})},\ \Eprint
  {http://arxiv.org/abs/1912.04721} {arXiv:1912.04721}\BibitemShut {NoStop}%
\bibitem [{\citenamefont {T{\"{o}}rm{\"{a}}}\ \emph {et~al.}(1996)\citenamefont
  {T{\"{o}}rm{\"{a}}}, \citenamefont {Jex},\ and\ \citenamefont
  {Stenholm}}]{Torma1996}%
  \BibitemOpen
  \bibfield  {author} {\bibinfo {author} {\bibfnamefont {P.}~\bibnamefont
  {T{\"{o}}rm{\"{a}}}}, \bibinfo {author} {\bibfnamefont {I.}~\bibnamefont
  {Jex}}, \ and\ \bibinfo {author} {\bibfnamefont {S.}~\bibnamefont
  {Stenholm}},\ }\emph {\enquote {\bibinfo {title} {{Beam splitter realizations
  of totally symmetric mode couplers}},}\ }\href {\doibase
  10.1080/09500349608232738} {\bibfield  {journal} {\bibinfo  {journal}
  {Journal of Modern Optics}\ }\textbf {\bibinfo {volume} {43}},\ \bibinfo
  {pages} {245} (\bibinfo {year} {1996})}\BibitemShut {NoStop}%
\bibitem [{\citenamefont {Barak}\ and\ \citenamefont
  {Ben-Aryeh}(2007)}]{Barak2007}%
  \BibitemOpen
  \bibfield  {author} {\bibinfo {author} {\bibfnamefont {Ronen}\ \bibnamefont
  {Barak}}\ and\ \bibinfo {author} {\bibfnamefont {Yacob}\ \bibnamefont
  {Ben-Aryeh}},\ }\emph {\enquote {\bibinfo {title} {{Quantum fast Fourier
  transform and quantum computation by linear optics}},}\ }\href {\doibase
  10.1364/JOSAB.24.000231} {\bibfield  {journal} {\bibinfo  {journal} {Journal
  of the Optical Society of America B}\ }\textbf {\bibinfo {volume} {24}},\
  \bibinfo {pages} {231} (\bibinfo {year} {2007})}\BibitemShut {NoStop}%
\end{thebibliography}%

\section*{Methods}

\subsection*{Experimental Details}

A continuous-wave grating-stabilised laser (Toptica DL Pro HP) at 405~nm is used to pump a periodically poled Pottasium Titanyl Phosphate (ppKTP) crystal (1~mm~$\times$~2~mm~$\times$~15~mm) at 125~mW to generate a pair of orthogonally polarised photons at 810~nm entangled in their transverse position-momentum degree-of-freedom (DoF) through the process of Type-II spontaneous parametric down conversion (SPDC). A telescope system of lenses is used to shape the pump beam and focus it on the crystal with a $1/e^2$ beam diameter of 1.2~mm. Phase-matching conditions are achieved via temperature-tuning the crystal in a custom-built oven that keeps it at 38$^\circ$C. For more details on the high-dimensional entanglement source, please see the Supplementary Information. 

After the crystal, the pump is filtered out using a dichroic mirror and a band-pass filter (F), while the pair of produced photons are separated by a polarising beam splitter (PBS). The reflected photon (corresponding to Alice) has its polarisation rotated from vertical to horizontal with a half-wave plate (HWP) and made incident on a phase-only SLM (SLM$_3$, Hamamatsu X10468-02, effective area size of 15.8 $\times$ 12 mm, pixel pitch of 20~$\mu$m, resolution of 792 $\times$ 600, reflection efficiency of approximately 90~\%, and diffraction efficiency of approximately 75~\%) that is placed in the Fourier plane of the crystal using a 400~mm lens. The transmitted photon (corresponding to Bob) is sent to a top-down programmable circuit constructed from a 2~m graded-index multimode fibre (MMF, Thorlabs M116L02) placed between two programmable phase-only spatial light modulators (SLM$_{1,2}$). After reflection from the SLMs, a telescope system and an aspheric lens are used for mode-matching the photons to either single-mode fibre (SMF) or MMF collection modes.

Local projective measurements of the transverse spatial photonic modes are made with computer-generated holograms (CGH) displayed by the parallel-aligned liquid-crystal-on-silicon (LCOS) layer of the SLMs. The CGH for a particular spatial mode is displayed on the SLM at each party. If the incident spatial mode is the complex conjugate of the target mode displayed on the hologram, it is converted into a Gaussian mode, which couples efficiently into an SMF positioned in the far-field of the first-order diffraction spot. The SMFs guide the filtered photons to superconducting nanowire single-photon detectors (Quantum Opus, Opus One, efficiency >90\% at 810~nm). Coincidence events between the detection of signal and idler photons in the selected modes are registered by a coincidence counting logic  (Swabian Time Tagger Ultra) within a coincidence window of 0.2~ns.



\subsection*{Acquisition of transfer matrix}
The optical apparatus used for constructing our circuits is described by the transfer matrix (TM): $\mathbf{T}=U_2 P_2 U_1 P_1$, where $U_2$ is a $2f$ lens system and $P_j$ is the $j$-th phase plane displayed on SLM$_j$. In order to construct the circuits, we need to know the transfer matrix $U_1$ of the 2~m-long graded-index multi-mode fiber (Thorlabs-M116L02, core diameter = $50.0\pm2.5$ $\mu m$, NA = $0.200\pm 0.015$). This is measured using the recently developed multi-plane neural network (MPNN) technique \cite{goel2023referenceless}, which uses random phase patterns displayed at both ends of the MMF to acquire the fibre TM without using an external reference field. In contrast with the main experiment, here we use a superluminiscent diode along with a $810 \pm 1.5$nm filter (Semrock LL01-810-12.5) for TM characterisation. The number of spatial modes ($n$) of a graded-index fiber  at wavelength $\lambda$ in both polarisations depends on its $V$-number as $V=2\pi a \text{NA}/\lambda$, which is a function of the core radius ($a$) and numerical aperture ($\text{NA}$). We search for the transfer matrix $U_1$ by optimising
\be
\text{min} \left| I_k - |U_2 P_{2_k} U_1 P_{1_k}|^2 \right|^2.
\ee
The optimisation is based on the gradient descent method fitting the acquired characterisation data consisting of random phase patterns on each plane $(P_{1_k}, P_{2_k})$ displayed on SLM$_1$ and SLM$_2$ and the corresponding measured output speckle intensity images $\{I_k\}_k$~\cite{goel2023referenceless}. The model of the optical apparatus is implemented in Keras using Tensorflow2 and complex-number layers developed in \cite{Caramazza2019}. The dataset is prepared in three parts. First, each input mode in a given basis that is supported by the fiber is displayed on SLM$_1$. This data provides accurate information of $|U_1|$. In the second part, random superpositions of these input modes are prepared on SLM$_1$ and sent through the fibre. The output intensity speckles in this part allow for the recovery of the relative phase and amplitude of the transmission coefficient corresponding to a particular output mode. Finally, both SLMs are used for displaying random superpositions of the input and output modes. This final part of the dataset allows for accurate reconstruction of $U_1$, including calibration of the unknown relative phases across the output modes. Note that $U_1$ includes the associated coupling optics between the first and second phase planes and our measurement is limited to one polarisation channel of the MMF. The entire measurement takes $\sim90$ minutes, primarily limited by the SLM refresh rate and hologram calculation time, which together take up 90\% of the total measurement time. The optimization time is $\sim1$ minute performed on a GPU (GeForce RTX 3060, CPU Intel Core i7-8700, RAM 16 GB).


\subsection*{Construction of linear circuits}
Primarily, programming a circuit $\mathbb{T}$ is achieved by calculating the phase solutions $\{P_j\}_{j=1}^{L}$ at each phase plane. The wavefront matching algorithm (WFM) can do so by iteratively matching the wavefronts of target input and output optical modes propagating through the device across all phase planes~\cite{Hashimoto2005,Sakamaki2007,Fontaine2017}. First, input arguments that contain a set of input spatial modes $\{\ket{\psi_a(\mathbf{q})}\}_{a=0}^{d-1}$ labelled in the logical basis by $\{\ket{a}_\text{in}\}_{a=0}^{d-1}$, a corresponding set of output spatial modes $\{\ket{\phi_a(\mathbf{q})}\}_{a=0}^{d-1}$ that is related to the inputs via $\ket{a_\text{out}}=\mathbb{T}\ket{a_\text{in}}$, and a set of transfer functions $\{U_j\}$ between phase planes are provided to the WFM algorithm. 

For each $i$-th iteration, a phase solution at the $p$-th plane is updated in a cyclic manner starting from the first to the last $L$-th plane and then back from the last plane to the first. At a particular reconfigurable phase plane $P_p$, the transfer matrix of the optical device $\mathbf{T}$ represented in the spatial $\mathbf{q}$ basis is decomposed into two sections: 
\be
    \mathbf{T}:=\prod_{j=1}^{L}U_j P_j = B_p P_p F_p,
    \label{Eq:wfm_sectioning}
\ee
where $B_p=\prod_{j=p}^{L} P_{j+1} U_j$ and $F_p=\prod_{j=1}^{p-1} U_j P_j$ such that the forward-propagating input mode onto the $p$-th phase plane is represented by $\ket{\psi_{a,(p)}} = F_p \ket{\psi_a}$ and the backward-propagating output mode onto the $p$-th phase plane is $\ket{\phi_{a,(p)}} = B_p^{\dagger}\ket{\phi_a}$. The phase mismatch between these input and output modes can then be adjusted by $P_p$:
\be
    \ket{\phi_a} = B_p P_p F_p \ket{\psi_a} \implies \ket{\phi_{a,(p)}}= P_p\ket{\psi_{a,(p)}}.
    \label{Eq:WFMsection}
\ee
Considering all $d$-target modes of interest, the \textit{matching} matrix: $M_{p}:=\sum_{a,a'=0}^{d-1} \bra{\phi_{a',(p)}}P_p\ket{\psi_{a,(p)}}\ket{a'}\bra{a}$ captures the mode mixing at each phase plane. The WFM algorithm maximise $\tr\left(M_{p}\right)$ by calculating a phase solution $P^{[i]}_{p}$ from the weighed average of overlapped fields over all $d$-target modes as follows:
\be
    P^{[i]}_{p}(\mathbf{q})= \exp\left(i\arg \left(\sum_{a=0}^{d-1}\phi_{a,(p)}^{[\text{latest}]}(\mathbf{q})\odot\psi_{a,(p)}^{*\text{[latest]}}(\mathbf{q})\right)\right),
\ee
where $\odot$ is an element-wise multiplication on the $\mathbf{q}$ coordinate (SLM pixels) and $\phi_{a,(p)}^{\text{[latest]}}(\mathbf{q})$ and $\psi_{a,(p)}^{*\text{[latest]}}(\mathbf{q})$ is the latest update of output and input optical fields at the $p$-th phase plane taking into account all other previous updated phase planes $\{P^{[i]}_{p}\}$ in the current iteration in both forward and backward directions. The algorithm is iterated until an appropriate value of gate fidelity (Eq.~\ref{Eq:Fpure}) is achieved or saturated.

In our experiment, the following gates are implemented and defined as:
\be
\begin{aligned}
\mathbb{I}&=\sum_{a=0}^{d-1} \ket{a}\bra{a}, \qquad &
\mathbb{Z}&=\sum_{a=0}^{d-1} \ket{a}\omega^a_d\bra{a},\qquad \\
\mathbb{X}&=\sum_{a=0}^{d-1} \ket{a\oplus1}\bra{a}, \qquad &
\mathbb{F}&=\frac{1}{\sqrt{d}}\sum_{a,b=0}^{d-1} \ket{b}\omega^{ab}_{d}\bra{a},\\
\end{aligned}
\label{Eq:GateDefns}
\ee
where $\omega_d=\exp(2\pi i /d)$ and $a\oplus1:=(a+1) \mod d$. $\mathbb{R}$ is the random unitary which is sampled from the Haar measure for each implementation.


\subsection*{Quantum state tomography (QST)}

The characterisation of the entangled state both before and after manipulation by the circuits is performed via quantum state tomography (QST). QST is implemented via an informationally complete set of measurements and numeric inversion of the data, subject to physical constraints. Our projective measurements are related to the ideal measurements by incorporating the detection efficiency $\eta^{\mu}_{a}$ of measuring a particular spatial mode, $\ket{\psi^{\mu}_a}$ (corresponding to the $a$-th element of the $\mu$-th basis). This is estimated through knowledge of the CGH and its effect on an incoming spatial mode. The resulting projective measurements are given by $\hat{\widetilde{\Pi}}^{\mu}_{a} = \eta^\mu_a  \hat{\Pi}^{\mu}_{a}$, where $\hat{\Pi}^{\mu}_{a}=\ketbra{\psi^{\mu}_a}{\psi^{\mu}_a}$ is the ideal projector.

We perform QST via semidefinite programming (SDP) on both input and output states after manipulation by the optical circuits. The SDP imposes data fitting of the non-normalised measurements subject to positive semi-definitness of the state $\rho$, and unit trace, and reads as
\be
\begin{aligned}
\label{eqs:QSTSDP}
\min_{\rho, R} \quad& |\mathcal{C}^{\mu\nu}_{ab}-R \tr(\hat{\widetilde{\Pi}}^{\mu}_{a}\otimes {\hat{\widetilde{\Pi}}^{\nu}_{b}} \rho)|^2\\\\
\textrm{s.t.} \quad & \rho\geq 0 \, , \, \tr[\rho]=1,
\end{aligned}
\ee
where $\mathcal{C}^{\mu\nu}_{ab}$ is the frequency of the outcome (coincidence count rate) and $R$ is the count rate per integration window. Our local measurement bases are complete sets of mutually unbiased bases (MUBs)~\cite{Wootters1989}, which are informationally complete for QST~\cite{Giovannini2013}, and constructed as : $\hat{\Pi}^{\mu}_{a}= \ketbra{M^\mu_{a}}{M^\mu_{a}}$ and $\hat{\Pi}^{\nu}_{b}= \ketbra{{M^\nu_{b}}^*}{{M^\nu_{b}}^*}$, where $\ket{M^\mu_{a}}=\frac{1}{\sqrt{d}}\sum^{d-1}_{m=0}\omega^{am+\mu m^2}_{d}\ket{m}$ on the basis of each implemented circuit, and $\omega_d=\exp(2\pi i /d)$ is a $d$-th root of unity. All SDPs are implemented in CVX, running the commercial solver MOSEK. In the experiments, the holograms corresponding to the local measurements made by Bob are displayed on SLM$_1$ in case of the QST of initial state, whereas in case of the QST of output state, the corresponding holograms are displayed on SLM$_2$. For all cases, holograms corresponding to projections made by Alice are displayed on SLM$_3$.


\subsection*{Quantum process tomography (QPT and ancilla-assisted QPT)}

A single, well-characterised, and sufficiently strongly correlated state supported on an extended Hilbert space, along with a tomographically complete measurement, can also achieve quantum process tomography. This process is known as ancilla-assisted quantum process tomography (AA-QPT) \cite{Altepeter2003,Valencia2020}. On average, our initial state is close but not exactly equal to a maximally entangled state (Extended Data Fig. 1). In order to determine the underlying quantum process we must invert the dependence on the initial state to recover the Choi state of the channel, i.e., an optical circuit. The initial state can be written as a linear operation, $\mathcal{A}$, acting only on party A of a maximally entangled state $\rho^+$,
\be
\begin{aligned}
\label{eqs:initialStateOperator}
\rho^\text{(in)}
&=\sum_n A^n\otimes \mathbb{I} \rho^+ A^{n\dagger}\otimes \mathbb{I} \\
&=\mathcal{A}\otimes \mathbb{I} (\rho^+ )
\end{aligned}
\ee
The output state we tomograph depends on both the initial state and the channel, $\xi$, which acts on party B. The channel and the operator generating the initial state thus commute, so the output state can be written as the linear operator, $\mathcal{A}$ (from Eqs.~\ref{eqs:initialStateOperator} above) acting on the Choi state~\cite{DAriano2003,Altepeter2003}, $\rho_\xi:=\mathbb{I}\otimes \xi (\rho^{+})$,
\be
\begin{aligned}
\label{eqs:RhoOut}
\rho^\text{(out)}_\xi &= \mathbb{I}\otimes \xi (\rho^\text{(in)})\\
&= \mathcal{A}\otimes \mathbb{I} \bigl(\rho_\xi \bigr)\\
\end{aligned}
\ee
Providing $\mathcal{A}$ is invertible, this linear equation can be inverted to recover the Choi state, $\rho_\xi$.

The channel's Choi state can be recovered by operating the inverse, $( \mathcal{A}\otimes \mathbb{I})^{-1}$, on the output state, however, the resultant Choi state may not correspond to a physical channel. In practise, the channels we program are necessarily completely positive, requiring that the Choi state is positive semi-definite, $\rho_\xi \geq 0$.

If our transformations were lossless, the channel would also be trace-preserving and the Choi state would obey $\tr_B[\rho_\xi]=\mathbb{I}$. However, in our case of non-trace preserving channels, we characterise the normalised Choi state, $\tr[\rho_\xi]=1$.

In order to impose positivity of the recovered Choi state in ancilla-assisted quantum process tomography, the semi-definite program is optimised directly over positive $\rho_\xi$,
\be
\begin{aligned}
\label{eqs:QPTSDP}
\min_{\rho_\xi, R} \quad& |\mathcal{C}^{\mu\nu}_{ab}-R \tr(\hat{\widetilde{\Pi}}^{\mu}_{a}\otimes {\hat{\widetilde{\Pi}}^{\nu}_{b}} \mathcal{A}\otimes \mathbb{I} \bigl(\rho_\xi \bigr))|^2\\
\textrm{s.t.} \quad & \rho_\xi \geq 0 \, , \, \tr[\rho_\xi ] = 1,
\end{aligned}
\ee
The fidelity of such a non-trace preserving process then corresponds to that of the process, excluding explicit dependence on global loss~\cite{Bongioanni2010}.

Alternatively, quantum process tomography (QPT) can be achieved by preparing and inputting a set of tomographically complete input states into a given process and performing measurements on the output states. This prepare and measure QPT can be accommodated in the same formalism by noting that measurement of a perfect maximally entangled input state is equivalent, up-to-normalisation, to state preparation. This allows classical characterisation of our gates via the same SDP Eqs.~\ref{eqs:QPTSDP}.


\subsection*{Fidelity, success probability, and optical losses}
We use two figures of merit to characterise the implemented circuits---fidelity $\mathcal{F}$ and success probability $\mathcal{S}$. The first is the Uhlmann-Josza fidelity between two density matrices:
\be
\mathcal{F}(\rho,\rho_\text{o}):=\left(\tr \left(\sqrt{ \sqrt{\rho} \rho_\text{o} \sqrt{\rho}} \right)\right)^2,
\label{Eq:F}
\ee
where $\rho_0$ and $\rho$ can refer to the experimentally recovered normalised but non-trace preserving Choi state, $\widetilde{\rho_{\xi*}}$, and the ideal target Choi state, $\rho_\xi$$=(\mathbb{I}\otimes\mathbb{T})\rho^+(\mathbb{I}\otimes\mathbb{T}^{\dagger})$. The fidelity thus implies an accuracy of the implemented circuits. Note that, in the case when the experimental and target Choi states are pure $(\mathcal{P}:=\tr\left( \rho^2 \right)=1)$, the implemented circuit can be represented by a rank-one Kraus operator $\widetilde{\mathbb{T}}$, and the fidelity reduces to:
\be
\mathcal{F}=\frac{\left(\tr\left(\widetilde{\mathbb{T}}^{\dagger} \mathbb{T}\right)\right)^2}{\tr\left(\widetilde{\mathbb{T}}^{\dagger} \widetilde{\mathbb{T}}\right) \tr\left(\mathbb{T}^{\dagger} \mathbb{T}\right)},
\label{Eq:Fpure}
\ee
which is normalised by the transmittance due to the scattering from the $d$-dimensional space of a circuit into other optical modes. The transmittance is quantified by the second figure of merit, the success probability, $\mathcal{S}$, of an implemented circuit:
\be
\mathcal{S}:=\frac{\tr\left(\widetilde{\mathbb{T}}^{\dagger} \widetilde{\mathbb{T}}\right)}{\tr\left(\mathbb{T}^{\dagger} \mathbb{T}\right)},
\label{Eq:S}
\ee
The implemented $d$-dimensional circuit $\widetilde{\mathbb{T}}$ is embedded in $\mathbf{T}$ which lives in the $n$-dimensional space of the apparatus such that: $\widetilde{\mathbb{T}} = \mathbf{P}_\text{o}\mathbf{T}\mathbf{P}_\text{i}$, where 
$\mathbf{P}_\text{o}$ is the output dimensional reduction and $\mathbf{P}_\text{i}$ is the input dimensional expansion, which map the circuit from $d$ inputs of interest to $n$ inputs of the optical device and from $n$ outputs of the optical device to the $d$ output modes of circuit, respectively. We aim to use $\mathcal{S}$ to measure the scattering loss which stems from the effect of the top-down circuit design, whereas another optical loss of the apparatus is analysed in the last part of this section. Experimentally, the success probability $\mathcal{S}$ (Eq.~\ref{Eq:S}) is estimated on the output $\mathbf{x}$ space as 
\be
\mathcal{S}=\frac{1}{d}\sum_{a=0}^{d-1}\int d^{2}\mathbf{x}\left|\sum_{b=0}^{d-1} \widetilde{t}_{ab}\phi_b(\mathbf{x})\right|^2=\frac{1}{d}\sum_{a=0}^{d-1}\int d^{2}\mathbf{x}\sum_{b=0}^{d-1}\frac{\mathcal{I}_{ab}}{\mathcal{I}_a}\left|\phi_b(\mathbf{x})\right|^2,
\ee
where $\widetilde{t}_{ab}$ is a transmission coefficient given that $\widetilde{\mathbb{T}}=\sum_{a,b=0}^{d-1} \widetilde{t}_{ab}\ket{b}\bra{a}$ and $\phi_b(\mathbf{x})$ is the $b$-th standard output optical field of circuit on the output $\mathbf{x}$ space. The normalisation of $\widetilde{t}_{ab}$ is measured by the ratio of optical flux inside the $b$-th output mode of a circuit, $\mathcal{I}_{ab}\propto|\widetilde{t}_{ab}|^2$, to total output optical flux transmitting through the system given the $a$-th input mode, $\mathcal{I}_a:=\int d^{2}\mathbf{x}\mathcal{I}_a(\mathbf{x})$ where $\mathcal{I}_a(\mathbf{x})\propto|\sum_{b=0}^{n-1} \widetilde{t}_{ab}\phi_{b}(\mathbf{x})|^2$. Noting that the optical flux $\mathcal{I}_{ab}$ can be moved outside the bracket because the target outputs are foci which are spatially separated and conveniently detected using a coherent light source. For a two-photon entangled state, the success probability is then calculated in a similar way using the outcomes of joint measurements between Alice and Bob, the latter of whom performs the measurements $\hat{\widetilde{\Pi}}(\mathbf{x}):=\eta(\mathbf{x})\ket{\mathbf{x}}\bra{\mathbf{x}}$ across the output spatial $\mathbf{x}$ space:
\be
\mathcal{S}^{\mu}= \frac{1}{d}  \sum_{a=0}^{d-1} \int d^{2}\mathbf{x} \sum_{b=0}^{d-1}\frac{ \mathcal{C}_{ab}^{\mu,\nu=0} }{\mathcal{C}_{a}^{\mu} (\mathbf{x})}\left|\phi_b(\mathbf{x})\right|^2,
\label{Eq:S_twophoton}
\ee
where the coincidence counts ($\mathcal{C}_{ab}^{\mu,\nu=0}$ and $\mathcal{C}_{a}^{\mu}(\mathbf{x})$) are calibrated by the detection efficiencies for both parties and $\mathcal{C}_{a}^{\mu}(\mathbf{x})$ is defined as
\be
\mathcal{C}_{a}^{\mu}(\mathbf{x}) = R  \frac{1}{\eta^{\mu}_a \eta(\mathbf{x})}\tr(\hat{\widetilde{\Pi}}^{\mu}_{a}\otimes \hat{\widetilde{\Pi}}(\mathbf{x})\rho_\text{o}). \\
\ee
Note that since we perform the experiment in all input spatial mode bases, the success probability is averaged over all input bases, $\mathcal{S} = \sum_{\mu=0}^{d} \mathcal{S}^{\mu}/(d+1)$. Moreover, one can show that Eq.~\ref{Eq:S_twophoton} can be reformulated back to Eq.~\ref{Eq:S} in the case that the input maximally entangled state and process (circuit) are pure.

Finally, the overall transmittance $\mathcal{T}$ of a circuit is measured using the two-photon entangled state at the input and output of a circuit,
\be
    \mathcal{T}^{\mu}_{a} = \sum\limits_{b=0}^{d-1} \frac{ \mathcal{C}^{\mu,\nu=0}_{ab} }{  \mathcal{C^{*}}^{\mu,\nu=0}_{ab}},
\ee
where $\mathcal{C^{*}}^{\mu,\nu=0}_{ab}$ is the coincidence counts at the initial state. The overall transmittance, $\mathcal{T}=\mathcal{S}\times\mathcal{T}_o$, includes the success probability, $\mathcal{S}$, of a circuit and other optical transmittance of the apparatus, $\mathcal{T}_o$.


\subsection*{High-dimensional entanglement certification with systematic errors}

The fidelity and purity of the gates (see Table~\ref{table:F} and Supplementary Information) show that our circuits are not perfect ($\mathcal{F}$<100\%). Recent work has shown that even slight imperfections in measurements can compromise entanglement witnesses that normally assume perfect measurements \cite{Morelli2022}. Here we investigate whether entanglement can be detected and quantified from the relative frequencies corresponding to our data measured in two approximately conjugate bases (red squares in Fig.~3). Focusing on the case of dimension five, we address the question of whether there exists some bipartite quantum state, $\rho_{AB}$, whose Schmidt number is no more than $s$, that can model the measured relative frequencies. Since the data must admit a quantum model with some quantum state, which has a corresponding Schmidt number, a negative answer implies that the state is entangled and that its Schmidt number is at least $s+1$. 

Knowledge of the precise noise present in our experiment can lead to stronger entanglement criteria~\cite{Jebarathinam2020,sadana2022} but since we do not know the relevant noise, we adopt an approach where the state is treated as uncharacterised. In our analysis, we assume that Alice's measurements correspond to the computational basis and the first MUB respectively. Thus we associate to her the measurement operators $A_{a|0}=\ketbra{a}{a}$ for the first setting and $A_{a|1}=\ketbra{M^1_a}{M^1_a}$, for the the second setting. Here, her outcome can take values $a=0,\ldots,d-1$. On Bob's side, the  POVMs $\{B_{b|\nu}\}$ for $\nu=0,1$ are inferred from tomographic data. Importantly, Bob's measurement operators corresponding to the relevant outcomes $b=0,\ldots,d-1$ are incomplete because the measurement features outcomes that are not accessible in the lab. We therefore ad-hoc complete his measurements by associating the additional measurement operator, $B_{d|\nu}=\openone -\sum_{b=0}^{d-1} B_{b|\nu}$, to the inaccessible outcome. In a quantum model, we expect the probabilities to follow Born's rule, 
\begin{equation}\label{pQ}
p_Q(a,b|k)=\tr \left(\mathcal{B}_{abk}\rho_{AB}\right),
\end{equation}
where $\mathcal{B}_{abk}=A_{a|k}\otimes B_{b|k}$.
Naturally, however, neither Alice's nor Bob's above measurements are characterised flawlessly. In other words, the probabilities $p_{\textrm{lab}}(a,b|k)$ inferred from the measured relative frequencies for the events $a,b\in\{0,\ldots,d-1\}$, will not exactly correspond to the given POVMs. Consequently, when addressing whether the underlying state must be entangled, we cannot even expect that there exists any quantum state such that $p_Q(a,b|k)=p_{\textrm{lab}}(a,b|k)$. This is the well-known drawback of standard entanglement witnessing. To address this issue, we must allow for $p_Q$ to reproduce $p_{\textrm{lab}}$ up to a reasonable accuracy, which accounts for the imperfections in the estimation of the lab POVMs of Alice and Bob and the photon counting statistics. Specifically, we introduce for each probability associated with the tuple $(a,b,k)$ a tolerance interval $t_{abk}$ such that
\begin{equation}\label{tol}
p_Q(a,b|k)-t_{abk}\leq p_{\textrm{lab}}(a,b|k)\leq p_Q(a,b|k)+t_{abk}.
\end{equation}
We determine the tolerances due to systematic and statistical errors in our experiment from both our measurement devices and our gates' tomography, via a Monte Carlo simulation of our experiment as detailed in Supplementary Information. From multiple simulations of the experiment, corresponding to $p_j(a,b|k)$, we compute for each $(a,b,k)$ the mean deviation from $p_{\textrm{lab}}$ and its standard deviation, i.e.~

\begin{align}
&\Delta_{abk}=\frac{1}{N}\sum_{j=1}^{N} X^{(j)}_{abk}\\
&\sigma_{abk}=\sqrt{\frac{1}{N}\sum_{j=1}^{N} \left(\Delta_{abk}-X_{abk}^{(j)}\right)^2},
\end{align}

where $X_{abk}^{(j)}=|p_{\textrm{lab}}(a,b|k)-p_j(a,b|k)|$ is the deviation in the sample $j$ for the  tuple $(a,b,k)$. We evaluate the standard deviation of ${p_j(a,b|k)}_{0<j<N}$ for all ${a,b,k}$ and find that this converges at $N\approx1500$, allowing us to safely choose $N=2000$. We then choose the tolerance for each and every tuple $(a,b,k)$ to be three standard deviations above the mean deviation from $p_{\textrm{lab}}$. Hence we put $t_{abk}=\Delta_{abk}+3\sigma_{abk}$.

With reasonable errors accounted for in the quantum model, we now wish to address whether entanglement is necessary to model the results of the experiment. Specifically, does there exist a $p_Q$ as given in \eqref{pQ} generated from a state $\rho_{AB}$ whose Schmidt number is at most $s$, such that it reproduces $p_{\textrm{lab}}$ within the tolerances \eqref{tol}? This question is both hard to solve and phrased as a binary decision. While we will soon address the former, we first emphasise that it is favourable to replace the latter with a quantitative statement because then one can also estimate the confidence in the falsification of the hypothesis (namely that the answer is positive). For that purpose, we introduce additional noise to our measured data $p_{\textrm{lab}}$. Specifically, we consider a mixture of our experimental results, $p_{\textrm{lab}}$,  with the probabilities obtained from probing the POVMs of Alice and Bob with a maximally mixed state. The mixing parameter is $v\in[0,1]$. This gives the artificially noisy probability distribution
\begin{equation}
p_{\textrm{noise}}(a,b|k)=vp_{\textrm{lab}}(a,b|k)+(1-v)\frac{ \text{Tr} B_{b|k}}{d^2}.
\end{equation}
Clearly, the experimental data corresponds to $v=1$. Now, we can use $v$ as a quantifier for a model in which $\rho_{AB}$ has Schmidt number at most $s$. That is, we solve
\begin{equation}\label{problem}
\begin{aligned}
\max \quad & v\\
\text{such that}\quad & p_Q(a,b|k)=\text{Tr}\left(\mathcal{B}_{abk} \rho_{AB}\right) \\
& p_{\textrm{noise}}(a,b|k) \geq p_Q(a,b|k)-t_{abk}\\
& p_{\textrm{noise}}(a,b|k) \leq p_Q(a,b|k)+t_{abk}\\
& \text{Tr}\left(\rho_{AB}\right)=1 \\
& \rho_{AB}\in \mathcal{S}_s\\
& \rho_{AB} \succeq 0,
\end{aligned}
\end{equation}
where $\mathcal{S}_s$ is the set of all bipartite states of local dimension $d$ with Schmidt number at most $s$. Note that any value $v<1$ implies that an amount of noise corresponding to a rate $1-v$ must be additionally added to the experimental data in order for it to admit a model for Schmidt number $s$. In other words, it implies that no model based on Schmidt number $s$ is possible and hence a Schmidt number at least $s+1$ is necessary.

As mentioned previously, it is difficult to solve \eqref{problem} because the characterisation of $\mathcal{S}_s$ is challenging. However, it is sufficient for our purposes to relax $\mathcal{S}_s$ into a larger set but over which computations can be done efficiently. It is well known that $\rho_{AB}\in \mathcal{S}_s$ implies that \cite{Tomiyama1985,Terhal2000};
\begin{equation}\label{sdpcon}
\left(\openone_d \otimes R_{\frac{1}{s}}\right)(\rho_{AB})\succeq 0,
\end{equation} 
where $R_{\alpha}(\rho)=\text{Tr}\left(\rho\right)\openone_d-\alpha \rho $ is the generalised reduction map. If we replace the condition $\rho_{AB}\in\mathcal{S}_s$ with \eqref{sdpcon}, we will obtain an upper bound on \eqref{problem} that is computable as a semidefinite program. We have evaluated this semidefinite program for Schmidt numbers $s=1,2,3,4,5$ and obtained the results 

\begin{equation}
\begin{aligned}
& s=1 \quad \Rightarrow \quad &  v\leq 0.7068\\
& s=2 \quad \Rightarrow \quad &  v\leq 0.8219\\
& s=3 \quad \Rightarrow \quad &  v\leq 0.9138\\
& s=4 \quad \Rightarrow \quad &  v\leq 0.9924\\
& s=5 \quad \Rightarrow \quad &  v\leq 1 .
\end{aligned}
\end{equation}

Note that $s=1$ corresponds to a separable model and that $s=5$ corresponds to a generic five-dimensional entangled state, as was used in the experiment. We conclude that within the estimated tolerances no quantum model for $p_{\textrm{lab}}$ exists that relies on less than five-dimensional entanglement. Thus, we certify Schmidt number $s=5$. Notably, only an additional noise rate of $0.0076$ is needed to make possible a model based on $s=4$. Thus, if we were to somewhat increase the tolerance interval, e.g.~from to $3\sigma$ above the mean to $4\sigma$ above the mean, then the certified Schmidt number would decrease to $s=4$. In either case, the data obtained by applying the $\mathbb{I}$ and $\mathbb{F}$ gates on our input state certify high-dimensional entanglement.

\section*{Data availability}
Data that support the plots within this paper and other findings of this study are available from the corresponding authors upon reasonable request. Source data are provided with this paper.

\section*{Code availability}

Code that supports simulations within this paper and other findings of this study are available in Ref~\cite{Goel_Simulation_Codes_for_2022}. The code used to measure and analyze the experimental data for this work is available on reasonable request.

\clearpage
\onecolumngrid
\appendix

\renewcommand{\thesubsection}{S.\arabic{section}.\arabic{subsection}}
\renewcommand{\thesection}{S.\arabic{section}}

\setcounter{equation}{0}
\setcounter{figure}{0}
\setcounter{table}{0}
\renewcommand{\theequation}{S.\arabic{section}.\arabic{equation}}
\renewcommand{\thetable}{S.\arabic{table}}
\renewcommand{\thefigure}{S.\arabic{figure}}
\renewcommand{\theHfigure}{S.\arabic{figure}}


\begin{center}\large{\textbf{Supplementary information for: Inverse-design of high-dimensional quantum optical circuits in a complex medium}} \end{center}

The following supplementary information is provided: details of the high-dimensionally entangled two-photon source are presented in \ref{Methods:source}, while the experimental estimation of success probability and optical losses is discussed in \ref{SI:exp_loss}. Detailed results on the manipulation of high-dimensionally entangled states in dimensions $(d=2,3,5,7)$ using various types of gates in the macro-pixel and the orbital-angular-momentum (OAM) bases are presented in \ref{SI:Results}. Furthermore, the independent characterization of our measurement apparatus and its effect on errors in QST and AA-QPT are reported in \ref{SI:PrepAndMeasure} and \ref{Methods:SystematicErrorsQST}. A brief discussion of the misestimation of the initial state in linear inversion is presented in \ref{Methods:AAQPTNoise}. Finally, numerical results on the programmability and scalability of our gates and the effect of randomness and symmetry of mode-mixers on the gate performance are reported in \ref{SI:Results_2} and \ref{SI:Results_RandomandSym}.

\section{\label{Methods:source}High-dimensional two-photon entanglement source}
\begin{figure*}[htp]
\centering\includegraphics[width=0.9\textwidth]{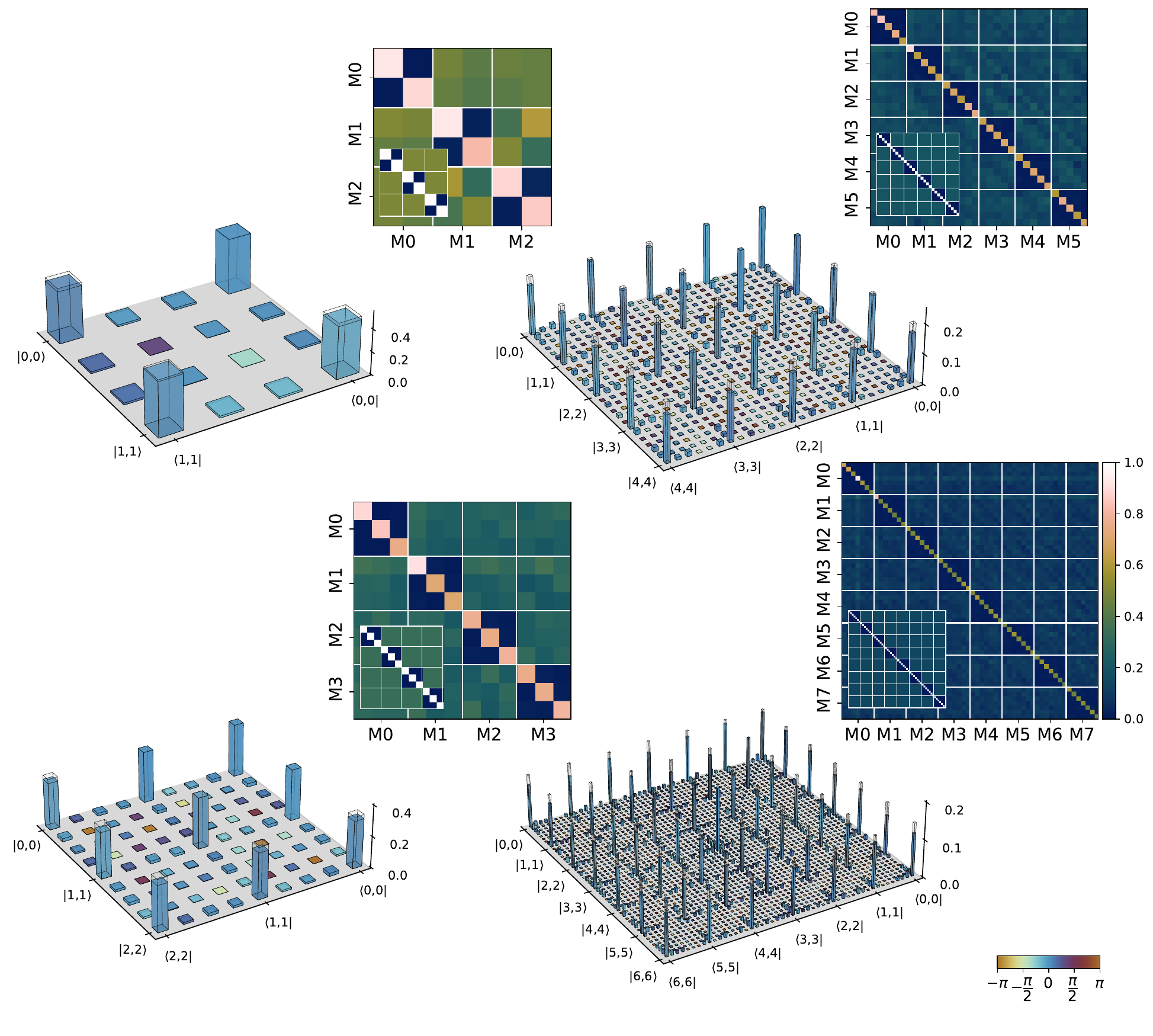}
\caption{\textbf{Generated high-dimensional entangled two-photon states in the macro-pixel basis.} Measured coincidence counts in all mutually unbiased bases (MUBs) and reconstructed density matrices via quantum state tomography in dimensions, $d=[2,3,5,7]$. The amplitude and phase of density matrix elements are represented by the height of the bars and their colour, respectively. The ideal amplitude is represented by a transparent bar overlaid on the experimental result.}
\label{fig:SPDCresultsPixel}
\end{figure*}

\begin{figure*}[htp]
\centering\includegraphics[width=0.9\textwidth]{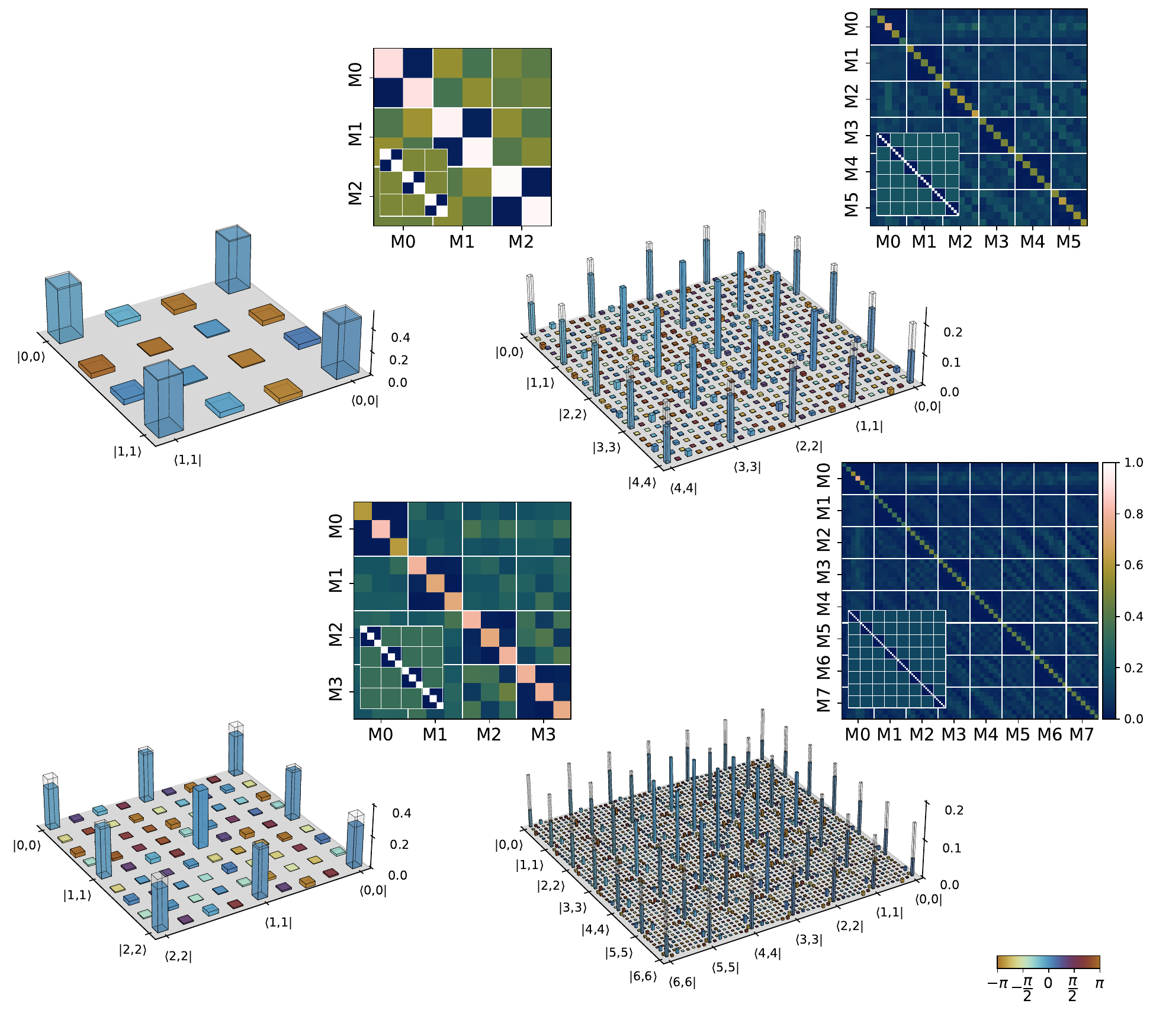}
\caption{\textbf{Generated high-dimensionally entangled two-photon states in the OAM basis.} Measured coincidence counts in all mutually unbiased bases (MUBs) and reconstructed density matrices via quantum state tomography in dimensions, $d=[2,3,5,7]$. The amplitude and phase of density matrix elements are represented by the height of the bars and their colour, respectively. The ideal amplitude is represented by a transparent bar overlaid on the experimental result.}
\label{fig:SPDCresultsOAM}
\end{figure*}

The high-dimensional spatially entangled two-photon state is generated via degenerate Type-II spontaneous parametric downconversion (SPDC) by pumping a 15-mm long periodically poled Potassium Titanyl Phosphate (ppKTP) crystal with a 405nm continuous-wave laser. After passing through the long-pass dichroic filter and the band-pass filter, the signal and idler photons are separated using a polarising-beam-splitter (PBS) and are mapped on to the liquid-crystal-on-silicon-based spatial light modulators (SLM) ($P_1$ and $P_3$) which are placed in the Fourier plane of the ppKTP crystal. The main characteristics of the entanglement source, i.e., the strength of transverse-momentum correlation, generated beam waist, and the position of the beams, are determined by using our developed $2D\pi$-measurement~\cite{Srivastav2021}, which involves the joint coincidence measurement of local $\pi$-phase step knife-edge scans across the SLMs at each party. The two-photon state is then characterised via quantum state tomography 
(See Methods)
and the entanglement dimensionality is certified~\cite{Bavaresco:2018gw} using a high-dimensional entanglement witness in two discrete spatial-mode bases---the Macro-pixel basis~\cite{HerreraValencia2020} and the orbital-angular-momentum (OAM) basis~\cite{Valencia2021}. We utilise these two bases as the set of input target modes for constructing the programmable circuits. For the macro-pixel basis, the size of pixels and their spacing are determined by the joint transverse momentum amplitude (JTMA)~\cite{Srivastav2021}. For the OAM basis, the set of modes are $\{\ket{-\ell},\dots,\ket{0},\dots,\ket{\ell}\}$ when the dimension of the circuit $d$ is odd, otherwise $\{\ket{-\ell},\dots,\ket{-1},\ket{1},\dots,\ket{\ell}\}$. The measured quantifiers of the generated high-dimensionally entangled two-photon states are reported in Table~\ref{table:inputstate} and the reconstructed density matrices are depicted in Fig.~\ref{fig:SPDCresultsPixel} and Fig.~\ref{fig:SPDCresultsOAM} for the macro-pixel and OAM bases, respectively.

\begin{smallboxtable}[float=h]{Generated high-dimensionally entangled two-photon states: State purity $\mathcal{P}$, Entanglement dimensionality $d_{ent}$, Fidelity to maximally entangled state $\mathcal{F}$, Entanglement of Formation ($E_{oF}$). The reported errors are due to misalignment of measurement apparatus as well as photon counting statistics as discussed in \ref{Methods:SystematicErrorsQST}.}{inputstate}
\begin{center}
\begin{tabular}{c|c|cccc}
    \hline
    Basis & Dimension ($d$) & $\mathcal{P}$ & $d_{ent}$ & $\mathcal{F}(\rho^\text{(in)},\rho^+)$ & $E_{oF}$ (ebits) \\
    \hline
\multirow{4}{*}{\rotatebox[origin=c]{90}{Macro-Pixel}}
& $2$  &  $96.7 \pm 0.2$ \%  & 2 & $ 96.8 \pm 1.4$ \% &   $ 0.798 \pm 0.042$ \\
& $3$  &  $92.6 \pm 0.2$ \%  & 3 & $ 95.3 \pm 0.4$ \% &   $ 1.231 \pm 0.013$ \\
& $5$  &  $88.8 \pm 0.1$ \%  & 5 & $ 91.6 \pm 0.6$ \% &   $ 1.510 \pm 0.021$ \\
& $7$  &  $75.1 \pm 1.5$ \%  & 6 & $ 83.0 \pm 1.2$ \% &   $ 1.194 \pm 0.012$ \\
\hline
\multirow{4}{*}{\rotatebox[origin=c]{90}{OAM}}
& $2$  &  $97.7 \pm 0.4$ \%  & 2 & $ 97.5 \pm 0.2$ \% &   $ 0.812 \pm 0.001$ \\
& $3$  &  $85.9 \pm 0.2$ \%  & 3 & $ 90.8 \pm 0.6$ \% &   $ 1.113 \pm 0.014$ \\
& $5$  &  $83.2 \pm 0.1$ \%  & 5 & $ 86.5 \pm 0.7$ \% &   $ 1.194 \pm 0.019$ \\
& $7$  &  $80.0 \pm 0.1$ \%  & 6 & $ 80.6 \pm 0.6$ \% &   $ 1.109 \pm 0.016$ \\
\hline
   \end{tabular}
  \color{black}
\end{center}
\end{smallboxtable}


\section{\label{SI:exp_loss}Experimental estimation of success probability and optical losses}

\begin{figure}[htp]
    \centering
       \includegraphics[width=0.4\linewidth]{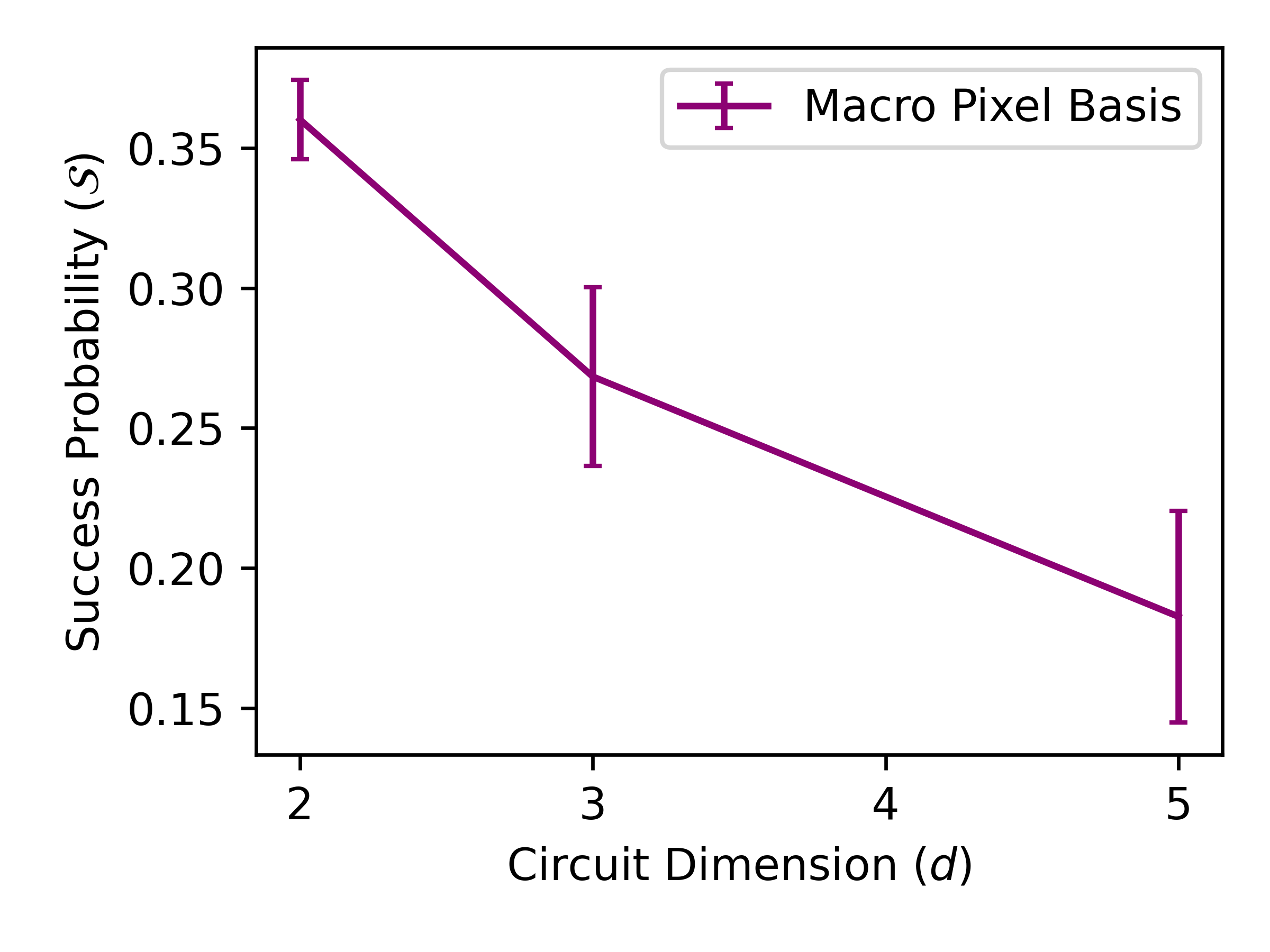}%
       \includegraphics[width=0.4\linewidth]{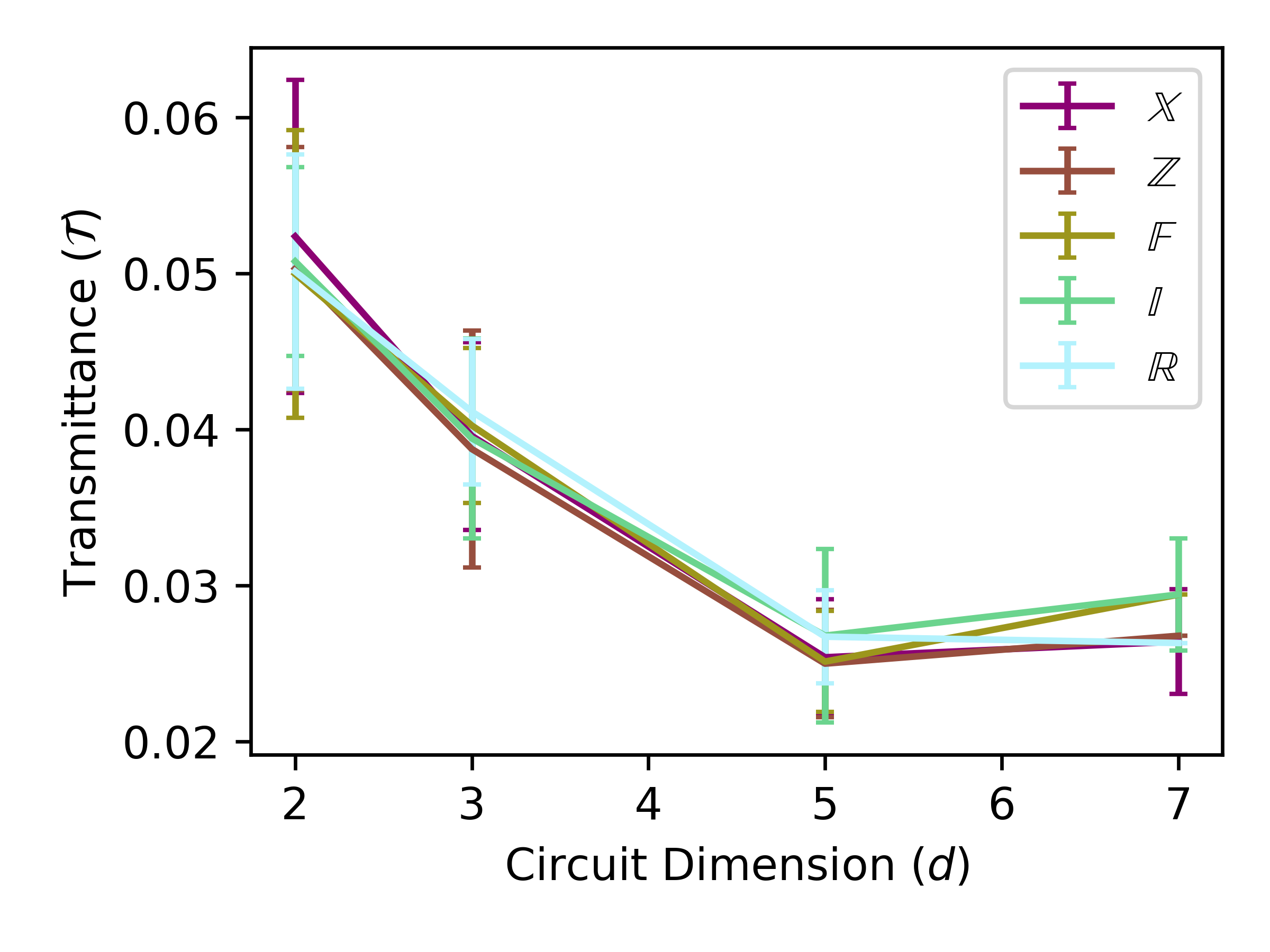}
    \caption{Success probability (left) and optical transmittance (right) in macro-pixel input basis.}
    \label{fig:SP}
\end{figure}

 In the experiment, the measurements of success probability and overall transmittance are performed in the macro-pixel input basis. The results are shown in Fig.~\ref{fig:SP} where the averaged optical transmittance is $0.051 \pm 0.008$, $0.040 \pm 0.006$, $0.026 \pm 0.004$,  and $0.028 \pm 0.003$ for 2, 3, 5, and 7-dimensional gates respectively. We measure the success probability for 3 randomly chosen implementations of Fourier-$\mathbb{F}$ gates in $2,3$, and $5$ dimensions and observe an $\mathcal{S}$ of $0.36 \pm 0.014$, $0.27 \pm 0.03$, and $0.18 \pm 0.04$, respectively. We thus infer an average optical transmittance of $\mathcal{T}_o= 0.14 \pm 0.02$ which is attributed to the optical insertion and propagation losses, diffraction efficiency of the two SLMs, and other interface losses at the lenses and mirrors. We note that the main contribution to the loss is due to the success probability arising from control of only a single polarisation channel of the multi-mode fiber, which can be further improved when all polarisation-spatial modes of the optical system are controlled.

\section{\label{SI:Results} Manipulation of high-dimensional spatially entangled states}

To study the programmabiltiy of our quantum circuits, we program many types of unitary gates, namely, the identity-$\mathbb{I}$, Pauli-$\mathbb{Z}$, Pauli-$\mathbb{X}$, Fourier-$\mathbb{F}$, and random unitaries-$\mathbb{R}$, in $2, 3, 5,$ and $7$ dimensions in both the macro-pixel and OAM input bases. For each implementation, we randomly appoint a set of target output foci to encode the circuit with. We then use these circuits to locally transform the $d$-dimensional spatially entangled two-photon state, which is then characterised by performing AA-QPT.

Table~\ref{table:best_FP_gates} shows the fidelity and purity of the best-case gates in each dimension. The measured tomography data and reconstructed density matrices of these processes are shown in Figs.~\ref{fig:SI_Pixel_res} and \ref{fig:SI_OAM_res} for the macro-pixel and OAM bases. The process fidelity quantifies how close to ideal our gates are, taking into account the non-maximally entangled nature of the input state. As a separate metric, the fidelities of the output state to an ideal transformed state allows independent verification that the gates preserve high-dimensional entanglement. Tables~\ref{table:best_output_states} and \ref{table:outputstate_dent} show the output state fidelity, purity, and entangement dimensionality for all best-case gates.

\begin{figure}[htp]
    \centering
        \includegraphics[width=0.45\linewidth]{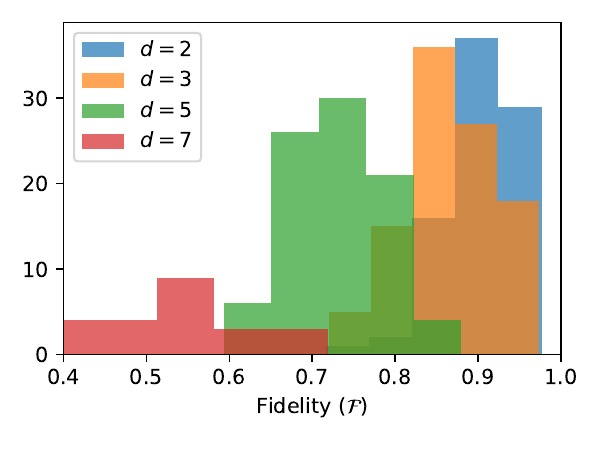}\\
    \caption{Histogram of process fidelities of  optical circuits implemented in different dimensions.}
    \label{fig:SI_gates_hist}
\end{figure}

\begin{figure}[htp]
    \centering
        \includegraphics[width=\linewidth]{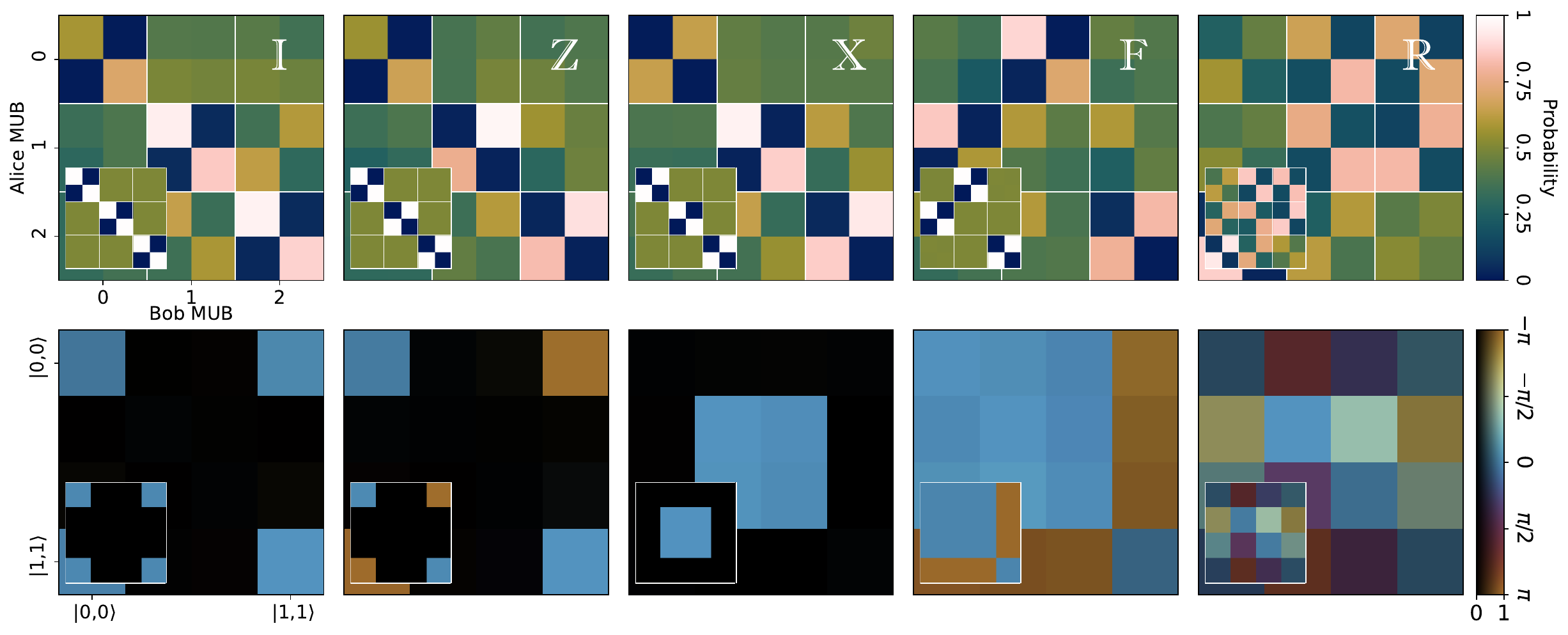}\\
        \includegraphics[width=\linewidth]{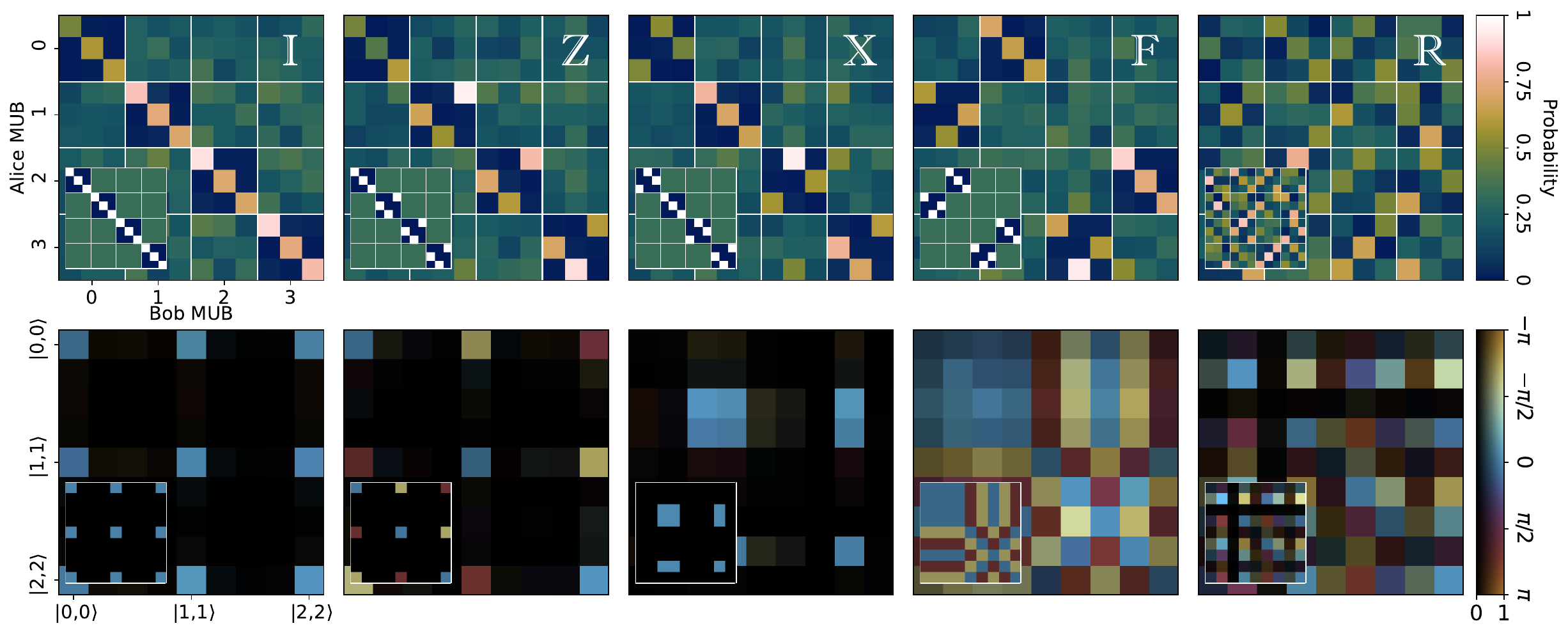}\\
        \includegraphics[width=\linewidth]{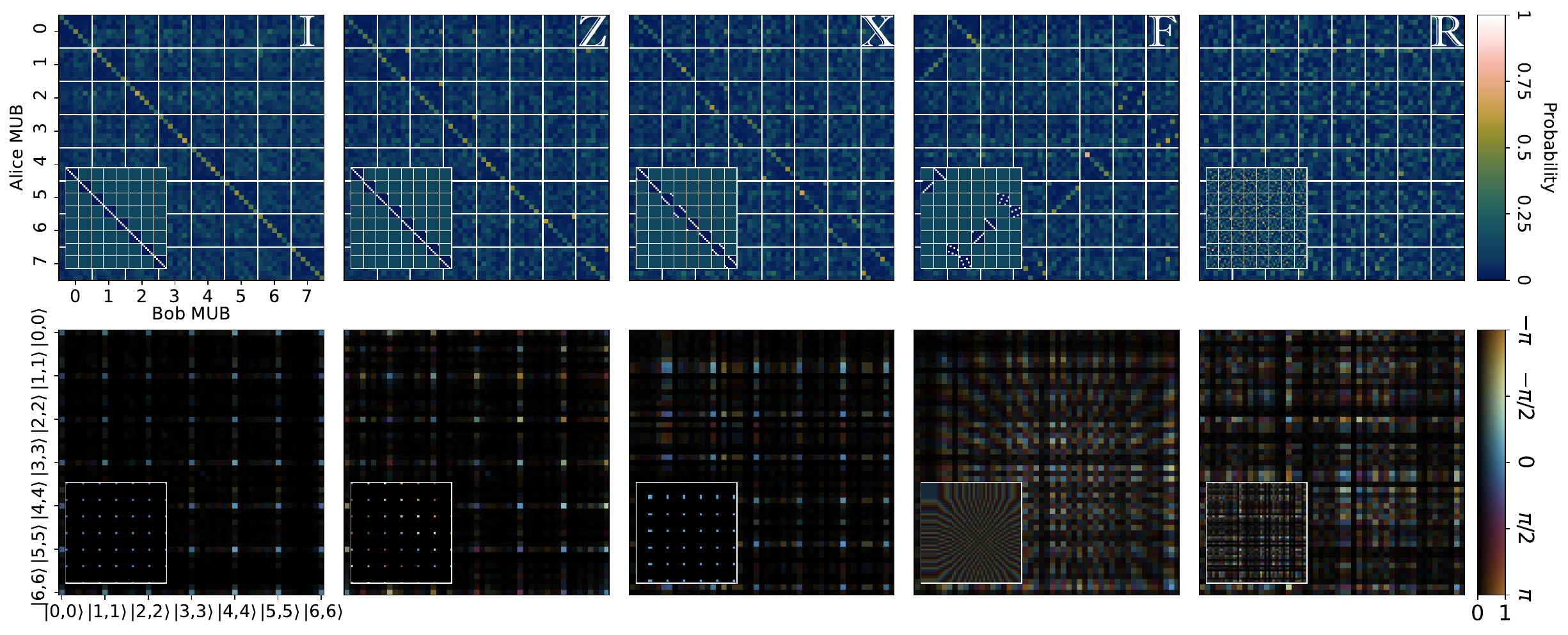}\\
    \caption{Manipulation of 2, 3, and 7-dimensional spatially entangled two-photon states in the macro-pixel basis using the $\mathbb{I}$, Pauli-$\mathbb{Z}$, Pauli-$\mathbb{X}$, Fourier $\mathbb{F}$, and random unitary $\mathbb{R}$ gates. In each panel, the upper part shows the two-photon coincidence counts in all MUBs and the lower part depicts reconstructed density matrices of the Choi states $\widetilde{\rho_{\xi*}}$.  The density matrix legend captures both the amplitude (brightness, normalised for clarity) and phase (colour).}
    \label{fig:SI_Pixel_res}
\end{figure}

In theory, the process fidelity of a circuit in a given dimension should be independent of the type of circuit and input/output modes used. In practice, however, experimental imperfections relating to the measurement apparatus and characterisation of the MMF result in a spread of fidelities with a lower average over all circuit instances, which are reported in Table.~\ref{table:mean_FP_gates}. This shows the average fidelities and purities of circuits for various types of gates sampled over different sets of output foci. The histogram of fidelities of all the implemented gates are shown in Fig.~\ref{fig:SI_gates_hist}.

\begin{smallboxtable}[float=h]{Fidelity $\mathcal{F}$ and purity $\mathcal{P}$ of the channel, $\widetilde{\rho_{\xi*}}$ implemented by a gate $\mathbb{T}$ in $d$ dimensions in the macro-pixel and OAM bases corresponding to the results presented in Fig.~\ref{fig:SI_Pixel_res} and~\ref{fig:SI_OAM_res}. The reported errors are due to misalignment of measurement apparatus as well as photon counting statistics as discussed in \ref{Methods:SystematicErrorsQST}.}{best_FP_gates}
\begin{center}

    \begin{tabular}{c|c|c|c|c|c|c|c|c|c}
    \hline
       \multirow{2}{*}{Basis} & \multirow{2}{*}{$\mathbb{T}$} &
    \multicolumn{4}{c|}{$\mathcal{F}(\widetilde{\rho_{\xi*}},\rho_\xi)$}&
    \multicolumn{4}{c}{$\mathcal{P}(\widetilde{\rho_{\xi*}})$} \\
   \cline{3-10}
   & & $d = 2$ & $d = 3$ & $d = 5$  & $d = 7$ & $d = 2$ & $d = 3$ & $d = 5$  & $d = 7$ \\
   \hline
 \multirow{5}{*}{\rotatebox[origin=c]{90}{Macro-Pixel}}
& $\mathbb{I}$& $96.7 \pm 0.9$ \%& $97.4 \pm 0.7$ \%& $88.0 \pm 0.7$ \%& $71.9 \pm 1.1$ \%& $94.5 \pm 0.4$ \%& $99.5 \pm 0.3$ \%& $95.4 \pm 0.4$ \%& $74.7 \pm 1.2$ \%\\
& $\mathbb{Z}$& $97.7 \pm 0.9$ \%& $96.1 \pm 0.5$ \%& $80.6 \pm 1.0$ \%& $65.2 \pm 1.0$ \%& $96.3 \pm 0.4$ \%& $99.5 \pm 0.3$ \%& $93.6 \pm 0.5$ \%& $77.5 \pm 1.0$ \%\\
& $\mathbb{X}$& $97.6 \pm 0.8$ \%& $95.0 \pm 0.7$ \%& $79.2 \pm 1.0$ \%& $60.1 \pm 1.0$ \%& $95.4 \pm 0.4$ \%& $97.5 \pm 0.4$ \%& $91.9 \pm 0.6$ \%& $82.6 \pm 1.3$ \%\\
& $\mathbb{F}$& $95.7 \pm 0.9$ \%& $89.3 \pm 0.8$ \%& $76.9 \pm 1.1$ \%& $58.9 \pm 0.7$ \%& $93.8 \pm 0.5$ \%& $88.3 \pm 0.5$ \%& $84.6 \pm 0.6$ \%& $75.1 \pm 1.5$ \%\\
& $\mathbb{R}$& $96.8 \pm 0.7$ \%& $91.8 \pm 0.8$ \%& $80.1 \pm 1.1$ \%& $63.5 \pm 0.7$ \%& $95.4 \pm 0.3$ \%& $91.3 \pm 0.3$ \%& $83.8 \pm 0.5$ \%& $77.3 \pm 1.4$ \%\\
\hline
\multirow{5}{*}{\rotatebox[origin=c]{90}{OAM}}
& $\mathbb{I}$& $91.2 \pm 0.4$ \%& $89.8 \pm 0.7$ \%& $76.6 \pm 1.0$ \%& $56.9 \pm 0.9$ \%& $92.2 \pm 0.2$ \%& $92.1 \pm 0.4$ \%& $83.1 \pm 0.3$ \%& $61.5 \pm 0.4$ \%\\
& $\mathbb{Z}$& $91.0 \pm 0.4$ \%& $87.0 \pm 0.6$ \%& $79.1 \pm 0.6$ \%& $54.2 \pm 1.0$ \%& $91.5 \pm 0.2$ \%& $87.3 \pm 0.3$ \%& $83.3 \pm 0.3$ \%& $61.2 \pm 0.3$ \%\\
& $\mathbb{X}$& $95.0 \pm 0.5$ \%& $91.0 \pm 0.5$ \%& $74.4 \pm 0.9$ \%& $53.0 \pm 0.9$ \%& $93.9 \pm 0.3$ \%& $94.3 \pm 0.3$ \%& $82.7 \pm 0.3$ \%& $61.4 \pm 0.3$ \%\\
& $\mathbb{F}$& $93.3 \pm 0.4$ \%& $87.9 \pm 0.5$ \%& $75.3 \pm 0.7$ \%& $57.4 \pm 0.7$ \%& $94.6 \pm 0.2$ \%& $87.9 \pm 0.2$ \%& $81.3 \pm 0.3$ \%& $62.3 \pm 0.3$ \%\\
& $\mathbb{R}$& $92.2 \pm 0.5$ \%& $84.9 \pm 0.8$ \%& $72.3 \pm 0.9$ \%& $51.7 \pm 1.2$ \%& $91.4 \pm 0.3$ \%& $88.2 \pm 0.3$ \%& $81.2 \pm 0.3$ \%& $64.1 \pm 0.3$ \%\\
\hline
    \end{tabular}
\end{center}
\end{smallboxtable}

\begin{smallboxtable}[float=h]{Fidelity $\mathcal{F}$ and purity $\mathcal{P}$ of the output state $\widetilde{\rho_{\xi*}^{(out)}}$, after being operated on by gate $\mathbb{T}$ in $d$ dimensions in the macro-pixel and OAM bases. These results correspond to the data and density matrices presented in Fig.~\ref{fig:SI_Pixel_res} and~\ref{fig:SI_OAM_res}. The reported errors are due to misalignment of measurement apparatus as well as photon counting statistics as discussed in \ref{Methods:SystematicErrorsQST}.}{best_output_states}

\begin{center}

    \begin{tabular}{c|c|c|c|c|c|c|c|c|c}
    \hline
      \multirow{2}{*}{Basis} & \multirow{2}{*}{$\mathbb{T}$} &
    \multicolumn{4}{c|}{$\mathcal{F}(\widetilde{\rho_{\xi*}^\text{(out)}},\rho_\xi)$}&
    \multicolumn{4}{c}{$\mathcal{P}(\widetilde{\rho_{\xi*}^\text{(out)}})$} \\
  \cline{3-10}
  & & $d = 2$ & $d = 3$ & $d = 5$  & $d = 7$ & $d = 2$ & $d = 3$ & $d = 5$  & $d = 7$ \\
  \hline
\multirow{5}{*}{\rotatebox[origin=c]{90}{Macro-Pixel}}
& $\mathbb{I}$& $95.6 \pm 1.2$ \%& $95.6 \pm 0.8$ \%& $85.3 \pm 0.7$ \%& $69.0 \pm 0.8$ \%& $96.4 \pm 0.4$ \%& $97.8 \pm 0.3$ \%& $91.4 \pm 0.4$ \%& $70.3 \pm 0.9$ \%\\
& $\mathbb{Z}$& $97.5 \pm 1.1$ \%& $95.9 \pm 0.7$ \%& $80.8 \pm 0.9$ \%& $58.1 \pm 0.5$ \%& $97.9 \pm 0.5$ \%& $99.5 \pm 0.3$ \%& $90.3 \pm 0.5$ \%& $68.0 \pm 0.6$ \%\\
& $\mathbb{X}$& $96.3 \pm 1.0$ \%& $93.8 \pm 0.8$ \%& $77.3 \pm 0.9$ \%& $53.9 \pm 0.8$ \%& $96.5 \pm 0.4$ \%& $97.1 \pm 0.4$ \%& $91.3 \pm 0.5$ \%& $70.0 \pm 0.4$ \%\\
& $\mathbb{F}$& $93.0 \pm 0.8$ \%& $87.6 \pm 0.8$ \%& $76.7 \pm 0.8$ \%& $55.2 \pm 0.6$ \%& $92.2 \pm 0.4$ \%& $85.5 \pm 0.4$ \%& $80.4 \pm 0.4$ \%& $69.5 \pm 0.6$ \%\\
& $\mathbb{R}$& $93.4 \pm 0.7$ \%& $89.2 \pm 0.9$ \%& $76.7 \pm 0.6$ \%& $61.8 \pm 0.6$ \%& $92.9 \pm 0.3$ \%& $88.7 \pm 0.3$ \%& $83.1 \pm 0.4$ \%& $70.8 \pm 0.5$ \%\\
\hline
\multirow{5}{*}{\rotatebox[origin=c]{90}{OAM}}
& $\mathbb{I}$& $92.0 \pm 0.3$ \%& $83.7 \pm 0.7$ \%& $75.6 \pm 1.0$ \%& $55.6 \pm 0.7$ \%& $91.4 \pm 0.2$ \%& $83.8 \pm 0.2$ \%& $77.5 \pm 0.3$ \%& $70.8 \pm 0.2$ \%\\
& $\mathbb{Z}$& $92.9 \pm 0.4$ \%& $82.2 \pm 0.5$ \%& $76.1 \pm 0.5$ \%& $51.8 \pm 0.5$ \%& $91.5 \pm 0.2$ \%& $81.1 \pm 0.2$ \%& $78.5 \pm 0.2$ \%& $71.6 \pm 0.2$ \%\\
& $\mathbb{X}$& $93.7 \pm 0.4$ \%& $86.2 \pm 0.5$ \%& $73.0 \pm 0.6$ \%& $51.1 \pm 0.8$ \%& $92.4 \pm 0.2$ \%& $84.1 \pm 0.2$ \%& $79.1 \pm 0.2$ \%& $74.5 \pm 0.2$ \%\\
& $\mathbb{F}$& $93.9 \pm 0.4$ \%& $84.3 \pm 0.5$ \%& $75.7 \pm 0.3$ \%& $58.6 \pm 0.4$ \%& $93.5 \pm 0.2$ \%& $79.7 \pm 0.2$ \%& $77.5 \pm 0.2$ \%& $71.6 \pm 0.2$ \%\\
& $\mathbb{R}$& $92.0 \pm 0.5$ \%& $83.4 \pm 0.3$ \%& $68.9 \pm 0.5$ \%& $50.4 \pm 0.3$ \%& $91.6 \pm 0.2$ \%& $80.8 \pm 0.2$ \%& $78.0 \pm 0.2$ \%& $74.2 \pm 0.2$ \%\\
\hline
    \end{tabular}
    \label{tab:best_output_states}
\end{center}
\end{smallboxtable}

\begin{smallboxtable}[float=h]{Entanglement dimensionality of the output state $\widetilde{\rho_{\xi*}^{(out)}}$, after being operated on by gate $\mathbb{T}$ in $d$ dimensions in the macro-pixel and OAM bases. These results correspond to the data and density matrices presented in Fig.~\ref{fig:SI_Pixel_res} and~\ref{fig:SI_OAM_res}. The reported errors in dimensionality are based on the errors in fidelity reported in Table~\ref{table:best_output_states}.}{outputstate_dent}

\begin{center}

    \begin{tabular}{c|c|c|c|c|c|c|c|c}
    \hline
      \multirow{3}{*}{$\mathbb{T}$} &
     \multicolumn{8}{c}{ $d_{\text{(ent)}} (\widetilde{\rho_{\xi*}^\text{(out)}})$ } \\
  \cline{2-9}
      &  \multicolumn{4}{c|}{ Macro-Pixel} & \multicolumn{4}{c}{OAM} \\
        \cline{2-9}
   & $d = 2$ & $d = 3$ & $d = 5$  & $d = 7$  & $d = 2$ & $d = 3$ & $d = 5$  & $d = 7$ \\
 \hline
$\mathbb{I}$& $2\substack{ +0\\ -0 } $ & $3\substack{ +0\\ -0 } $ & $5\substack{ +0\\ -0 } $ & $5\substack{ +1\\ -0 } $ & $2\substack{ +0\\ -0 } $ & $3\substack{ +0\\ -0 } $ & $4\substack{ +0\\ -0 } $ & $4\substack{ +1\\ -0 } $ \\
$\mathbb{Z}$& $2\substack{ +0\\ -0 } $ & $3\substack{ +0\\ -0 } $ & $5\substack{ +0\\ -1 } $ & $5\substack{ +0\\ -1 } $ & $2\substack{ +0\\ -0 } $ & $3\substack{ +0\\ -0 } $ & $4\substack{ +1\\ -0 } $ & $4\substack{ +0\\ -0 } $ \\
$\mathbb{X}$& $2\substack{ +0\\ -0 } $ & $3\substack{ +0\\ -0 } $ & $4\substack{ +1\\ -0 } $ & $4\substack{ +1\\ -0 } $ & $2\substack{ +0\\ -0 } $ & $3\substack{ +0\\ -0 } $ & $4\substack{ +0\\ -0 } $ & $4\substack{ +0\\ -0 } $ \\
$\mathbb{F}$& $2\substack{ +0\\ -0 } $ & $3\substack{ +0\\ -0 } $ & $4\substack{ +1\\ -0 } $ & $4\substack{ +1\\ -0 } $ & $2\substack{ +0\\ -0 } $ & $3\substack{ +0\\ -0 } $ & $4\substack{ +1\\ -0 } $ & $5\substack{ +0\\ -1 } $ \\
$\mathbb{R}$& $2\substack{ +0\\ -0 } $ & $3\substack{ +0\\ -0 } $ & $4\substack{ +1\\ -0 } $ & $5\substack{ +0\\ -0 } $ & $2\substack{ +0\\ -0 } $ & $3\substack{ +0\\ -0 } $ & $4\substack{ +0\\ -0 } $ & $4\substack{ +0\\ -0 } $ \\
\hline
\end{tabular}
\label{tab:outputstate_dent}
\end{center}
\end{smallboxtable}

\begin{smallboxtable}[float=h]{Average fidelity $\mathcal{F}$ and purity $\mathcal{P}$ of the channel, $\widetilde{\rho_{\xi*}}$ implemented by a gate $\mathbb{T}$ in $d$ dimensions in the macro-pixel and OAM bases. The standard deviation is calculated from an ensemble of implemented gates with different output foci, and for many realisations for the $\mathbb{R}$ gates. }{mean_FP_gates}
\begin{center}
    \begin{tabular}{c|c|c|c|c|c|c|c|c|c}
    \hline
       \multirow{2}{*}{Basis} & \multirow{2}{*}{$\mathbb{T}$} &
    \multicolumn{4}{c|}{$\mathcal{F}(\widetilde{\rho_{\xi*}},\rho_\xi)$}&
    \multicolumn{4}{c}{$\mathcal{P}(\widetilde{\rho_{\xi*}})$} \\
   \cline{3-10}
   & & $d = 2$ & $d = 3$ & $d = 5$  & $d = 7$ & $d = 2$ & $d = 3$ & $d = 5$  & $d = 7$ \\
   \hline
\multirow{5}{*}{\rotatebox[origin=c]{90}{Macro-Pixel}}
& $\mathbb{I}$& $91.7 \pm 4.7$ \%& $89.4 \pm 5.7$ \%& $78.8 \pm 6.3$ \%& $70.0$ \%$^\star$& $95.1 \pm 1.0$ \%& $97.0 \pm 2.4$ \%& $89.7 \pm 3.7$ \%& $77.4$ \%$^\star$\\
& $\mathbb{Z}$& $91.8 \pm 3.8$ \%& $91.7 \pm 4.3$ \%& $75.2 \pm 4.2$ \%& $65.2$ \%$^\star$& $94.7 \pm 1.7$ \%& $97.1 \pm 1.3$ \%& $90.2 \pm 2.6$ \%& $77.5$ \%$^\star$\\
& $\mathbb{X}$& $91.7 \pm 3.6$ \%& $90.5 \pm 4.9$ \%& $74.0 \pm 5.2$ \%& $59.0$ \%$^\star$& $94.7 \pm 1.3$ \%& $97.5 \pm 1.4$ \%& $88.7 \pm 3.5$ \%& $78.7$ \%$^\star$\\
& $\mathbb{F}$& $88.8 \pm 3.9$ \%& $86.5 \pm 1.9$ \%& $69.6 \pm 4.5$ \%& $58.9$ \%$^\star$& $89.8 \pm 2.5$ \%& $88.3 \pm 2.2$ \%& $80.5 \pm 2.0$ \%& $75.1$ \%$^\star$\\
& $\mathbb{R}$& $89.2 \pm 6.5$ \%& $85.1 \pm 3.1$ \%& $72.8 \pm 4.6$ \%& $63.5$ \%$^\star$& $92.6 \pm 2.1$ \%& $88.7 \pm 1.8$ \%& $83.1 \pm 1.4$ \%& $77.3$ \%$^\star$\\
\hline
\multirow{5}{*}{\rotatebox[origin=c]{90}{OAM}}
& $\mathbb{I}$& $88.3 \pm 2.0$ \%& $83.5 \pm 4.5$ \%& $69.3 \pm 4.0$ \%& $54.2 \pm 3.0$ \%& $91.2 \pm 0.7$ \%& $89.1 \pm 2.3$ \%& $81.3 \pm 2.2$ \%& $63.6 \pm 3.6$ \%\\
& $\mathbb{Z}$& $88.4 \pm 3.4$ \%& $84.5 \pm 2.4$ \%& $71.8 \pm 5.4$ \%& $48.3 \pm 7.6$ \%& $90.2 \pm 1.0$ \%& $89.7 \pm 1.5$ \%& $81.1 \pm 1.2$ \%& $61.5 \pm 1.7$ \%\\
& $\mathbb{X}$& $92.3 \pm 2.5$ \%& $82.7 \pm 4.8$ \%& $71.9 \pm 2.6$ \%& $46.2 \pm 4.8$ \%& $93.6 \pm 1.5$ \%& $88.4 \pm 3.1$ \%& $81.0 \pm 1.8$ \%& $61.7 \pm 1.3$ \%\\
& $\mathbb{F}$& $90.3 \pm 2.4$ \%& $81.0 \pm 5.2$ \%& $69.3 \pm 3.0$ \%& $51.7 \pm 4.2$ \%& $92.2 \pm 1.7$ \%& $85.9 \pm 1.8$ \%& $78.8 \pm 2.4$ \%& $62.8 \pm 2.0$ \%\\
& $\mathbb{R}$& $90.1 \pm 2.6$ \%& $81.4 \pm 3.6$ \%& $69.3 \pm 3.0$ \%& $47.9 \pm 4.4$ \%& $90.5 \pm 1.1$ \%& $88.7 \pm 4.0$ \%& $79.7 \pm 3.3$ \%& $62.0 \pm 4.1$ \%\\
\hline
    \end{tabular}
    
\end{center}
\small{$^\star$The standard deviation is not reported since we only have one realization of the macro-pixel gates in $d=7$.}
\end{smallboxtable}

\begin{figure}[htp]
    \centering
    \includegraphics[width=\linewidth]{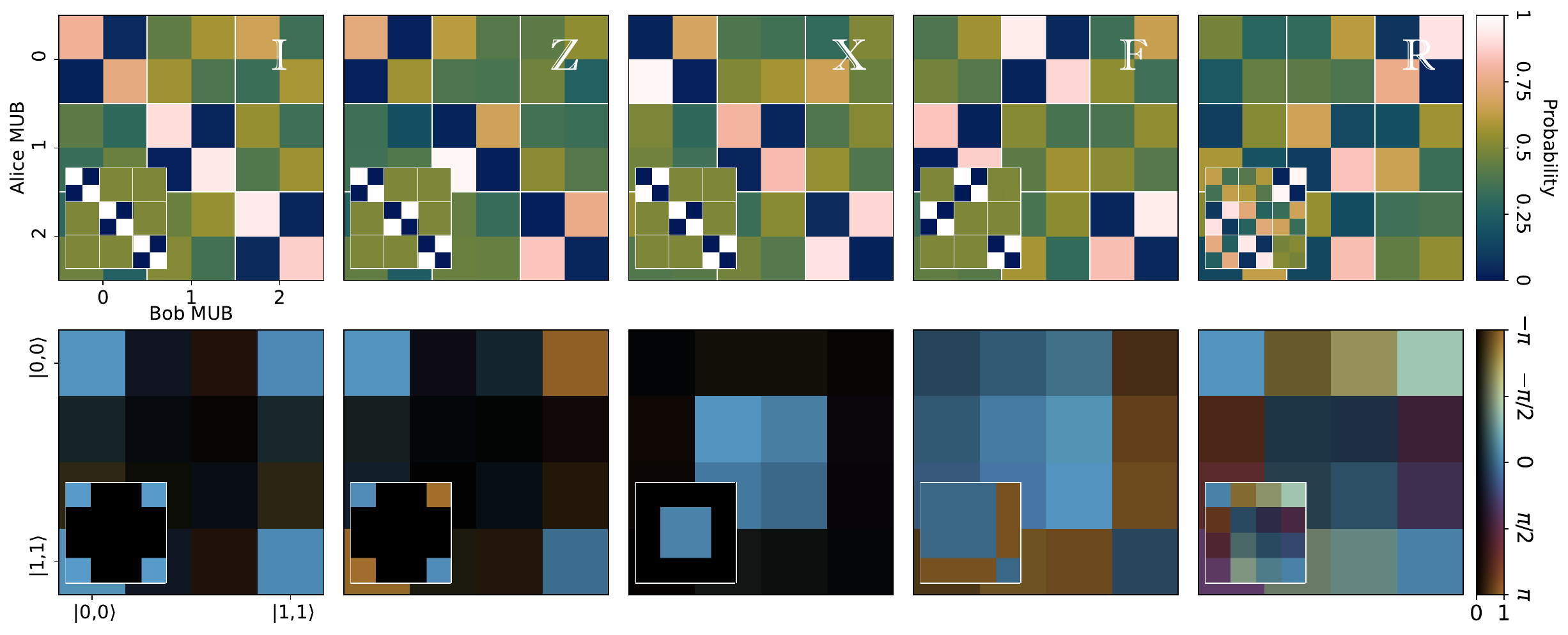}\\
    \includegraphics[width=\linewidth]{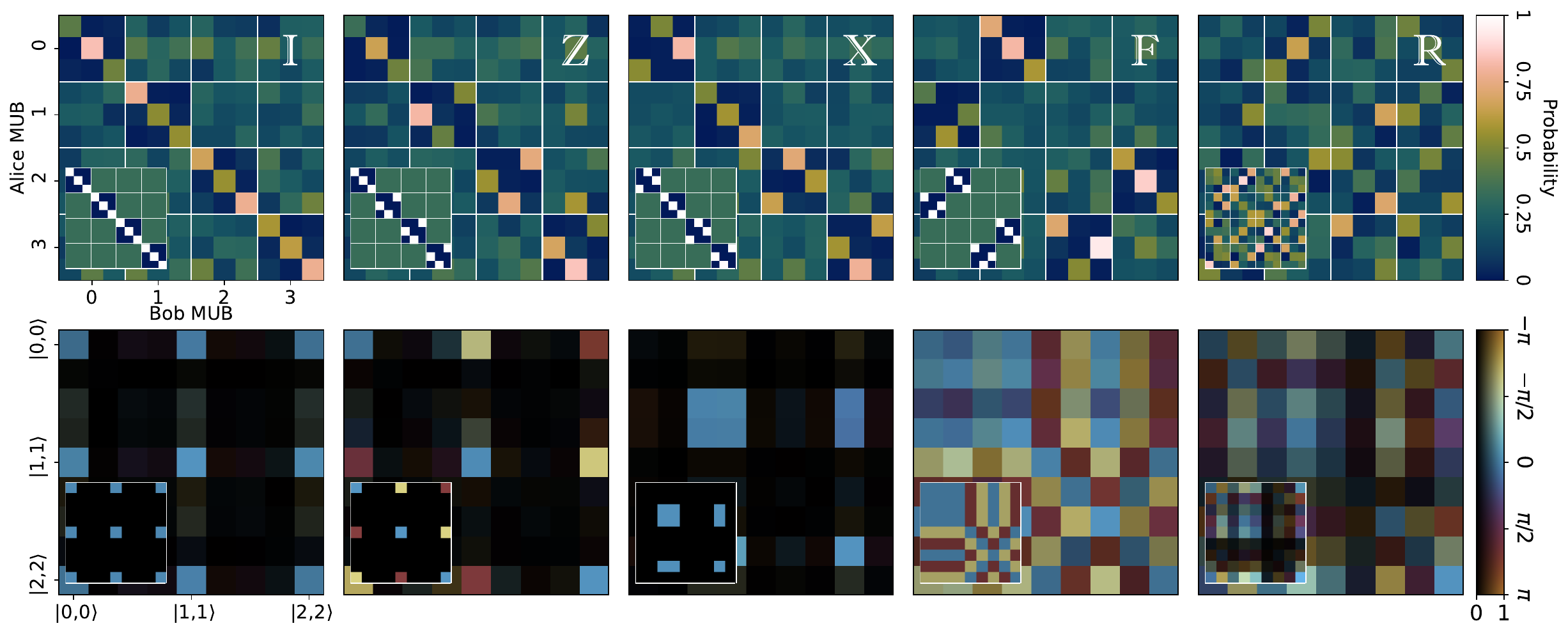}\\
    \includegraphics[width=\linewidth]{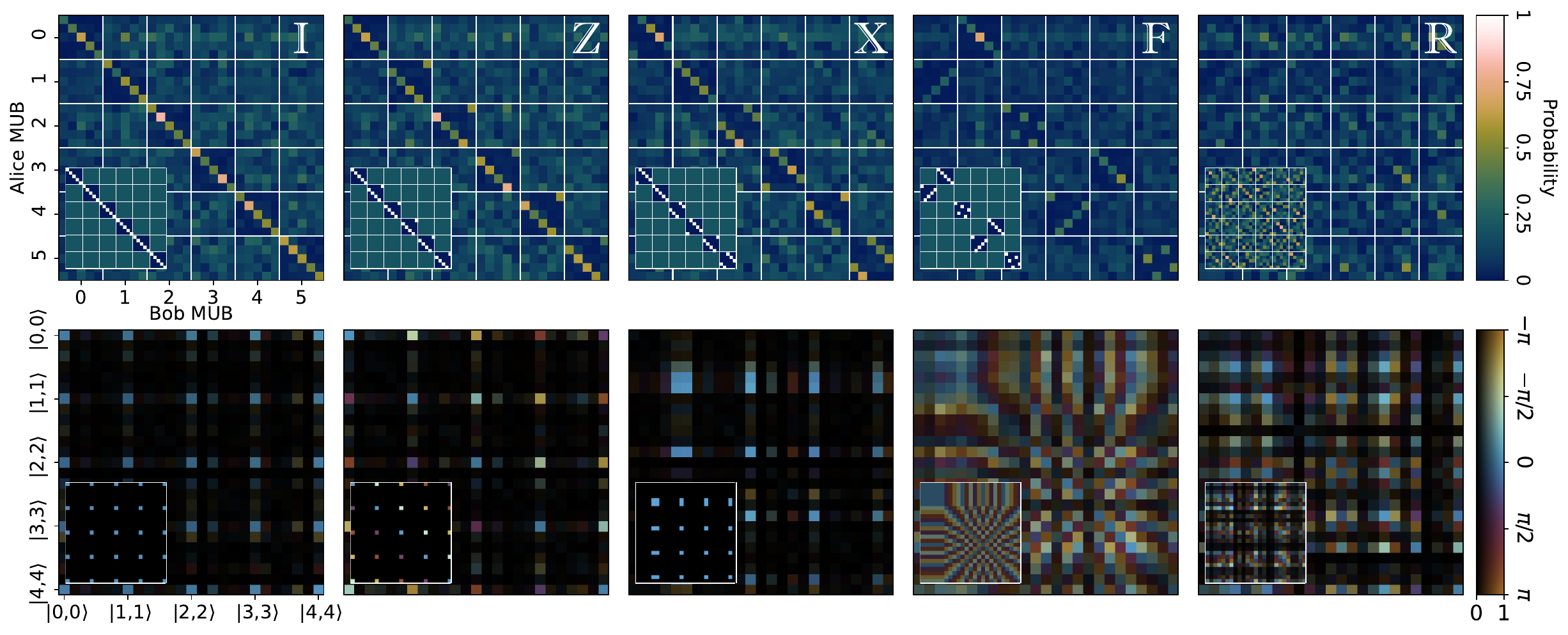}\\
    \end{figure}
\clearpage
\begin{figure}[H]
        \centering
        \includegraphics[width=\linewidth]{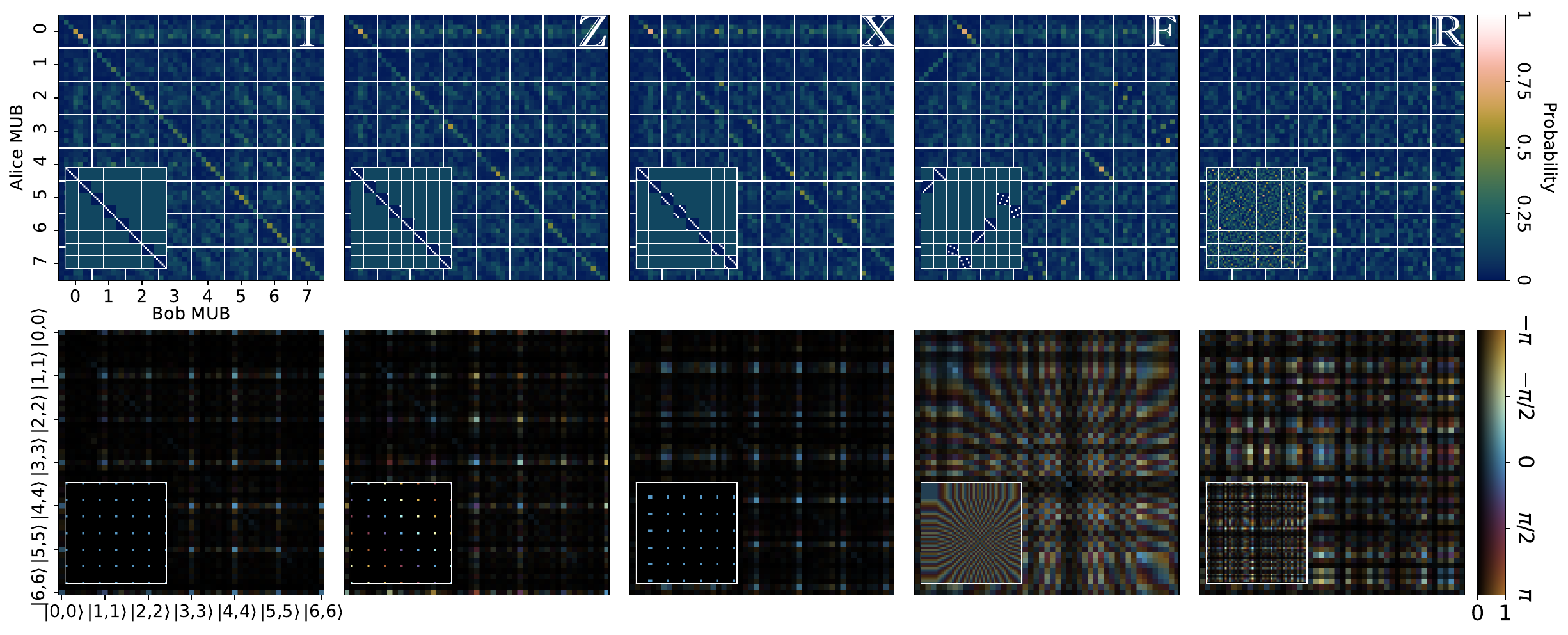}%
    \caption{Manipulation of 2, 3, 5, and 7-dimensional spatially entangled two-photon states in the OAM basis using the $\mathbb{I}$, Pauli-$\mathbb{Z}$, Pauli-$\mathbb{X}$, Fourier $\mathbb{F}$, and random unitary $\mathbb{R}$ gates. In each panel, the upper part shows the two-photon coincidence counts in all MUBs and the lower part depicts reconstructed density matrices of the Choi states $\widetilde{\rho_{\xi*}}$.  The density matrix legend captures both the amplitude (brightness, normalised for clarity) and phase (colour).}
    \label{fig:SI_OAM_res}
\end{figure}

\section{\label{SI:PrepAndMeasure} Independent characterisation of measurement apparatus and inverse-designed gates}

\begin{figure*}[htp]
\centering\includegraphics[width=0.9\textwidth]{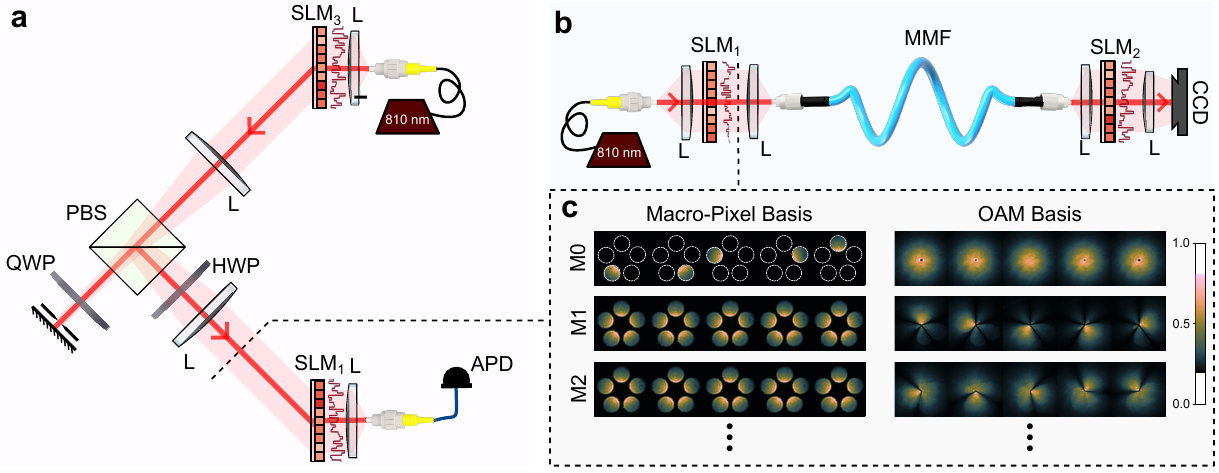}
\caption{\textbf{Independent characterisation of experiment: }(a) Prepare-and-measure setup implemented in an unfolded two-photon experiment used to characterise our measurement apparatus. (b) Prepare-and-measure setup used to characterise our inverse-designed gates. (c) CCD images of the classical input modes in the macro-pixel and OAM bases used for characterisation.}
\label{fig:prep_measure_fig}
\end{figure*}

In this section, we characterise our measurement apparatus and gate transformations independent of our entangled input source. This allows us to study how imperfections in the measurement apparatus limit the quality of our measured states and transformations.

Our measurements are characterised by performing a classical prepare-and-measure experiment in situ in the experimental setup. We unfold our setup about the nonlinear crystal by replacing it with a mirror and analyse it in the Klyshko advanced-wave picture~\cite{klyshko1988simple}.
As shown in Fig.~\ref{fig:prep_measure_fig}(a), a weak coherent state (superluminiscent diode filtered at $810 \pm 1.5 nm$ ) with the identical wavelength as the entangled photons is back-propagated from the idler photon measurement stage and prepared in a chosen state (precisely those used in our QST measurements). This state is then directed to the signal measurement stage by a mirror placed at the crystal plane, and subsequently projected by the signal photon measurement stage and recorded by an avalanche photo-diode. The measurements performed with the unfolded classical setup are identical to those performed in the main experiment, and allow us to gauge the reduction in the quality of entangled state and gates due to the imperfect measurement apparatus.

\begin{figure*}[htp]
\centering\includegraphics[width=0.9\textwidth]{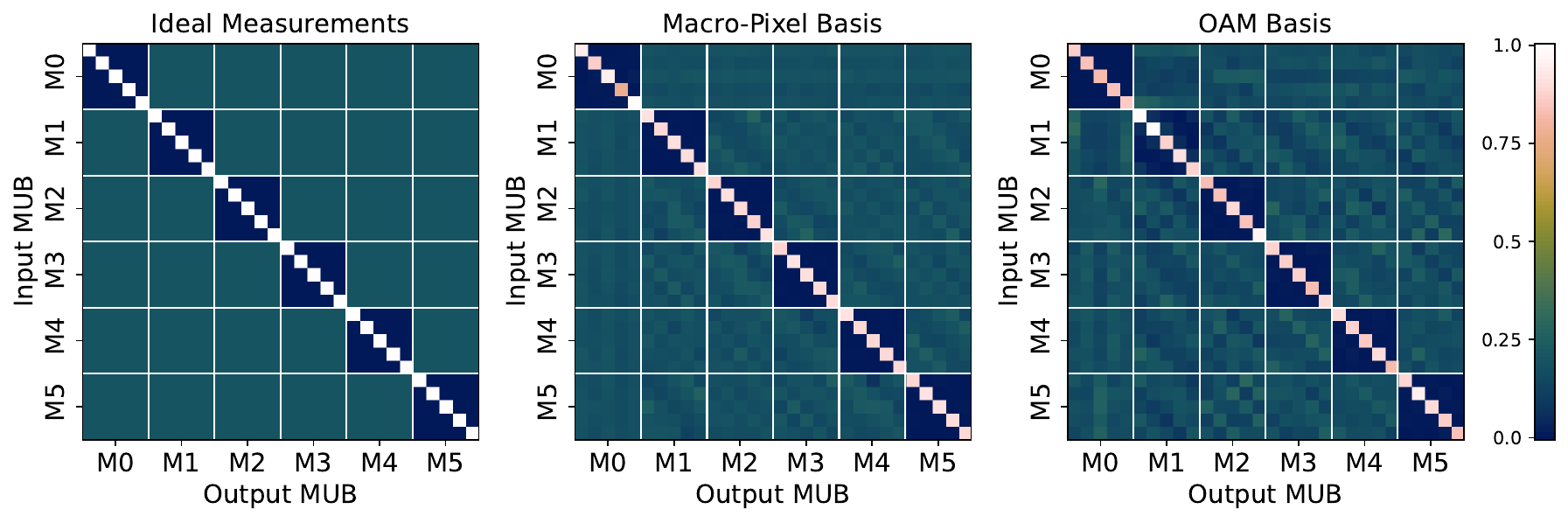}
\caption{\textbf{Characterisation of measurement apparatus:} Cross-talk matrices corresponding to (a) ideal measurements in the macro-pixel/OAM bases, and imperfect measurements in the (b) macro-pixel basis and (c) OAM basis characterised with a classical prepare-and-measure setup.}
\label{fig:misal_cmats}
\end{figure*}

Fig.~\ref{fig:misal_cmats} shows the measured cross-talk matrices for five-dimensional state preparations and measurements for every MUB in the macro-pixel and OAM basis. Fig.~\ref{fig:prep_measure_fig}(c) shows CCD images of the prepared input modes (image plane denoted by dotted lines in Fig~\ref{fig:prep_measure_fig}). The imperfections in the coupling matrices can be attributed to misalignments in the experimental setup, as well as imperfections in the computer generated holograms (CGHs) due to pixelisation and non-unit fill fraction. The effect of CGH-induced imperfections can be seen to be more pronounced in the OAM basis since superpositions of modes in this basis require amplitude modulation, while macro-pixel measurements can be performed with phase-only modulation. 

Treating these measurements as analogous to (folded) two-photon measurements allow us to upper-bound the quality of a maximally entangled state that can be measured with this setup. We use this procedure to obtain cross-talk matrices in the macro-pixel and OAM bases in $2,3,5$, and $7$ dimensions. We then perform QST on this data to calculate an upper bound on the fidelity ($\mathcal{F}$) to a maximally entangled state, purity ($\mathcal{P}$), and entanglement dimensionality that can be measured with this setup as shown in Table~\ref{table:PRinputstate}.

\begin{smallboxtable}[float=h]{Expected reduction in state purity $\mathcal{P}$, entanglement dimensionality $d_{ent}$, and fidelity $\mathcal{F}$ of a $d$-dimensional maximally entangled state due to imperfect measurements characterised using a classical prepare-and-measure experiment. The measured input state fidelities $\mathcal{F}_\text{exp}(\rho^\text{(in)},\rho^+)$ from Table~\ref{table:inputstate} are shown for comparison.}{PRinputstate}
\begin{center}
\begin{tabular}{c|c|ccccc}
    \hline
    Basis & Dimension ($d$) & $\mathcal{P}$ & $d_{ent}$ & $\mathcal{F}(\rho^\text{(in)},\rho^+)$ & $\mathcal{F}_\text{exp}(\rho^\text{(in)},\rho^+)$\\
    \hline
\multirow{4}{*}{\rotatebox[origin=c]{90}{Macro-Pixel}}
& $2$  &  $98.99 $ \%  &  $ 2 $  & $ 99.22$ \%   & $ 96.8 \pm 1.4$ \% \\
& $3$  &  $98.28 $ \%  &  $ 3 $  & $ 98.77$ \%  & $ 95.3 \pm 0.4$ \% \\
& $5$  &  $94.26 $ \%  &  $ 5 $  & $ 96.38$ \%  & $ 91.6 \pm 0.6$ \% \\
& $7$  &  $94.76 $ \%  &  $ 7 $  & $ 90.40$ \%   & $ 83.0 \pm 1.2$ \% \\
\hline
\multirow{4}{*}{\rotatebox[origin=c]{90}{OAM}}
& $2$  &  $100.0 $ \%  &  $ 2 $  & $ 99.80$ \% & $ 97.5 \pm 0.2$ \%  \\
& $3$  &  $93.06 $ \%  &  $ 3 $  & $ 96.11$ \% & $ 90.8 \pm 0.6$ \%  \\
& $5$  &  $90.15 $ \%  &  $ 5 $  & $ 94.26$ \% & $ 86.5 \pm 0.7$ \%  \\
& $7$  &  $89.91 $ \%  &  $ 7 $  & $ 93.64$ \% & $ 80.6 \pm 0.6$ \%  \\
\hline
    
   \end{tabular}
\end{center}
\end{smallboxtable}

Next, we perform a similar prepare-and-measure experiment on the gate transformations to verify their performance independent of the input entangled state. As shown in Fig.~\ref{fig:prep_measure_fig}(b), a filtered superluminiscent diode is sent through the programmable circuit consisting of the multimode fiber sandwiched between two SLMs. In this experiment, $\text{SLM}_1$ is used to prepare the classical input states sent into the gates while also contributing to the construction of the gate operation. The output states are measured on a camera, which allows multiple outcomes to be measured simultaneously. 

We construct various optical circuits as described in Methods
and perform quantum process tomography (QPT) on these gate operations. In order to perform QPT on the gate operation, SLM$_2$ is used to display the output-mode holograms that correspond to local measurements made by Bob. The resultant data is then processed in the same manner as in AA-QPT, while choosing the input state to be the ideal maximally entangled one, as described in Methods.

Table~\ref{table:best_classical_gates} shows the performance of these gates. When compared with results shown in Table~\ref{table:best_FP_gates}, one sees that these gates perform better on classical input states than on an entangled input state. This can be attributed to the fact that we use the same classical source (identical wavelength and bandwidth) to characterise our fiber transmission matrix. In addition, slight differences in alignment can result in imperfect mode-matching between our input entangled state and classical light source.

\begin{smallboxtable}[float=h]{Measured fidelity $\mathcal{F}$ and purity $\mathcal{P}$ of inverse-designed gates using a classical prepare-and-measure experiment.}{best_classical_gates}
\begin{center}
    \begin{tabular}{c|c|c|c|c|c|c|c|c|c}
    \hline
      \multirow{2}{*}{Basis} & \multirow{2}{*}{$\mathbb{T}$} &
    \multicolumn{4}{c|}{$\mathcal{F}(\widetilde{\rho_{\xi*}},\rho_\xi)$}&
    \multicolumn{4}{c}{$\mathcal{P}(\widetilde{\rho_{\xi*}})$} \\
  \cline{3-10}
  & & $d = 2$ & $d = 3$ & $d = 5$  & $d = 7$ & $d = 2$ & $d = 3$ & $d = 5$  & $d = 7$ \\
  \hline
\multirow{5}{*}{\rotatebox[origin=c]{90}{Macro-Pixel}}
& $\mathbb{I}$& $98.4 $ \%& $98.7 $ \%& $94.7 $ \%& $84.7 $ \%& $97.4 $ \%& $100.0 $ \%& $98.2 $ \%& $86.0 $ \%\\
& $\mathbb{Z}$& $98.4 $ \%& $99.1 $ \%& $94.5 $ \%& $84.0 $ \%& $98.2 $ \%& $99.6 $ \%& $94.5 $ \%& $88.9 $ \%\\
& $\mathbb{X}$& $98.6 $ \%& $99.4 $ \%& $95.3 $ \%& $83.3 $ \%& $100.0 $ \%& $100.0 $ \%& $98.5 $ \%& $83.3 $ \%\\
& $\mathbb{F}$& $96.9 $ \%& $90.5 $ \%& $89.3 $ \%& $81.4 $ \%& $96.1 $ \%& $90.8 $ \%& $88.8 $ \%& $79.2 $ \%\\
& $\mathbb{R}$& $97.4 $ \%& $93.3 $ \%& $88.1 $ \%& $79.0 $ \%& $96.3 $ \%& $92.0 $ \%& $86.4 $ \%& $73.7 $ \%\\
\hline
\multirow{5}{*}{\rotatebox[origin=c]{90}{OAM}}
& $\mathbb{I}$& $96.8 $ \%& $91.3 $ \%& $81.3 $ \%& $66.7 $ \%& $93.8 $ \%& $87.5 $ \%& $82.5 $ \%& $71.8 $ \%\\
& $\mathbb{Z}$& $97.0 $ \%& $89.9 $ \%& $82.3 $ \%& $68.6 $ \%& $94.8 $ \%& $87.8 $ \%& $79.9 $ \%& $67.1 $ \%\\
& $\mathbb{X}$& $95.5 $ \%& $89.9 $ \%& $77.5 $ \%& $71.4 $ \%& $92.4 $ \%& $86.6 $ \%& $71.2 $ \%& $67.5 $ \%\\
& $\mathbb{F}$& $97.0 $ \%& $90.6 $ \%& $82.1 $ \%& $68.8 $ \%& $96.0 $ \%& $86.2 $ \%& $82.2 $ \%& $68.3 $ \%\\
& $\mathbb{R}$& $96.6 $ \%& $90.7 $ \%& $80.5 $ \%& $67.3 $ \%& $94.0 $ \%& $90.0 $ \%& $78.2 $ \%& $64.1 $ \%\\
\hline
    \end{tabular}
\end{center}
\end{smallboxtable}


\section{\label{Methods:SystematicErrorsQST} Errors in QST and AA-QPT due to the measurement apparatus}

\begin{figure}[htp]
    \centering
    \includegraphics[width=\linewidth]{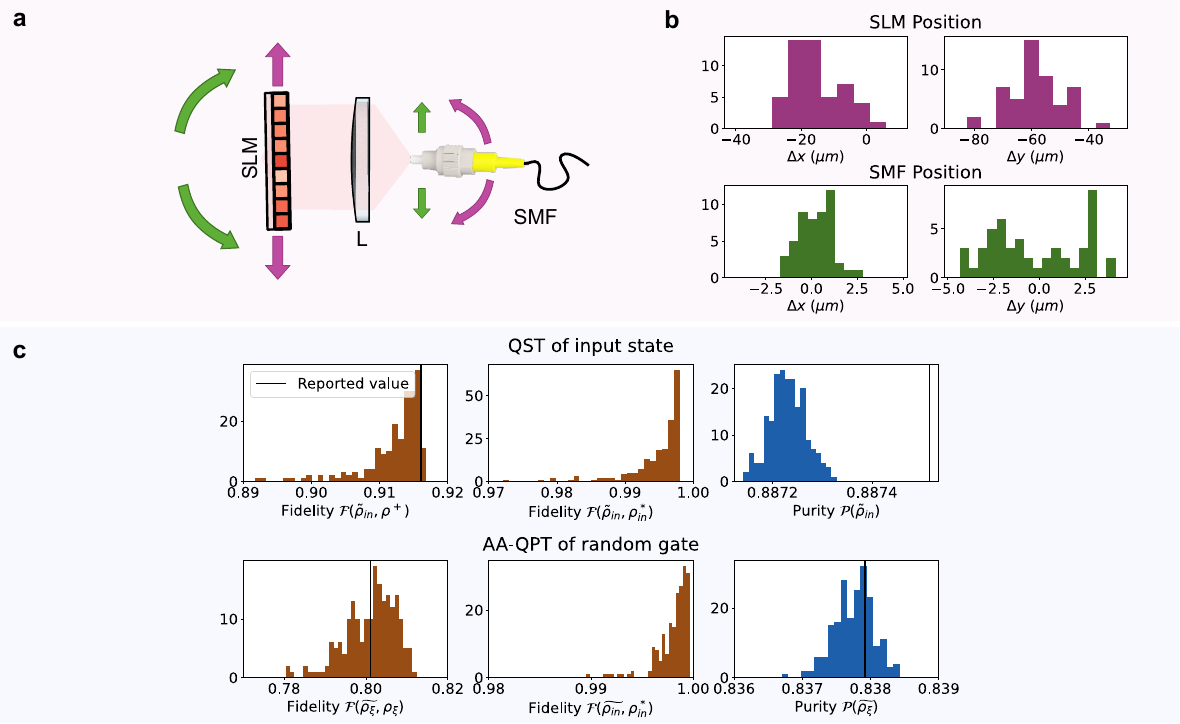}
    \caption{\textbf{Misalignment-induced systematic errors}: (a) Sources of misalignment in spatial single-outcome projective measurements performed with a spatial light modulator (SLM), lens (L), and a single-mode fiber (SMF). (b) Histograms of errors in positions of holograms displayed on the SLM and positions of the SMF. The misalignments are characterised using multiple prepare-and-measure experiments. (c) Histograms of errors propagating through QST and AA-QPT in fidelity and purity of states and processes. Here we show the example of a 5-dimensional entangled state and a random-unitary gate in the macro-pixel basis. The black lines indicate the values reported in the main experiment.}
      \label{fig:misal_err_concept}
\end{figure}

When performing QST and AA-QPT, errors in the measurement apparatus can lead to erroneous state reconstruction~\cite{Rosset2012}, which often tends to be the systematic type. As demonstrated above in~\ref{SI:PrepAndMeasure}, the measurements in our laboratory conditions are not perfect, so here we address the problem by using the above prepare-and-measure experiment to estimate the distribution of errors in our measurement settings, and use these to estimate the consequent errors in state and process reconstruction using QST and AA-QPT.

The prepare-and-measure experiment described in~\ref{SI:PrepAndMeasure} is repeated 50 times, following realignment procedures used in the actual experiment. The realignment consists of performing a knife-edge scan to find the center of incident beams on the SLM, as well as performing a raster-scan on the fiber facet by displaying different gratings on the SLM. In each trial, the tomography of measurements is used to fit model parameters pertaining to the experimental sources of misalignment, i.e.~tilt and displacement of projected optical fields on the SMF, as shown in  Fig.~\ref{fig:misal_err_concept}(a). The results of these fitted experimental parameters are presented in histograms  Fig.~\ref{fig:misal_err_concept}(b), which provide distributions of the experimental misalignments.
Given these distributions, we perform error propagation by numerically sampling from the ensemble of misalignments and calculating the resultant imperfect measurement projectors ($\widetilde{\hat{\Pi}}_i$), which correspond to the actual measurement errors that can occur in our experiment. 

To assess the resultant estimation errors on our QST/AA-QPT fidelity, fidelity to target state, and state purity, we simulate the effects of these measurement errors on our experimentally estimated state, $\widetilde{\rho_*}$, in the following manner. For QST, given an estimated state $\widetilde{\rho_*}$, in the $i$th trial we generate a sample measurement, $\widetilde{\hat{\Pi}}_i$, from the ensemble of misaligned measurements containing a set of tomographic measurement projectors.
We calculate the measured detection statistics of $\widetilde{\hat{\Pi}}_i$ on $\widetilde{\rho_*}$, and perform the QST reconstruction via SDP to obtain a recovered state estimate, $\widetilde{\rho_i}$. For AA-QPT, we first perform this sampling procedure to generate the input state, then repeat the procedure for Choi state estimation using that particular input state sample.
We repeat these steps 200 times, and record the QST/AA-QPT fidelities, $\{\mathcal{F}(\widetilde\rho_i,\widetilde\rho_*)\}_i$, the fidelities to target state $\{\mathcal{F}(\widetilde{\rho_i},\rho^{\text{target}})\}_i$, and the purities $\{\mathcal{P}(\widetilde{\rho_i})\}_i$. 
These distributions then correspond to the measurement errors of our actual state estimation procedure, providing that the ground-truth state (which we cannot know) undergoes similar reconstruction errors to the state estimate $\widetilde{\rho_*}$.

\begin{figure}[htp]
    \centering
    \includegraphics[width=\linewidth]{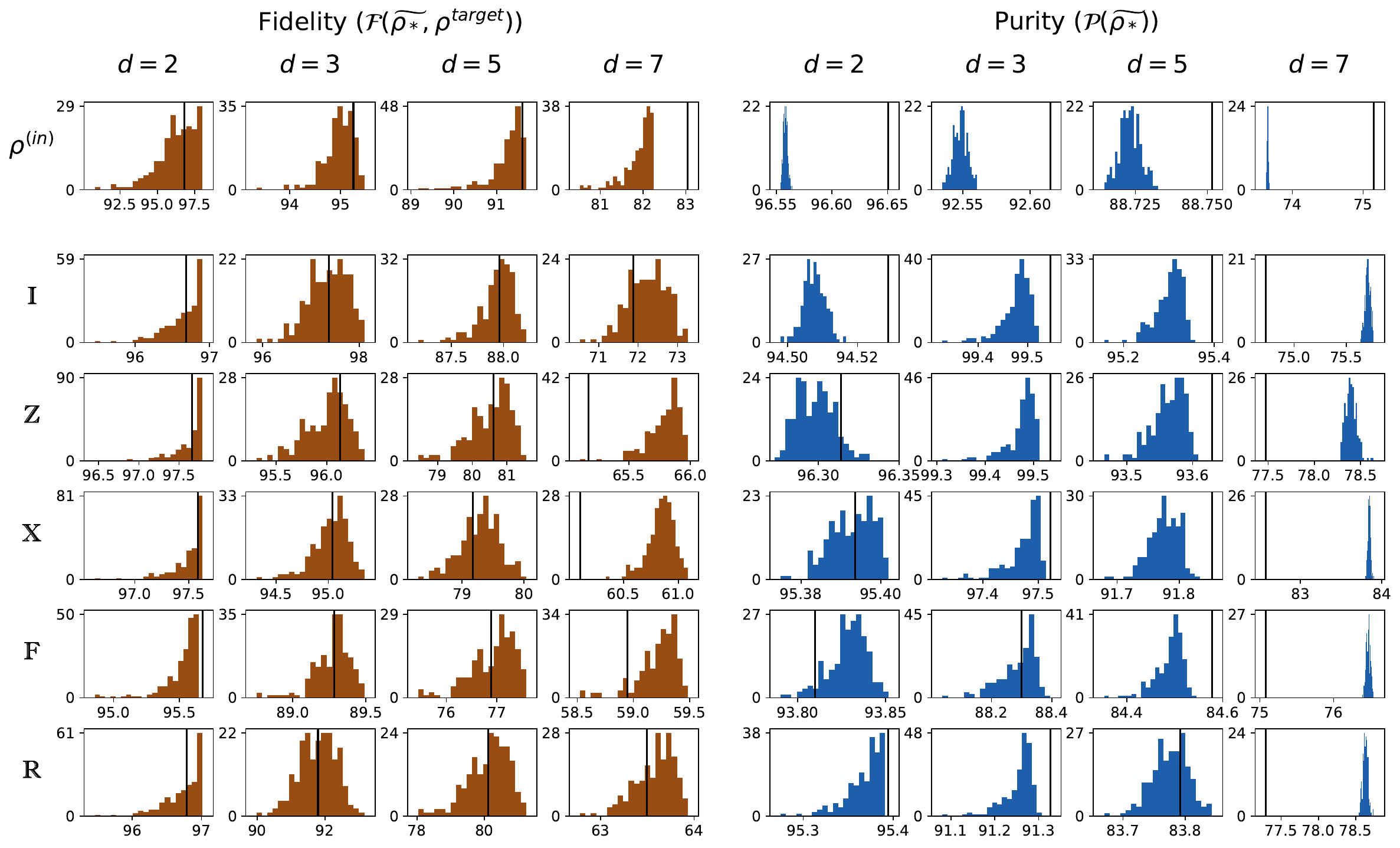}
    \caption{\textbf{Measurement error analysis for macro-pixel basis}: Histogram of fidelity and purity of reconstructed input state $\rho^\text{(in)}$ using QST as well as quantum gates reported in Table~\ref{table:best_FP_gates} using AA-QPT, with respect to the reported values (black line).}
      \label{fig:errors_fp}
\end{figure}

\begin{figure}[htp]
    \centering
    \includegraphics[width=\linewidth]{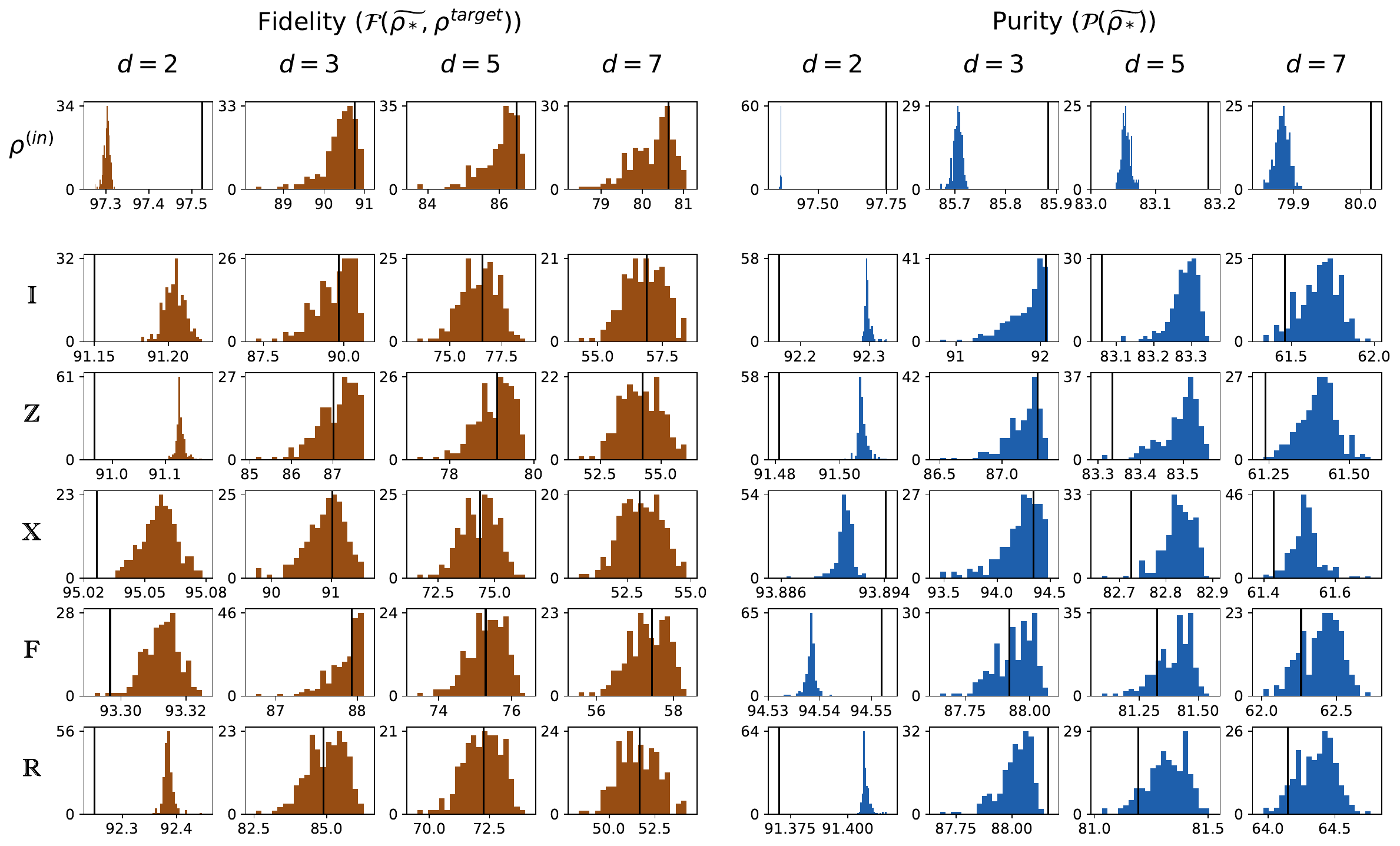}
    \caption{\textbf{Measurement error analysis for OAM basis}: Histogram of fidelity and purity of reconstructed input state $\rho^\text{(in)}$ using QST as well as quantum gates reported in Table~\ref{table:best_FP_gates} using AA-QPT, with respect to the reported values (black line).}
      \label{fig:errors_oam}
\end{figure}

The fidelity of QST (AA-QPT), the fidelity to the target (target Choi) state, and the purity are presented in Fig.~\ref{fig:misal_err_concept}(c), for QST of the 5-dimensional macro-pixel entangled input state and AA-QPT of the 5-dimensional Random gate.
These show representative behaviour of the sampling procedure. Complete results for each estimated state are presented in Fig.~\ref{fig:errors_fp} and Fig.~\ref{fig:errors_oam}. 
We also mark the corresponding property of the underlying experimental state estimate, $\widetilde{\rho_*}$, as black lines in Fig.~\ref{fig:misal_err_concept}(c).

For the 5-dimensional input state, we find that the fidelity of the QST procedure, exemplified by $\{\mathcal{F}(\widetilde{\rho_i},\widetilde{\rho_*})\}_i$, has sample mean fidelity 
$99.48\%$ indicating a slight infidelity of QST caused by the measurement error.
For the fidelity to the maximally entangled target state ($\{\mathcal{F}(\widetilde{\rho_i},\rho^{\text{target}})\}_i$), we observe that measurement errors almost exclusively lead to a reduction, demonstrating that the measurement errors likely cause systematically an underestimation of the true initial state quality. The standard error of this fidelity to target state, $\Delta_M=0.57\%$, is calculated with respect to the underlying state fidelity ($\mathcal{F}(\widetilde{\rho_*},\rho^{\text{target}})$), and we conservatively cite this as the symmetric error on our recorded fidelities in the main text. Similarly, the purity of the recovered state under these measurement errors ($\{\mathcal{P}(\widetilde{\rho_i})\}_i$) almost exclusively decreases (albeit very slightly), leading to a likely underestimation of state purity, though we again cite this standard deviation in the main text and Table~\ref{table:inputstate}, and these results are summarised in Table~\ref{table:allSystematicErrors}.

For AA-QPT, where both the input state and the recovered Choi state are sampled as above, we observe somewhat different behaviours. The fidelity of the AA-QPT procedure for a 5-dimensional random gate,$\{\mathcal{F}(\widetilde{\rho_{\xi i}},\widetilde{\rho_{\xi*}})\}_i$ has a sample mean of 
$99.81\%$ indicating reasonably stable process recovery given our measurement errors. 
The fidelity to target Choi state shows a broadened distribution featuring both over- and under-estimation of fidelity to target Choi state (unlike the QST case), with a standard deviation
$\Delta_M = 0.62\%$. The purity of the recovered Choi states is also broadened about the underlying Choi state purity $\mathcal{P}(\widetilde{\rho_{\xi*}})$. 
We cite the standard deviation with respect to the measured values in the main text and Table~\ref{table:best_FP_gates} and summarise in Table~\ref{table:allSystematicErrors}.

In some cases, particularly where the input state basis suffers more detrimental impurity caused by measurement errors than the output state, we can observe a slight systematic tendency to over-estimate Choi state fidelities and purities. This discrepancy between QST and AA-QPT behaviours can be explained by noting that an over-estimation of noise in an input state would lead to misattribution of output state noise to the input state and not to the Choi state, thus under-estimating Choi state noise. 
We demonstrate this effect symbolically when performing a direct inversion to solve for the Choi state in the case of white noise mixing of the input state in Supplementary~\ref{Methods:AAQPTNoise}.

\begin{smallboxtable}[float=h]{Results of errors from photon counting statistics and imperfect measurement settings on the input state, $\rho^\text{(in)}$ in QST and the Choi states of gates, $\rho_{\xi}$, in AA-QPT. We report the sample mean of the QST/AA-QPT fidelities ($\overline{\mathcal{F}(\widetilde{\rho_i},\widetilde{\rho_*})}$), and the one standard deviation Poisson error ($\Delta_P$) and measurement error ($\Delta_M$) of the fidelity to target state ($\mathcal{F}(\widetilde{\rho_*},\rho^\text{target})$) and purity ($\mathcal{P}(\widetilde{\rho_*})$).}{allSystematicErrors}

\begin{center}

    \begin{tabular}{c|c|c|c|c|c|c|c|c|c|c|c|c|c|c|c|c|c|c|c|c|c}
    \hline
       \multirow{3}{*}{\rotatebox[origin=c]{90}{Basis}} & \multirow{3}{*}{\rotatebox[origin=c]{90}{State}} &
       
        \multicolumn{4}{c|}{$\overline{ \mathcal{F}(\widetilde{\rho_i},\widetilde{\rho_*})}$ (\%)} &
          \multicolumn{8}{c|}{$\Delta \mathcal{F}(\widetilde{\rho_*},\rho^\text{target})$ (\%)} &
        \multicolumn{8}{c}{ $\Delta \mathcal{P}(\widetilde{\rho_*})$ (\%)} \\
        \cline{3-22}
     & & \multirow{2}{*}{$d = 2$} & \multirow{2}{*}{$d = 3$} & \multirow{2}{*}{$d = 5$}  & \multirow{2}{*}{$d = 7$} & 
     \multicolumn{2}{c|}{$d = 2$} & \multicolumn{2}{c|}{$d = 3$} & \multicolumn{2}{c|}{$d = 5$}  & \multicolumn{2}{c|}{$d = 7$} &
     \multicolumn{2}{c|}{$d = 2$} & \multicolumn{2}{c|}{$d = 3$} & \multicolumn{2}{c|}{$d = 5$}  & \multicolumn{2}{c}{$d = 7$} \\
    
    \cline{7-22}
    
    & & & & & & $\Delta_P$ & $\Delta_M$  & $\Delta_P$ & $\Delta_M$ & $\Delta_P$ &$\Delta_M$  & $\Delta_P$ &$\Delta_M$ & $\Delta_P$ & $\Delta_M$  & $\Delta_P$ & $\Delta_M$ & $\Delta_P$ &$\Delta_M$  & $\Delta_P$ &$\Delta_M$ \\
    \hline
\multirow{6}{*}{\rotatebox[origin=c]{90}{Macro-Pixel}} & $\rho^\text{(in)}$ 
  & $99.66$   & $99.72$   & $99.48$   & $96.64$   &  $0.12$ & $1.42$    &  $0.08$ & $0.41$    &  $0.06$ & $0.57$    &  $0.06$ & $1.20$  & $0.22$  & $0.09$    & $0.15$  & $0.07$    & $0.11$  & $0.03$    & $0.10$  & $1.50$    \\   \cline{2-22}
& $\rho_\mathbb{I}$  
& $99.89$ 
& $99.88$ 
& $99.80$ 
& $99.63$ 
& $0.85$  & $0.25$   
& $0.52$  & $0.41$   
& $0.68$  & $0.16$   
& $0.96$  & $0.60$   
& $0.45$  & $0.02$   
& $0.26$  & $0.07$   
& $0.39$  & $0.10$   
& $0.64$  & $0.98$   
\\
& $\rho_\mathbb{Z}$  
& $99.89$ 
& $99.89$ 
& $99.82$ 
& $99.59$ 
& $0.85$  & $0.24$   
& $0.45$  & $0.24$   
& $0.78$  & $0.61$   
& $0.75$  & $0.63$   
& $0.44$  & $0.02$   
& $0.25$  & $0.07$   
& $0.50$  & $0.07$   
& $0.48$  & $0.92$   
\\
& $\rho_\mathbb{X}$  
& $99.89$ 
& $99.89$ 
& $99.80$ 
& $99.56$ 
& $0.78$  & $0.18$   
& $0.67$  & $0.17$   
& $0.95$  & $0.33$   
& $0.67$  & $0.75$   
& $0.41$  & $0.01$   
& $0.36$  & $0.06$   
& $0.58$  & $0.08$   
& $0.38$  & $1.27$   
\\
& $\rho_\mathbb{F}$  
& $99.87$ 
& $99.90$ 
& $99.83$ 
& $99.70$ 
& $0.93$  & $0.21$   
& $0.84$  & $0.12$   
& $0.97$  & $0.44$   
& $0.66$  & $0.32$   
& $0.51$  & $0.02$   
& $0.49$  & $0.06$   
& $0.60$  & $0.09$   
& $0.43$  & $1.39$   
\\
& $\rho_\mathbb{R}$  
& $99.87$ 
& $99.89$ 
& $99.81$ 
& $99.70$ 
& $0.64$  & $0.27$   
& $0.54$  & $0.55$   
& $0.94$  & $0.62$   
& $0.71$  & $0.22$   
& $0.33$  & $0.03$   
& $0.29$  & $0.08$   
& $0.53$  & $0.03$   
& $0.39$  & $1.34$   
\\
\hline
\multirow{6}{*}{\rotatebox[origin=c]{90}{OAM}} & $\rho^\text{(in)}$ 
  & $99.78$   & $99.44$   & $99.19$   & $98.78$   &  $0.04$ & $0.22$    &  $0.04$ & $0.55$    &  $0.03$ & $0.69$    &  $0.02$ & $0.65$  & $0.08$  & $0.39$    & $0.06$  & $0.18$    & $0.06$  & $0.13$    & $0.05$  & $0.13$ \\
    \cline{2-22}
& $\rho_\mathbb{I}$  
& $99.98$ 
& $99.57$ 
& $99.65$ 
& $99.51$ 
& $0.39$  & $0.05$   
& $0.38$  & $0.61$   
& $0.41$  & $0.90$   
& $0.47$  & $0.79$   
& $0.19$  & $0.13$   
& $0.21$  & $0.32$   
& $0.24$  & $0.22$   
& $0.28$  & $0.24$   
\\
& $\rho_\mathbb{Z}$  
& $99.98$ 
& $99.63$ 
& $99.62$ 
& $99.55$ 
& $0.31$  & $0.16$   
& $0.38$  & $0.47$   
& $0.45$  & $0.46$   
& $0.53$  & $0.82$   
& $0.17$  & $0.03$   
& $0.20$  & $0.18$   
& $0.28$  & $0.17$   
& $0.30$  & $0.17$   
\\
& $\rho_\mathbb{X}$  
& $99.98$ 
& $99.56$ 
& $99.64$ 
& $99.53$ 
& $0.45$  & $0.03$   
& $0.39$  & $0.35$   
& $0.40$  & $0.81$   
& $0.44$  & $0.80$   
& $0.25$  & $0.00$   
& $0.22$  & $0.23$   
& $0.24$  & $0.11$   
& $0.26$  & $0.10$   
\\
& $\rho_\mathbb{F}$  
& $99.98$ 
& $99.73$ 
& $99.66$ 
& $99.53$ 
& $0.43$  & $0.02$   
& $0.43$  & $0.22$   
& $0.46$  & $0.50$   
& $0.44$  & $0.53$   
& $0.24$  & $0.01$   
& $0.24$  & $0.08$   
& $0.27$  & $0.10$   
& $0.25$  & $0.19$   
\\
& $\rho_\mathbb{R}$  
& $99.98$ 
& $99.68$ 
& $99.65$ 
& $99.58$ 
& $0.51$  & $0.14$   
& $0.42$  & $0.66$   
& $0.44$  & $0.78$   
& $0.48$  & $1.10$   
& $0.28$  & $0.04$   
& $0.22$  & $0.16$   
& $0.23$  & $0.15$   
& $0.24$  & $0.26$   
\\
\hline
    \end{tabular}
\end{center}
\end{smallboxtable}

\color{black}


\section{Misestimation of initial state in linear inversion}
\label{Methods:AAQPTNoise}
Perfect recovery of the underlying Choi state of a channel requires accurate estimation of the initial state. Suppose the true initial state given by Eqs.
8
were misestimated by,
\be
\begin{aligned}
\label{eqs:initialStateOperatorNoises}
\rho^\text{(in, noise)}(p)
&=\bigl(p \mathcal{A}\otimes \mathbb{I} +(1-p) \mathcal{B}\otimes \mathbb{I}\bigr) (\rho^+ )
\end{aligned}
\ee
where $\mathcal{B}$ is some additive noise. When attempting the linear inversion of the Choi state, $\rho_\xi$, it would be misestimated by, 
\be
\begin{aligned}
\label{eqs:ChoiEstimateinitialStateNoise}
\rho_{\xi*}(p)
&=\bigl(p \mathcal{A}\otimes \mathbb{I} +(1-p) \mathcal{B}\otimes \mathbb{I}\bigr)^{-1} \bigl(\mathcal{A}\otimes \mathbb{I}\bigr)\rho_\xi\\
\end{aligned}
\ee
which may be unphysical. In the case of additive white noise, $\mathcal{B}\otimes\mathbb{I}$ is rank one (in the sense that its matrix representation when acting on vectorised states has rank one). We use the Sherman–Morrison formula to arrive at,
\be
\begin{aligned}
\label{eqs:ChoiEstimateinitialStateNoiseWhite}
\rho_{\xi*}(p)
&=\bigl(p \mathcal{A}\otimes \mathbb{I} +(1-p) \mathcal{B}\otimes \mathbb{I}\bigr)^{-1} \bigl(\mathcal{A}\otimes \mathbb{I}\bigr) (\rho_\xi )\\
&=\biggl(\tfrac{1}{p} \mathcal{A}^{-1}\otimes \mathbb{I} 
-\tfrac{(1-p)}{p^2}g
( \mathcal{A}^{-1}\otimes \mathbb{I})
(\mathcal{B}\otimes \mathbb{I})
(\mathcal{A}^{-1}\otimes \mathbb{I})
\biggr) \bigl(\mathcal{A}\otimes \mathbb{I}\bigr) (\rho_\xi )\\
&=\tfrac{1}{p} \rho_\xi 
-\biggl(\tfrac{(1-p)}{p^2} g
 (\mathcal{A}^{-1}\otimes \mathbb{I})
(\mathcal{B}\otimes \mathbb{I})
\biggr)  (\rho_\xi )\\
\text{where}\quad g&:= \tfrac{1}{1+\tfrac{1-p}{p}\tr[\mathcal{B}\otimes \mathbb{I}\mathcal{A}^{-1}\otimes \mathbb{I}]}
\end{aligned}
\ee
If the initial state were maximally entangled, $\mathcal{A}=\mathbb{I}$, then the second term in Eqs.~\ref{eqs:ChoiEstimateinitialStateNoiseWhite} subtracts (potentially filtered) white noise, $\mathcal{B}\otimes \mathbb{I} (\rho_\xi )$, from the true Choi state, leading to a resultant over-estimation of the channel purity. Assessing the systematic errors presented by misestimation of our initial state in our tomographic protocols is the subject of Supplementary~\ref{Methods:SystematicErrorsQST}

\color{black}


\section{\label{SI:Results_2} Programmability and scalability of top-down designed circuits}
To investigate the programmability and scalability of our approach, we numerically construct many instances of the high-dimensional circuits by varying our design parameters, i.e., dimension of circuit $d$, dimension of mode-mixers $n$, depth of circuit $L$, using the model
\be
    \mathbf{T}:=U_{L+1}\prod_{j=1}^{L}P_j U_j,
    \label{Eq:model}
\ee
where the $n$ reconfigurable phase elements at each $P_j$ are found using the wavefront-matching algorithm as explained in Methods
given a set of $n\times n$ unitary matrices $\{U_j\}_j$ and a set of $d$ input and output modes of interest. The codes for these simulations are available in~\cite{Goel_Simulation_Codes_for_2022}. Once a circuit is implemented, we classify its performance by measuring fidelity (Eq.~12
) and success probability (Eq.~13
). For a given set of design parameters $n$, $L$ and $d$, we implement 
multiple 
realisations of the Identity $\mathbb{I}$, Pauli-$\mathbb{X}$, Pauli-$\mathbb{Z}$, Fourier-$\mathbb{F}$, and random unitary $\mathbb{R}$ gates (Eq.~6
), by varying the set of $d$ input and output modes of interest which are randomly selected. We simulate these circuits by treating the mode-mixers $\{U_j\}_j$ as either a set of random unitaries or discrete Fourier transforms (DFT). 

\begin{figure}[htp]
    \centering
    \includegraphics[width=\linewidth]{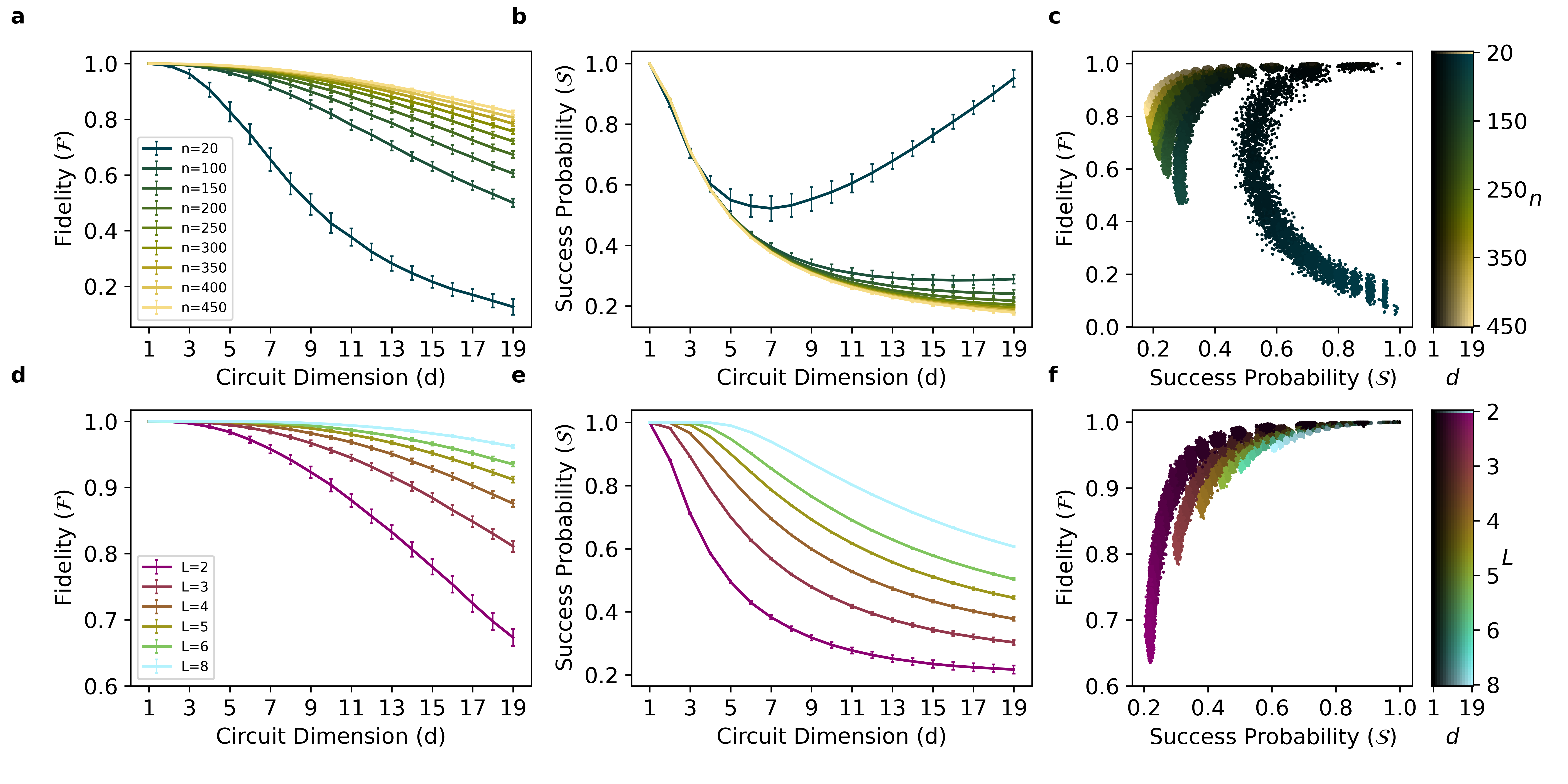}
    \caption{\textbf{Simulated fidelity $\mathcal{F}$ and success probability $\mathcal{S}$ for all gates implemented in a circuit with random unitary mode-mixers}: (a-c) $\mathcal{F}$ and $\mathcal{S}$ as a function of the dimension of mode-mixers, $n$, with a circuit depth $L=2$ and (d-f) $\mathcal{F}$ and $\mathcal{S}$ as function of circuit depth $L$ while using mode-mixers with dimension $n=200$. A sample size of 200 is used to derive the statistics of each data-point. The data is presented as mean values $\pm1$ standard deviation.}
      \label{fig:scaling_si}
\end{figure}

As presented in the main text, when the dimension of the circuit, $d$, is less than the dimension of mode-mixers, $n$ ($d/n<0.1$), both fidelity and success probability reduce as the circuit dimension increases. Fig.~\ref{fig:scaling_si} conveys these trends in detail. For all these realisations, all random mode-mixers $\{U_j\}$ correspond to the same random unitary matrix, and the positions of input/output modes of interest are randomly selected for each implementation. The behaviour reported in Fig.~\ref{fig:scaling_si} is also observed either when different random mode-mixers are used in each layer, or input/output modes are specifically assigned to the first $d$ modes of the $U_j$. Interestingly, the same trends for fidelity and success probability are observed for all gate types, indicating that there is no advantage when programming specific types of gates using this method. The same holds for both DFT and random mode-mixers with randomly selected inputs and outputs, indicating that a DFT is similar to a random unitary in this regime.

For a circuit described by Eq.~\ref{Eq:model}, the total number of degrees of freedom (DOFs) is given by the number of reconfigurable elements $nL$. On the other hand, the constraint on the number of DOFs required to program a $d$-dimensional circuit $\mathbb{T}$ is $\mathcal{O}(d^2)$~\cite{Borevich1981,Miller2013a}. A larger $d$ leads to loss of performance because the number of DOFs required to program the circuit increases. Thus, as $d$ increases, the number of reconfigurable elements does not change. Improving the figures of merit of a $d$-dimensional circuit can therefore be addressed in two primary ways: increasing the dimension of mode-mixers, $n$, or the depth of circuit, $L$. 

\begin{figure}[htp]
    \centering
    \includegraphics[width=\linewidth]{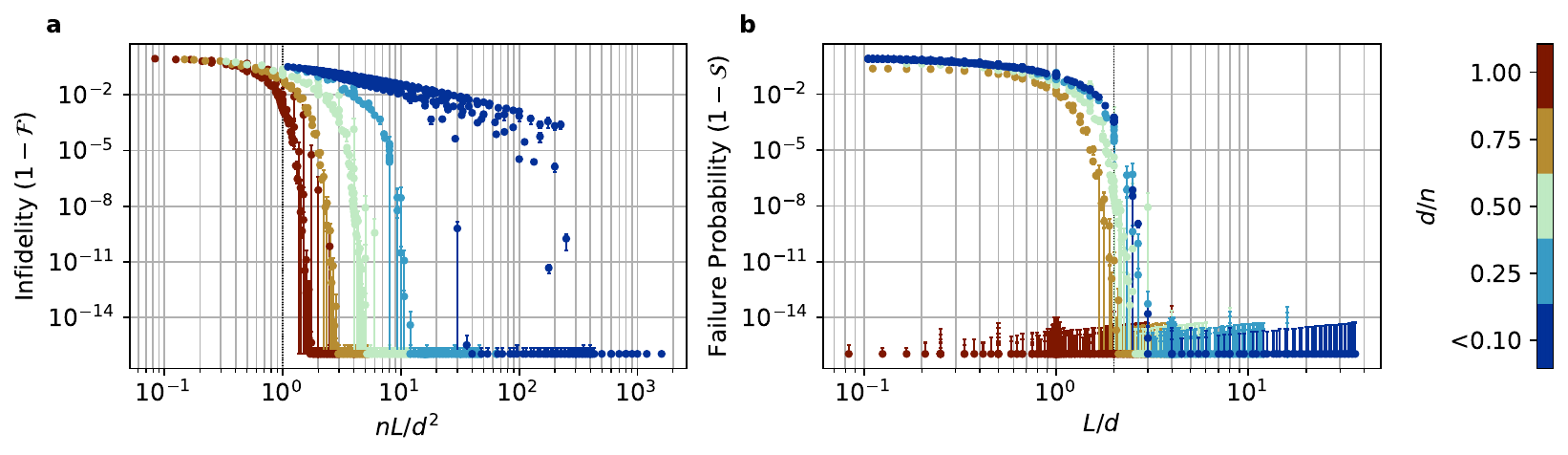}
    \caption{Infidelity $1-\mathcal{F}$ (a) and failure probability $1-\mathcal{S}$ (b) for simulated gates implemented in a circuit with random-unitary mode-mixers. Minimum values are capped at $10^{-16}$ since this corresponds to the floating-point precision in our simulation.A sample size of 96 is used to derive the statistics of each data-point. The data is presented as mean values $\pm1$ standard deviation.}
      \label{fig:scaling_si2}
\end{figure}

To study the region where the fidelity and success probability converge to unity, we have plotted our simulation data on a log-log plot (Fig.~\ref{fig:scaling_si2}), replacing fidelity with infidelity ($1-F$) and success probability with failure probability ($1-S$). As shown in Fig.~\ref{fig:scaling_si2}a, the infidelity converges to zero (capped at $10^{-16}$ due to the floating-point precision used) when the total number of reconfigurable elements exceeds the number required to program a unitary operation ($\mathcal{O}(d^2)$). The point at which $(1-F)\xrightarrow[]{}0$ depends on the ratio of $d/n$. In the case where $d=n$, the number of reconfigurable elements required is optimal, and the infidelity rapidly converges to zero. In the regime where $d/n<1$ however, the number of elements necessary for full programmability increases as $d/n$ decreases. For instance, when $d/n<0.1$ the infidelity converges to zero slower as a function of the number of elements (normalised by dimension). In Fig.~\ref{fig:scaling_si2}b, we observe that the failure probability decreases as a function of the circuit depth and rapidly converges to zero  (capped at $10^{-16}$) when $L\approx\mathcal{O}(d)$ in all cases of $d\leq n$. In this regime, full programmability is achieved and is seen to be relatively independent of $d/n$. Note that the failure probability for $d=n$ is always zero.

These results show that in order to achieve both high fidelity and success probability, one does not need to solely rely on the number of reconfigurable elements, but also on their distribution across the circuit. Ideally, one would want to implement a fully programmable circuit using the full dimension of mode-mixers ($d=n$) constructed from $2\times2$ beam splitters or $n\times n$ mode-mixers, since this achieves the minimum number of reconfigurable elements~\cite{Reck1994,Clements2016}. However, it is known that in the presence of imperfections, an increase in the depth of the circuit is required to improve the fidelity in such schemes~\cite{Miller2015,Burgwal2017,Pai2018}. Scaling the depth of circuit in practice may nonetheless not be the most viable option due to a variety of experimental reasons, namely, propagation and interface losses and accumulation of errors. Instead of considering errors and scattering losses as a problem, the top-down approach presents a reasonable alternative ($d<n$ and $L\lesssim \mathcal{O}(d)$) where randomness is exploited to implement a circuit with a low depth. Here, both fidelity and success probability are acceptably high, while $n$ can increase to improve the fidelity without reducing the success probability from unity when $L>2d$, or with a trade-off on the success probability when $L\leq2d$. Many optical devices, for instance, computer-generated holograms and MPLC devices~\cite{Morizur2010,Boucher2020} can be considered to fall in this category where mode-mixing can be implemented by free-space propagation, a $2f$ lens system, or another transformation such as a circulant matrix~\cite{Schmid2000,Huhtanen2015}.

Finally, while the deterministic construction of programmable circuits from discrete Fourier transforms is known for $d=n$~\cite{Idel2015,LopezPastor2021}, a proof of universal programmability of unitary circuits using random-unitary mode-mixers remains open to the best of our knowledge, and the deterministic construction of such circuits is yet to be discovered.

 \section{\label{SI:Results_RandomandSym}Effects of randomness and symmetry} 
A pertinent point to address is the role that randomness plays in the implementation of an optical circuit using the top-down approach. For instance, instead of having a random unitary evolution between phase planes, one could have another transformation such as a discrete Fourier transform (DFT) or a circulant matrix, which normally have a symmetric structure~\cite{Schmid2000,Huhtanen2015}. In practice, such transformations can be difficult to implement on an integrated platform, normally requiring a high degree of accurate control. Due to inherent imperfections and bending, a commercial multi-mode fiber is much closer to a random matrix in our basis of interest than a symmetric transformation. In this section, we investigate the effect of randomness on the gates performance, by comparing circuit implementations using random unitary mode-mixers ($\{U_j\}_j=R$) with ones using discrete Fourier transforms (DFT) ($\{U_j\}_j=F$), given that the set of input/output modes of interest are fixed to be the first $d$ modes of the $U_j$. This is to be contrasted with the randomly selected input/output modes in~\ref{SI:Results_2}, where there is no difference in performance between circuits with random and DFT mode-mixers. 

\begin{figure}[htp]
     \centering
     \includegraphics[width=\linewidth]{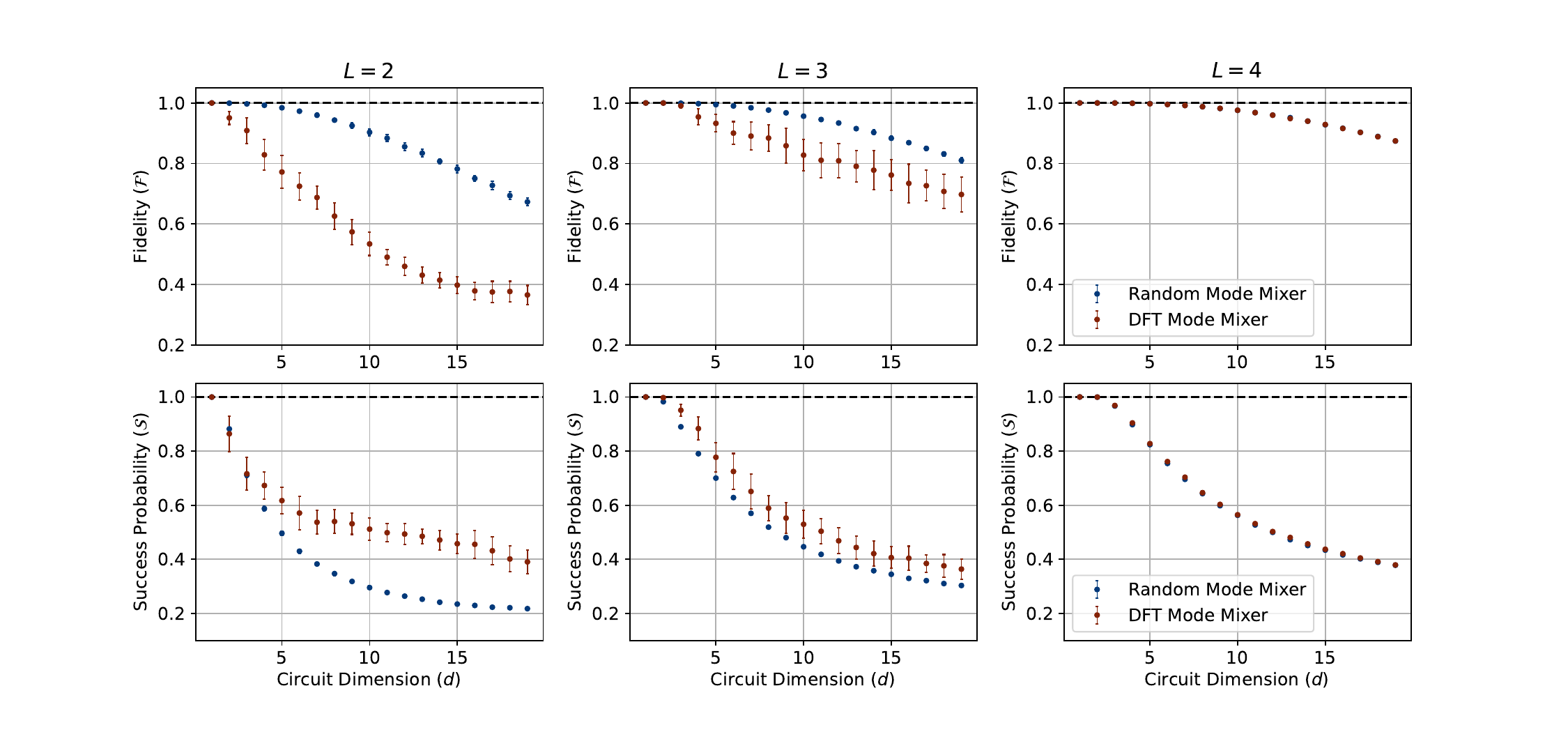}
     \caption{\textbf{Comparison of fidelity ($\mathcal{F}$) and Success probability ($\mathcal{S}$) of random gates implemented with random-unitary mode-mixers ($U_j=R$) and DFT mode-mixers ($U_j=F$) with dimension $n=200$, for different number of layers ($L=2,3,4$).} A sample size of 27 is used to derive the statistics of each data-point. The data is presented as mean values $\pm1$ standard deviation.}.
     \label{fig:symmetry_si}
 \end{figure}

The fidelity and success probability of random gates of different circuit dimension using a random unitary mode-mixer and a DFT mode-mixer are compared in Fig.~\ref{fig:symmetry_si}. We observe that for a low number of circuit layers ($L<4$), the random unitary mode-mixer performs better in terms of fidelity. However, this comes with a lower success probability than achievable with a DFT mode-mixer. At $L\geq 4$, there is no difference in performance between the two types of circuit, as in this regime, the combination of several DFTs and phase planes starts approaching a random structure itself. In general, one can expect that implementing circuits with symmetry will be better with a mode-mixer that has some inherent symmetry. For example, a DFT transformation is certainly easier to perform with a DFT mode-mixer. However, as pointed out above, such mode-mixers can be difficult to implement in practice as compared with random unitary mode-mixers.

It is known that a symmetry present on a unitary $U$ reduces the number of free parameters, consequently resulting in fewer reconfigurable elements needed in their implementation. For example, the number of optical elements required for $(d=2^{k})$-dimensional discrete Fourier transform is $(d/2)\log_{2}d$ where $k$ is a positive integer~\cite{Torma1996,Barak2007}. The use of symmetric mode-mixers thus brings both pros for implementing a particular circuitry and cons for full programmablity. Regardless, random mode-mixers are easier to find and use in practice, simply requiring a system with some inherent randomness such as an imperfect multi-mode fiber. In addition, they are more suitable for implementing a fully programmable circuit, and provide the added benefit of a large ambient mode-mixer dimension, which further increases circuit fidelity.

\color{black}

\end{document}